\documentclass{article}
\usepackage{graphicx,epsfig}
\usepackage{graphicx,subfigure}
\usepackage{arxiv}
\usepackage{amsmath}
\usepackage{amssymb}
\usepackage[utf8]{inputenc} 
\usepackage[T1]{fontenc}    
\usepackage{url}            
\usepackage{booktabs}       
\usepackage{amsfonts,bm}       
\usepackage{microtype}      
\usepackage[mathscr]{euscript}
\usepackage[dvipsnames]{xcolor}
\pdfoutput=1
\usepackage{hyperref}
\hypersetup{
    colorlinks=true,
    linkcolor=black,
    citecolor=black,
    filecolor=black,
    urlcolor=black,
}
\makeatother
\usepackage[numbers, compress]{natbib}
\title{Odd-viscosity induced surfactant-laden shear-imposed viscous film over a slippery incline}
\author{
  Md. Mouzakkir Hossain\\
  Department of Mathematics\\
  Indian Institute of Technology Jodhpur\\
  Rajasthan-342037, India\\
  \texttt{mouzakkir123@gmail.com} \\
\And
  Sukhendu Ghosh\\
  Department of Mathematics\\
  Indian Institute of Technology Jodhpur\\
  Rajasthan-342037, India\\
  \texttt{sukhendu.math@gmail.com} \\
     \And
  Harekrushna Behera*\\
  Center of Excellence for Ocean Engineering\\
  National Taiwan Ocean University\\
  Keelung 202301, Taiwan\\
  \texttt{hkb.math@gmail.com} \\
}

\begin{document}
\maketitle

\begin{abstract}
This research focuses on the stability analysis of an odd viscosity-induced shear-imposed Newtonian fluid flowing down an inclined slippery bed having an insoluble surfactant at the top of the liquid surface. 
The Orr-Sommerfeld boundary value problem is developed by applying the normal mode approach to the infinitesimal perturbed fluid flow and solved using the numerical method Chebyshev spectral collocation. The numerical results confirm the existence of Yih mode  and Marangoni mode  in the longwave zone. For the clean/contaminated surface of the film flow, the presence of an odd or Hall viscosity coefficient reduces the surface wave energy and delays the transition from laminar to perturbed flow. Also, it has stabilizing nature on the unstable Marangoni mode as well. The growth rate of both clean and contaminated liquid surfaces becomes more/less when the stronger external shear acts along the downstream/upstream direction of fluid flow. Further, the slip parameter leads to a lower critical Reynolds number and makes the liquid surface more unstable.
An increase in the critical Reynolds number due to the stronger Marangoni force ensures that the insoluble surfactant has the potential to dampen the Yih mode instability. 
Moreover, the unstable shear mode occurs in the finite wavenumber regime with very high inertial force and a small angle of inclination. The two-fold variation of the shear mode instability is possible with respect to the imposed shear. However, the inclusion of the odd viscosity coefficient in the viscous falling film may advance the shear mode instability.

\end{abstract}


\section{Introduction}
The dynamics and stability of a gravity-driven falling liquid film have received considerable attention in the last few decades due to its significant application in coating \cite{weinstein2004coating}, manufacture of photographic films in different technological applications \cite{han1980multiphase}, etc.
Hydrodynamic stability analysis of a viscous falling film over an inclined substrate was first developed by \citet{yih1963stability} and \citet{benjamin1957wave}. Theoretically, they derived a marginal boundary of the primary instability of the surface mode by using the longwave expansion method in the Orr-Sommerfeld boundary value problem. Later on, \citet{lin1967instability} and \citet{de1974stability} further extend the work of \citet{yih1963stability} and \citet{benjamin1957wave}, and identified the existence of shear mode with very high inertia force and small inclination angle.
They observed the competition between surface and shear modes to generate the primary instability of the film flow system.
In the last few decades, so many researchers have shown a strong interest in developing control systems to either improve or reduce the free surface flow instabilities of a single-layer film due to their vast applications in geophysical, industrial, and technical set-ups (\cite{tuck1983continuous}, \cite{allen1984collocation}, \cite{alekseenko1994wave}, \cite{weinstein2004coating}, \cite{zhao2019faraday}). Surface-active substances usually play an influential part in these applications in enhancing the stability of the liquid layer (see \cite{grotberg2004biofluid}, \cite{levy2007gravity}, \cite{anjalaiah2015effect}, \cite{blyth2004effect}). The attention is brought on by the numerous ways that single-layer fluid flow might be used to solve problems in the real world.  

The earlier studies mainly focused on the even viscosity part of the viscosity tensor in the hydrodynamic stability analysis. Even though the non-vanishing odd viscosity part in the viscosity tensor has major importance in the film flow dynamics such as in two-dimensional quantum Hall
fluid with an external magnetic field (\cite{avron1998odd}) or in swimming strategies (\cite{samanta2022role}). In the past, these arguments had been made in relation to plasmas ( for more information regarding this, see   \citet{berestetskii1982quantum}). In demand to stabilize the flow of a system, odd viscosity is essential. Even though odd viscosity is necessary to stabilize a flow system, there are fewer studies on it in the literature than there are on even viscosity. \citet{lapa2014swimming} were the first to analyze the nearly circular swimmers in two-dimensional liquids induced by odd viscosity in the case of a small Reynolds number. They observed that the odd viscosity part has a significant effect on the hydrodynamic stress tensor. 
Based on the previous investigation on the quantum Hall impact, \citet{avron1998odd} investigated the fundamental properties of odd viscosity by taking into account the solutions of the wave and Navier-Stokes equations for fictional fluids where the stress is dominated by the odd viscosity. He confirmed that the viscosity tensor can be expressed as the summation of symmetric (known as even viscosity) and antisymmetric (renamed as odd viscosity) components in a classical liquid, which breaks the reversal symmetry of time. 
Once both time-reversal and parity are broken, odd viscosity is universally non-vanishing generically (see \cite{kaminski2014nonrelativistic} and \cite{jensen2012parity}). For example, a nonzero odd viscosity can be produced by microscopic Coriolis or Lorentz forces (see \cite{chapman1990mathematical}). \citet{souslov2019topological}  illustrated how the topological characteristics of linear disturbance in the flow system are substantially influenced by the odd viscosity coefficient. The authors depicted that the odd viscosity and the external magnetic field on either side of an interface precisely rely on the net number of chiral edge states (\cite{jia2022effect}, \cite{gjevik1970occurrence}). 
Also,  the time-reversal symmetry of a classical liquid is frequently broken in nature, for example, in biological processes (see  \cite{sumino2012large}, \cite{klein2005complement}, \cite{tsai2005chiral}, \cite{livingston2017dementia} for more information and the other references therein). It is believed that this new behaviour will have a considerable impact on the dynamics of such systems.

In the recent time, so many researchers (\cite{kirkinis2019odd}, \cite{chattopadhyay2021influence}, \cite{bao2021odd}, \cite{mukhopadhyay2021thermocapillary}, \cite{chattopadhyay2022effect}, \cite{jia2022effect}, \cite{ chu2022effect}, \cite{mukhopadhyay2022surface}, \cite{paul2023hydrodynamic}) explored related works to odd viscosity induced viscous falling film under several fields of studies. \citet{kirkinis2019odd} considered viscous thin film surrounded by ambient gas phase over a uniformly heated rigid bed driven by surface shear stress. They derived an evolution equation of the liquid-gas interface and confirmed that the odd-viscosity suppresses the  thermocapillary instability. 
Later, \citet{bao2021odd} modeled a viscous liquid system with an odd viscosity coefficient having an externally imposed uniform normal electric field. Their results based on the linear and weakly nonlinear stability analysis predicted that the destabilization promoted by odd viscosity can be weakened by the external electric field. 
Further, \citet{mukhopadhyay2021thermocapillary} explored the characteristic of thermocapillary instability generated by an odd viscosity instigated viscous fluid over a uniformly heated inclined plane. A constant temperature gradient was presumed at the liquid surface. By performing the linear and weakly nonlinear stability analyses, they deduced that the odd viscosity incipient waves have the potential to suppress the thermocapillary instability and yield stabilization of the film flow. 
\citet{chattopadhyay2022effect} studied the effect of odd viscosity on the surface wave instability of a falling film on a moving vertical wall. They derived the nonlinear evolution equation of liquid thickness using the asymptotic expansion method and confirmed the significant stabilizing impact of odd viscosity on the liquid surface. 
Recently, \citet{samanta2022role} studied the dynamics of surface and shear waves triggered by the gravity-driven viscous liquid having the non-vanishing odd viscosity stress. He solved the Orr-Sommerfeld eigenvalue problem of the fluid flow model and demonstrated that the odd viscosity coefficient can reduce the surface and shear wave instability. Also, he confirmed that the odd viscosity coefficient can delay the transition process from the unidirectional base flow to the nonlinear perturbed wave, which is due to the reduction in the nonlinear wave speed and maximum amplitude.

In this paper, our idea is to extend the work of \citet{samanta2022role} by introducing an insoluble surfactant attached to the liquid surface to reduce the instability generated by the externally imposed shear force and also to explore the impact of the odd viscosity coefficient on the Marangoni instability other than the surface and shear waves instability. The extensive applications of surfactant-laden film flow in diverse domains of traditional technology and bio-medical flows (\cite{lin1970stabilizing}, \cite{whitaker1966stability}, \cite{ji1994instabilities}, \cite{halpern1998theoretical}) have encouraged our research. In the last few decades, several researchers used the insoluble surfactant and external shear at the liquid surface as a control strategy of a viscous fluid flow instability (\cite{wei2005effect}, \cite{samanta2014shear}, \cite{bhat2019linear}, \cite{samanta2021instability}, \cite{hossain2022shear}, \cite{hossain2022linear}) to study the wave dynamics of surfactant-laden falling liquid in an airway as occurring in airway occlusion process \cite{otis1993role, halpern1993surfactant} or bolus-dispersal surfactant replacement therapy \cite{halpern1998theoretical, espinosa1999bolus}.
The insoluble surfactants are extremely useful in delaying the transition from unperturbed unidirectional parallel flow to perturbed flow generated by imposed shear (\citet{bhat2019linear}). On the other hand, external shear force plays a significant role in controlling the instability mechanisms of fluid flow by altering the net driving force when the gravitational force acting on the bulk of the fluid becomes very weak. 
In the case of external shear-imposed liquid down an inclined plane with an insoluble surfactant at the liquid surface, \citet{bhat2019linear} found that the external shear causes the primary instability of Marangoni mode (surfactant mode). They observed three various temporal modes and investigated the physical mechanism of shear-induced surface mode instability based on the Energy budget analysis method (\cite{hooper1983shear}, \cite{kelly1989mechanism}).
\citet{hossain2022linear} used external shear at the free surface falling film over a porous medium, whereas an insoluble surfactant is considered at the fluid surface. The imposed shear in the downstream/upstream direction contributes to the surface mode instability/stability, and the clean surface is more unstable than the contaminated surface. Also, they confirmed that a thicker porous bed down to the fluid flow may advance the surface wave instability. More recently, \citet{paul2023hydrodynamic} studied an externally shear-imposed, viscous falling film over an inclined plane with broken time-reversal symmetry. They used both analytic and numerical techniques to solve the fluid flow model. It was shown that the odd-viscosity coefficient helps to stabilize the surface wave instability and has the capability to control the sturdy waves instigated by the external shear force. 
      
Due to the significant importance of the insoluble surfactant and imposed shear, we have been motivated to model a surfactant-laden externally shear imposed falling film down a slippery incline when reversal symmetry of time breaks in the viscous stress. The choice of a slip boundary condition at the liquid-bottom interface mainly comes from many
practical circumstances \cite{jeevahan2018superhydrophobic} in which the bottom bed is not perfectly smooth (i.e., slippery). For a more precise prediction of fluid flow instability, the addition of surface roughness is required. Darcy's law (\cite{beavers1967boundary}, \cite{pascal1999linear}) is used to model the fluid flow over a slippery bottom. 
The presence of odd viscosity introduces a new parameter in the pressure gradient of the fluid. In this considered fluid flow model, three modes of different characteristics are identified such as surface mode related to the deflection of the liquid surface, surface surfactant mode (Marangoni mode) related to the deflection of surfactant concentration at the fluid surface, and the shear mode, which exits due to high Reynolds number along with small inclination angle. Our main objective is to examine the induced odd viscosity coefficient's impact on the wave dynamics generated by the different modes and also to explore whether the imposed shear behaves differently in the different modes if the odd viscosity coefficient is present in the viscous falling film. 
To do this, we have derived the Orr-Sommerfeld boundary value problem (OS BVP) using normal mode form to the perturbed variables in the linear perturbation equation of the fluid flow problem. Then, the numerical technique of Chebyshev collocation is used to solve the OS BVP, and different types of results are plotted for different ranges of flow parameters by fixing the other parameters.

In view of these, the present study is arranged as follows: A set of the exact average equations for the odd viscosity induced by externally imposed shear surfactant-laden falling film is described in Section \ref{MF}. In Section \ref{OS BVP}, the OS BVP is computed using normal mode analysis in the perturbation equation. The possible characteristic behaviour of different types of modes is elaborately discussed in Section \ref{RAD}. Finally, the important outcomes are concluded in Section \ref{CON}. 

\section{Mathematical Formulation}  \label{MF}
\begin{figure}[ht!]
    \centering
   \includegraphics[width=12cm]{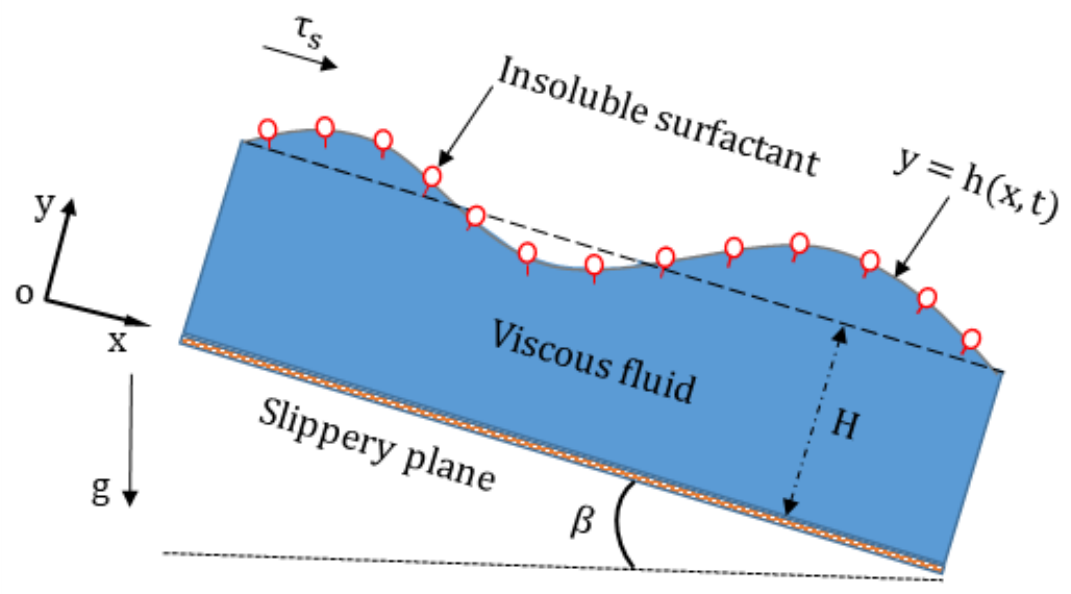} 
    \caption{Schematic diagram of an odd viscosity induced shear imposed falling film over a slippery  incline having an insoluble surfactant at the liquid surface.}
    \label{fig1}
\end{figure}

We have considered a fluid flow model in a two-dimensional Cartesian coordinate system, as shown in  Fig.~\ref{fig1}, where a gravity-driven incompressible and viscous Newtonian fluid flows over a slippery inclined plane of angle $\beta$ with the horizontal line. Here, the $x-$ axis is placed at the inclined liquid-bottom interface, and the $y-$ axis is taken perpendicular to the $x-$ axis, where $x$ and $y$ axes, respectively, indicate the streamwise
and cross-stream directions of fluid flow direction. Also, it is assumed that the interaction of the fluid surface with external shear stress $\tau_s$ acts along both directions of the fluid flow. Here, $H$ is assumed as the thickness of the unperturbed fluid of density $\rho$. In this work, we have assumed the viscous fluid breaks the reversal symmetry of time, which yields another viscosity coefficient, renamed as odd or Hall viscosity, to the Cauchy stress tensor.

As a result, the Cauchy stress tensor $ \mathcal{T}$ can be expressed as the sum of two components: 
\begin{equation}
    \mathcal{T}= \mathcal{T}^{e}+\mathcal{T}^{o}. \label{eq1}
\end{equation}
where $ \mathcal{T}^{e}$ defines the usual (even) component of the Cauchy stress tensor with the typical (even) viscosity $\mu_{e}$ and $ \mathcal{T}^{o}$ represents the odd component of the Cauchy stress tensor with the odd (Hall) viscosity $\mu_o$ that arises when time-reversal symmetry-breaking emerges. The above Eq.~\eqref{eq1} can be rewritten in two dimensions (\cite{avron1998odd}, \cite{samanta2022role}) as follows: 
\begin{align}
& \mathcal{T}^{e}_{ij}=-p\delta_{ij}+\mu_{e}\left( \frac{\partial u_{i}}{\partial x_{j}}+\frac{\partial u_{j}}{\partial x_{i}} \right) ~~~i,j=1,2, \label{even}
\\
& \mathcal{T}^{o}_{ij}=-\mu_{o}\left[ (\delta_{i1}\delta_{j1}-\delta_{i2}\delta_{j2} )\left( \frac{\partial u_{1}}{\partial x_{2}}+\frac{\partial u_{2}}{\partial x_{1}} \right)-(\delta_{i1}\delta_{j2}+\delta_{i2}\delta_{j1} )\left( \frac{\partial u_{1}}{\partial x_{1}}-\frac{\partial u_{2}}{\partial x_{2}} \right)\right] ~~~i,j=1,2,\label{odd}
\end{align}
where $\delta_{ij}$ refers to the Kronecker delta. Here $u_i$ is the velocity component in the $x_i$ direction. In this problem, $(x_1, x_2)$  and $(u_1,u_2)$ are chosen as $(x,y)$ and $(u,v)$, respectively.

\noindent The dimensionless form of the modified equation of motion of the considered fluid flow system may be represented as (\cite{samanta2022role}, \cite{chattopadhyay2022effect},   \cite{paul2023hydrodynamic}): 
\begin{align}
&u_x+v_y=0, \label{ee1}\\
&Re~(u_t+u~u_x+v~u_y)=-p_x+2+(u_{xx}+u_{yy})-\mu(v_{xx}+v_{yy}), \label{ee2}\\
&Re~(v_t+u~v_x+v~v_y)=-p_y-2\cot\beta+(v_{xx}+v_{yy})+\mu(u_{xx}+u_{yy}), \label{ee3}
\end{align}
where stream-wise and cross-stream velocity components of fluid flow are symbolized by $u$ and $v$, respectively, $p$ denotes the fluid pressure, and $g$ signs the acceleration due to gravity. The mean film thickness $H$ and the maximum average velocity $\displaystyle V = \frac{\rho g H^2 \sin\beta}{2\,\mu_e}$ are selected as the scales for the nondimensionalization of length and velocity, respectively, $\frac{H}{V}$ as the time scale, and $\frac{\mu_e V}{H}$ is taken as the pressure scale. Further, the new term $\displaystyle \mu=\frac{\mu_o}{\mu_e}$ in the equations of motion is the ratio of odd to even viscosity coefficient and renamed it as odd or Hall viscosity coefficient, whereas the $Re =\frac{\rho H V}{\mu_e}$ marks the Reynolds number.  
The kinematic boundary condition at the fluid surface yields
\begin{align}
    &v=h_{t}+u~h_{x} ~~~\text{at}~~~~y=h(x,t),\label{ee4}
\end{align}
The dynamic condition possesses the stress balance at the fluid surface. The tangential stress (\cite{wei2005effect}, \cite{samanta2014shear}, \cite{ samanta2022role}) balances the external shear stress and results 
\begin{align}
&\biggl[(1-h^2_{x})(u_{y}+v_{x})-2(u_x-v_y)h_x\biggr]+\mu\biggl[(u_x-v_y)(1-h^2_x)+2(u_{y}+v_{x})h_x\biggr] \nonumber\\ &~~~~~~~~~~~~~~~~~~~~~~~~~~~~~~~~~~~~~~~~~~~~~=(-Ma/Ca\,\Gamma_x+\tau)\sqrt{1+h_x^2} ~~~\text{at}~~~~y=h(x,t),
\end{align}
where  $\displaystyle \tau=\frac{ \tau_s H}{\mu_e V} $ is the dimensionless imposed shear. The positive value of $\tau$ ($+\tau$) marks the external shear directed towards the fluid flow, while the negative $\tau$ ($-\tau$) signifies the imposed shear directed towards the backflow. Moreover, the form of the normal stress jump boundary condition (\cite{anjalaiah2013thin}, \cite{bhat2020linear}, \cite{sani2020effect}) at the liquid surface is 
 \begin{align}
&p=\frac{1}{(1+h^2_{x})}\biggl[2\left\{u_{x}h^2_x-(u_y+v_x)h_{x}+v_y\right\}+\mu\left\{(u_y+v_x)(1-h_x^2)-2(u_x-v_y)h_x\right\}\biggr] \nonumber\\
&~~~~~~~~~~~~~~~~~~~~~~~~~~~~~~~~~~~~~~~~~~~~~-1/Ca\biggl[1-Ma(\Gamma-1)\biggr]\frac{h_{xx}}{(1+h^2_{x})^{\frac{3}{2}}}~~~\text{at}~~~~y=h(x,t). 
\end{align}
Here the dimensionless parameters $Ma=\frac{E~T_0}{\sigma_0}$ and $Ca=\frac{\mu V}{\sigma_0}$, respectively, represent the Marangoni and Capillary number, where $\sigma_0$ denotes the surface tension of the base flow. 
At the free surface of the fluid, an insoluble surfactant of concentration $\Gamma(x,t)$ is presumed to occupy the free surface of the fluid.
\noindent The evolution equation of the surfactant concentration $\Gamma (x,t)$ is described by the  concentration equation (\cite{blyth2004effect}, \cite{stone1990simple})
\begin{equation}
\Gamma_t+u~\Gamma_x+\Gamma\biggl(u_x+u_yh_x\biggr)=\frac{1}{Pe}\Gamma_{xx}.
\end{equation}
Here the dimensionless parameter $\displaystyle Pe=\frac{VH}{D_s}$ is called the P\'eclet number, where $D_s$ is the surface surfactant diffusivity.
The local surface tension $\sigma (x,t)$ fluctuates according to the local surface concentration at each specific position on the free surface of the fluid and is dependent on the surfactant concentration. It is important to note here that the surface tension and surfactant concentration are linearly related (\cite{bhat2019linear}, \cite{bhat2020linear}) with $\sigma(x,t)=\sigma_0-RT(\Gamma-\Gamma_0)$ (called, Langmuir isotherm linear relation), where $\Gamma_0$ represents the base surfactant concentration, $R$ is the universal gas constant, and $T$ is the absolute temperature \cite{frenkel2017surfactant}. 
Further, the bottom substrate is assumed to be slippery, so there is the Navier slip and no penetration of the liquid at the liquid-bottom interface (\cite{pascal1999linear}, \cite{samanta2011falling}). This yields the boundary conditions at $y=0$ as
\begin{align}
    &u=\alpha\,u_y \quad \text{and}\quad v=0.\label{ee6}
\end{align}
where $\alpha=\frac{\kappa}{H}$ is the slip parameter with $\kappa$ is the dimensional velocity slip length.

\subsection{The Base flow and stability solutions} 
\begin{figure}[ht!]
    \begin{center}
        \subfigure[]{\includegraphics*[width=7.4cm]{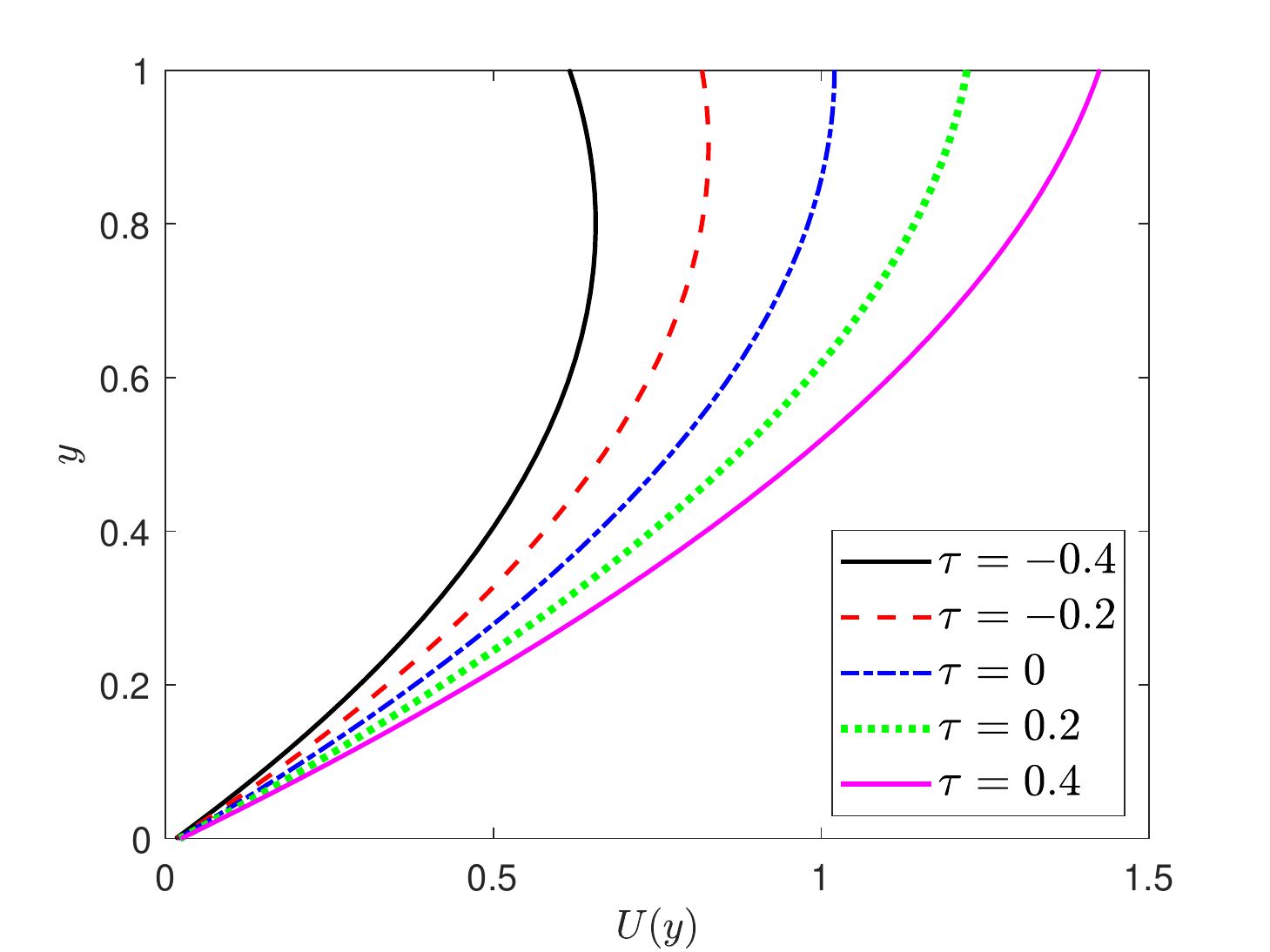}}
         \subfigure[]{\includegraphics*[width=7.4cm]{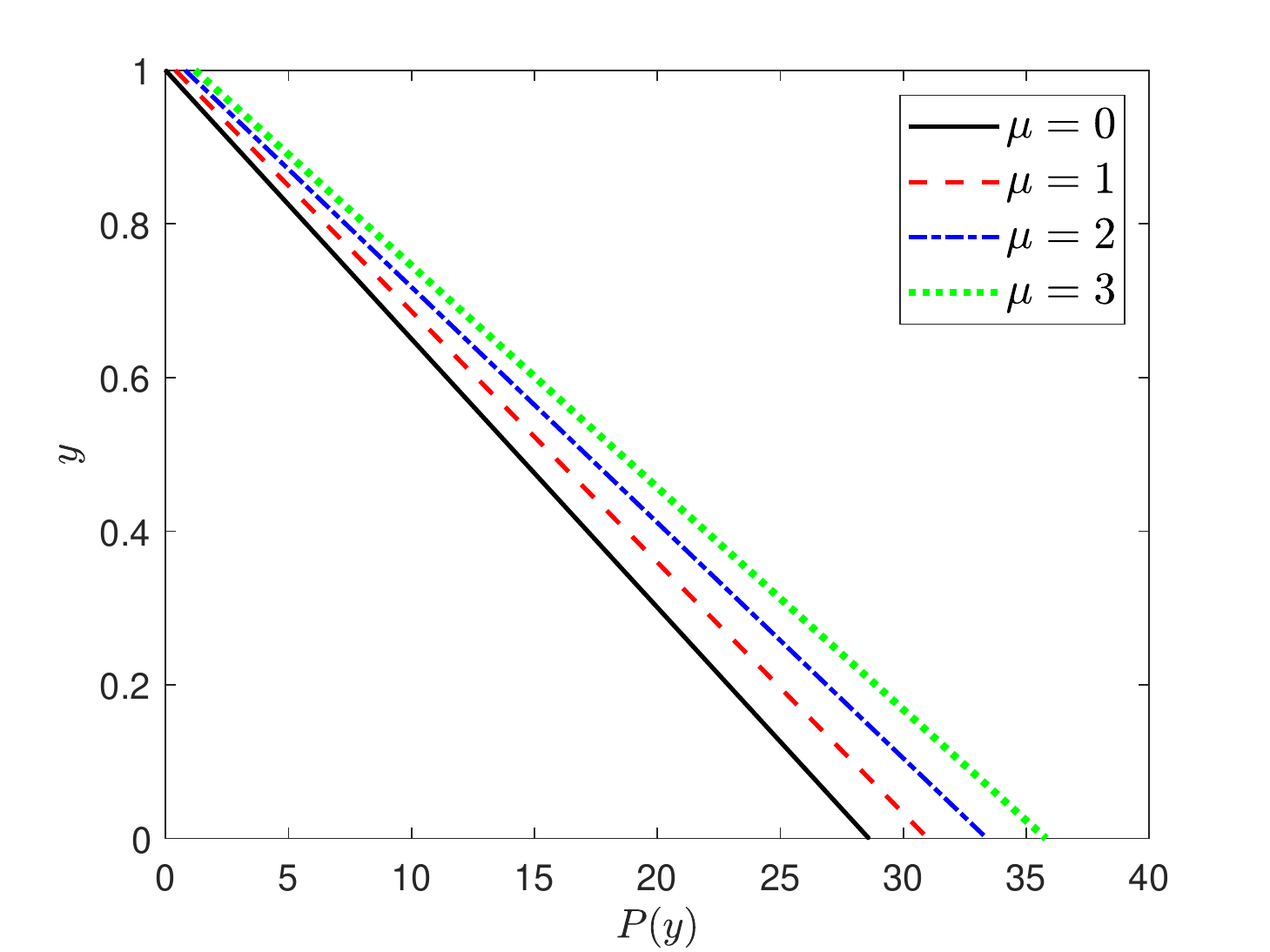}}
       \end{center}\vspace{-0.5cm}
    \caption{ The variation of (a) base velocity with $\mu=1$ when $\tau$ alters and (b) base pressure  with $\tau=0.4$ when $\mu$ alters. The other fixed parameters are $\alpha=0.01$ and $\beta=4^{\circ}$. }\label{fig2}
\end{figure}

To examine the linear instability behaviour of the considered flow system, the unidirectional parallel flow (base flow) solution is mandatory. The basic state solution is computed by considering the steady variables $ u=U(y)$, $v=0$, and $p=P(y)$ in the dimensionless equations of motion as well as the boundary conditions (Eqs.~\eqref{ee1}-\eqref{ee6}). Here, $U(y)$ and $P(y)$ designate the base velocity and pressure, respectively. The following base state solutions are derived by solving the resulting equations of motion associated with the boundary conditions corresponding to the base flow, which are of the form 
\begin{align}
&U(y)=-y^2+(2+\tau)( y+\alpha) \quad \text{and} \quad P(y)=2(\mu+\cot\beta)(1-y)+\mu\tau.\label{base}
\end{align}
Notably, the base velocity of the liquid film depends on both external force $\tau$ and slip parameter $\alpha$ but not on the odd viscosity coefficient $\mu$, while the base pressure depends on both $\tau$ and $\mu$. It is important to highlight that the imposed shear has no effect on the base pressure in the case of viscous fluid flow when the reversal symmetry of time does not break. (i.e., the absence of odd viscosity term in the viscosity (\cite{samanta2014shear}, \cite{ bhat2019linear}, \cite{sani2020effect}). But, the base pressure of the odd viscosity induced viscous falling film is significantly influenced by the external shear (see Eq.~\eqref{base}).
The variation of base velocity $U(y)$ against imposed shear (Fig.~\ref{fig2}(a)) displays that the external shear significantly affects the base velocity near the free surface. The externally imposed shear along the downstream/upstream direction advances/restricts the base velocity of the liquid. On the other hand, Fig.~\ref{fig2}(b) depicts that the basic pressure becomes maximum near the bottom substrate, and the higher odd viscosity randomly enhances the hydrostatic pressure of the fluid. Also, the presence of external shear causes some changes in base pressure at the liquid surface with respect to the odd viscosity (see Eq.~\eqref{base}).   

\section{The modified Orr-Sommerfeld equation} \label{OS BVP}
\noindent The linear stability analysis of the fluid flow model is conducted to understand the linear response of the infinitesimal perturbation of the base flow with respect to different fluid characteristic parameters. To do this, the base flow solution is perturbed with an infinitesimal disturbance, where each dimensionless flow variables      
 ($u,~ v,~ p,~ h, \Gamma$)=   ($ U(y) + \tilde{u},~ \tilde{v},~ P(y) +  \tilde{p}, ~1 +  \tilde{h},~ 1+\tilde{\Gamma}$), where each perturbed variables $\tilde{u}$, $\tilde{v}$, $\tilde{p}$, $\tilde{h}$, $\tilde{\Gamma}$ $<<1$ . The normal mode form of the perturbed variables are ($\tilde{u}, \tilde{v}, \tilde{h}, \tilde{\Gamma}$)= ($\phi'(y), -\mathrm{i}k\phi(y), \eta, \gamma$)$ \exp{[\mathrm{i}k(x-ct)]}$, where $\phi$ defines the stream function amplitude, the symbol $'$ signifies the derivative operator $\frac{d}{dy}$, $k$ represents the wavenumber in the streamwise direction, and $c=c_r+ic_i$ is the complex wave speed. The linear instability/stability can be examined by checking the exponential growth/decay of the perturbation.   
 The following Orr-Sommerfeld system is obtained by substituting the normal mode form of the perturbed variables in the perturbed fluid flow equation (for more details, see \cite{yih1963stability}, \cite{anjalaiah2013thin}, \cite{samanta2014effect} ) as
 \begin{align}	
&(\phi^{''''}-2k^2\phi^{''}+k^4\phi)-\mathsf{i}kRe\big[(U-c)~(\phi^{''}-k^2\phi)-U^{''}\phi\big]=0,\label{j1}\\
&\phi+(U-c)\eta=0~~~\mbox{at}~~y=1,\\
&(\phi^{''}+k^2\phi+2\mathsf{i}k\mu\phi^{'})+(U^{''}+2\mathsf{i}k \mu U^{'})\eta+\mathsf{i}k\, Ma/Ca\,\gamma=0~~~~\mbox{at}~~~~y=1,\\
&\phi^{'''}-3k^2\phi^{'}-\mathsf{i}kRe\big[(U-c)\phi^{'}-U^{'}\phi\big]-2\mathsf{i}k^3\mu\,\phi+\mathsf{i}k\bigg[2\mathsf{i}kU^{'}-2\cot\beta-k^2/Ca\bigg]\eta=0 ~~~~\mbox{at}~~~~y=1,\\
&\phi^{'}=\alpha\,\phi^{''} \quad \text{and} \quad \phi=0~~~~\mbox{at}~~~~y=0,\\
& \phi^{'}+\biggl[U-c-\mathsf{i}k/Pe\biggr]\gamma+U^{'}\eta=0~~~~\mbox{at}~~~~y=1.\label{j2} 
\end{align}
Here, $'$ symbol defines the derivative with respect to $y$. This system of ordinary differential Eqs.~\eqref{j1}--\eqref{j2} describes the generalised eigenvalue problem, named the OS BVP, where the generalised eigenvector $\phi$ corresponds to the generalised eigenvalue $c$. Moreover, the temporal frequency of the two-dimensional perturbation is defined by $\omega=\omega_r+\mathrm{i}\omega_i=k(c_r+ic_i)$, where $\omega_i=kc_i$ defines the temporal growth rate of the linear disturbance. The fluid flow becomes unstable if $\omega_i$ is positive, stable if $\omega_i$ is negative, and marginal (i.e., neither stable nor unstable) if $\omega_i$ is equal to zero. Notably, if the viscosity ratio is absent and the bottom surface is assumed to be smooth, then the above OS BVP (Eqs.~\eqref{j1}--\eqref{j2}) reduces to the OS BVP for an external shear-imposed contaminated viscous film down an inclined plane considered by \citet{bhat2019linear}. Also, the above Eqs.~\eqref{j1}--\eqref{j2} exactly match with the corresponding Orr-Sommerfeld eigenvalue problem  derived by \citet{samanta2022role}, when the liquid surface is free from surfactant (i.e., $Ma=0$), the imposed shear is absent (i.e., $\tau=0$), and the bottom substrate is considered as smooth plane (i.e., $\alpha=0$).

\section{Results and discussion} \label{RAD}
\begin{figure}[ht!]
    \begin{center}
        \subfigure[]{\includegraphics*[width=7.4cm]{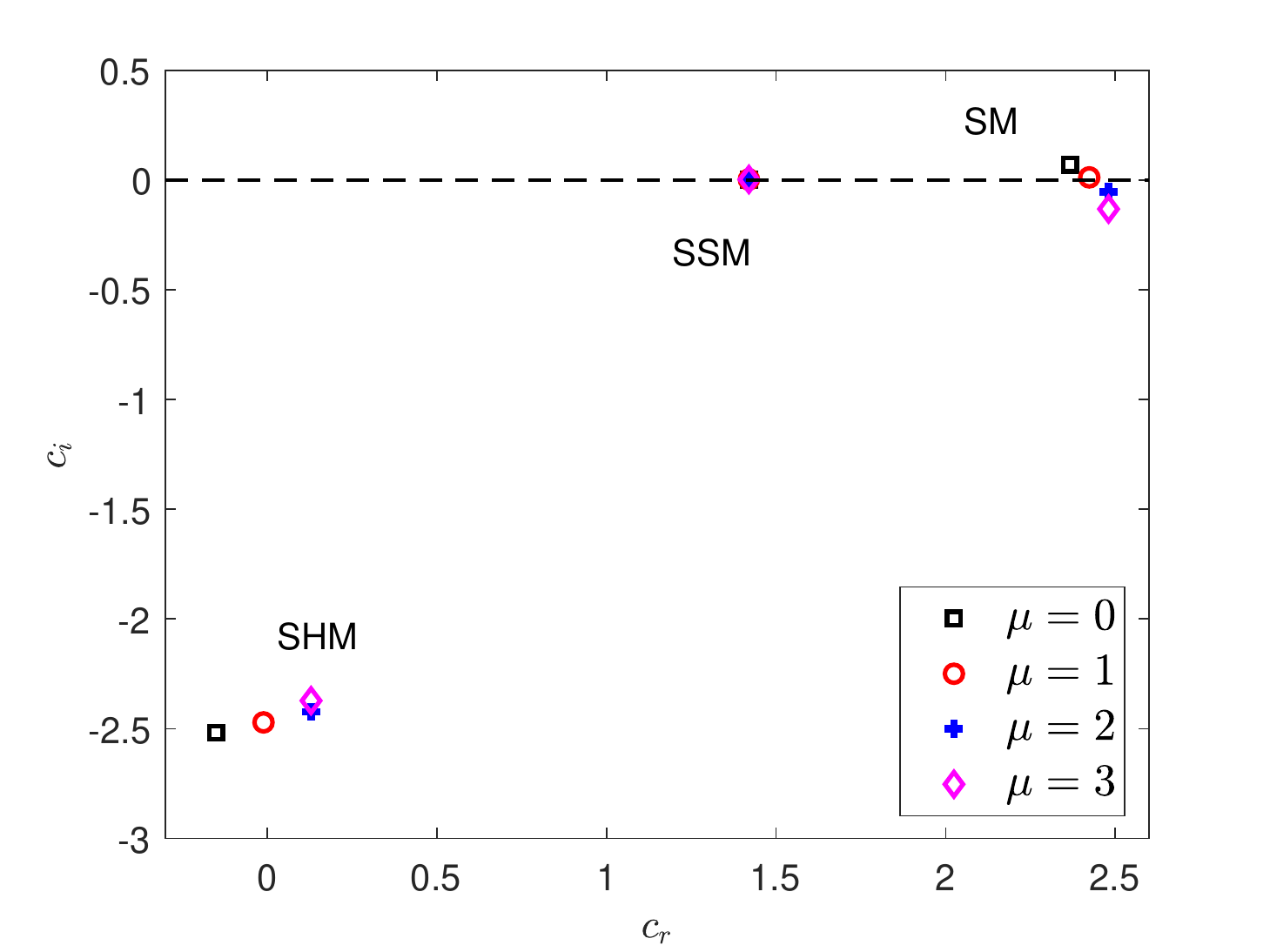}}
         \subfigure[]{\includegraphics*[width=7.4cm]{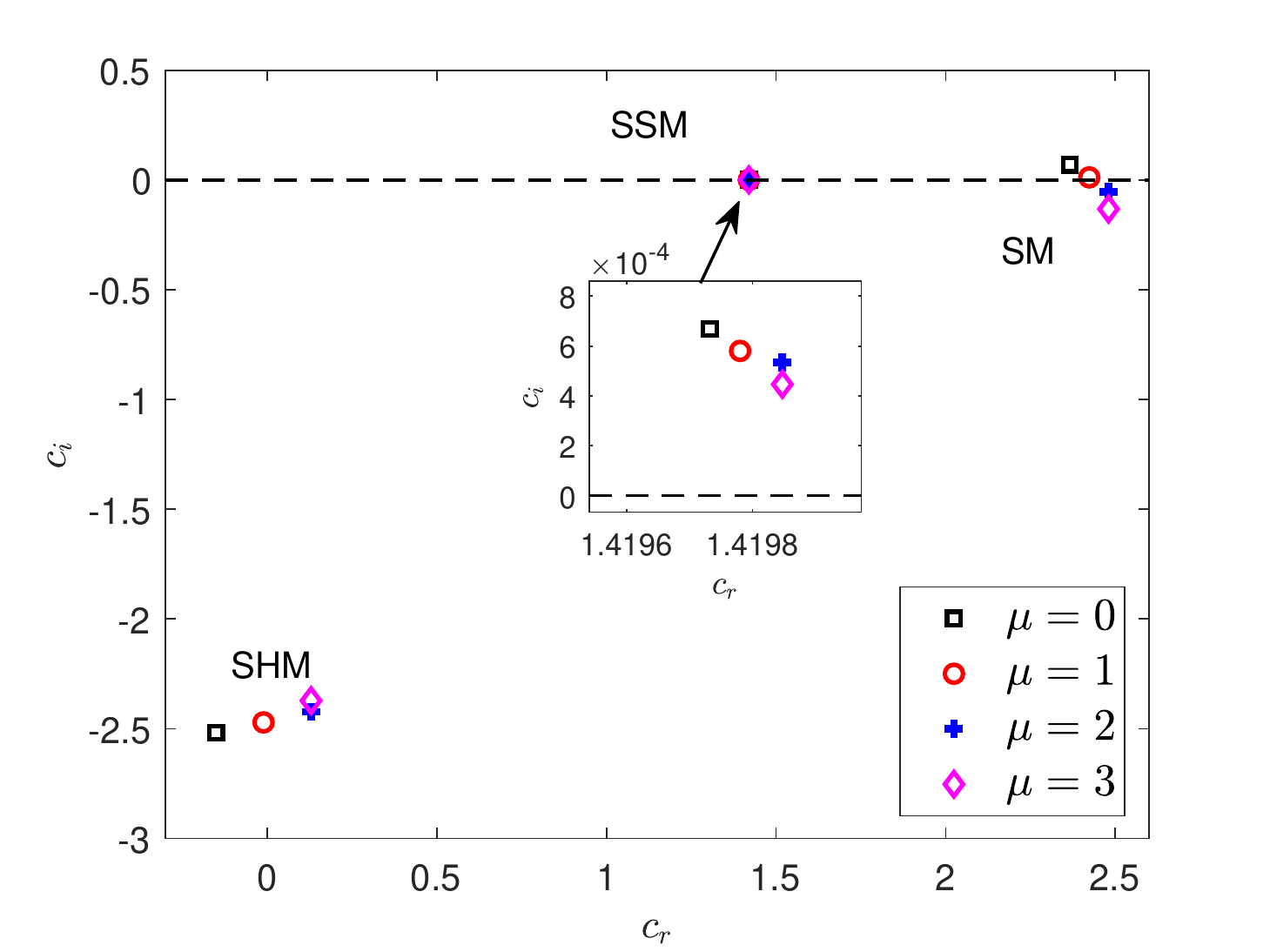}}
         \subfigure[]{\includegraphics*[width=7.4cm]{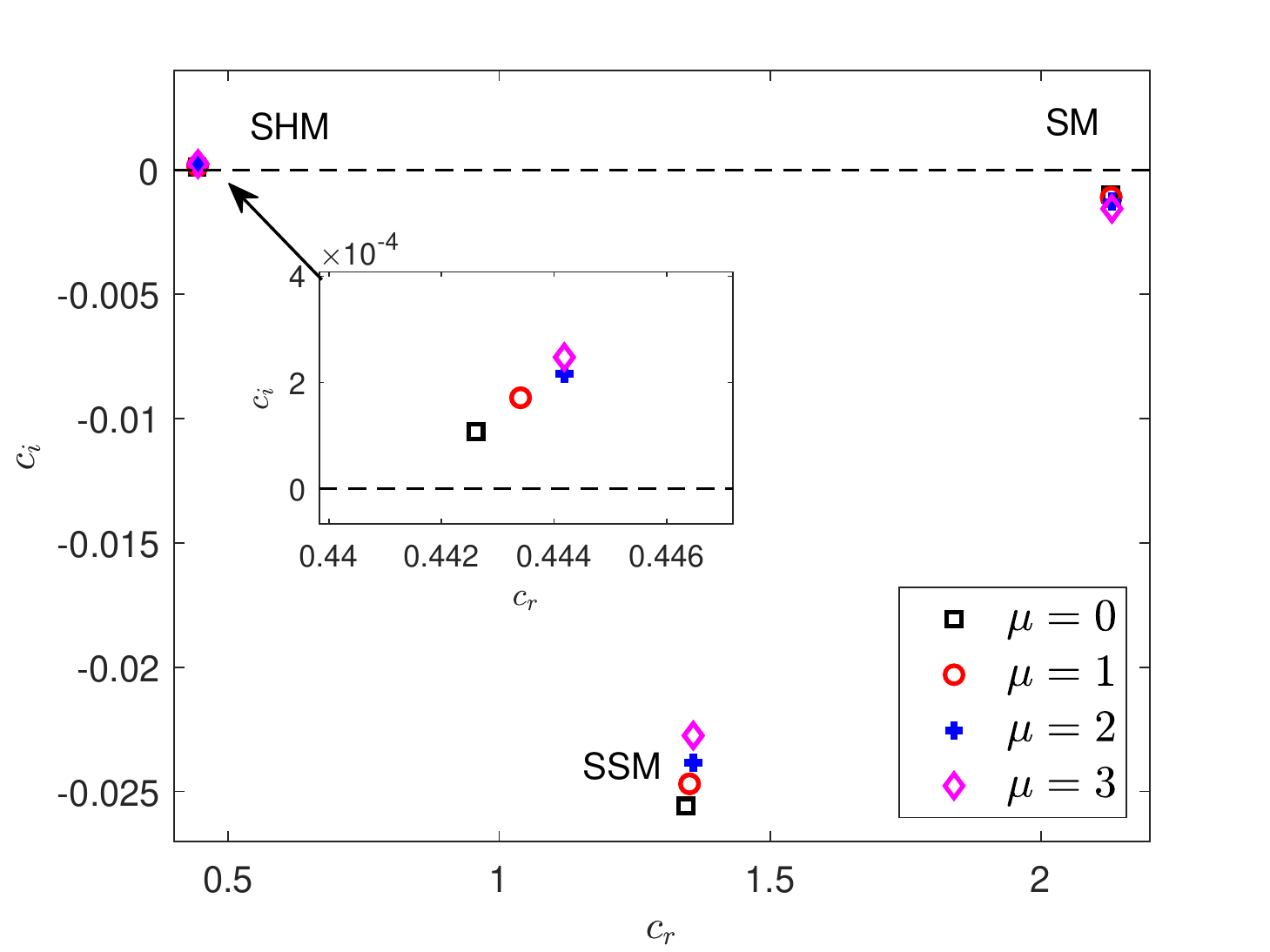}}
          \end{center}\vspace{-0.5cm}
    \caption{ The influence of odd viscosity $\mu$ on the eigenspectrum corresponding to the  unstable (a) surface mode with $Pe=1000$, (b) surface surfactant mode with $Pe=25$ when $k=0.05$, $Re=20$, $\tau=0.4$, and $\beta=4^{\circ}$, and (c) shear mode with $k=2.32$, $\tau=0.2$, $Re=4400$, $\beta=1^{'}$, and $Pe=1000$. The other fixed parameters  are  $\alpha=0.01$, $Ca=2$, and $Ma=1$.}\label{fig3}
\end{figure}

In this section, we have revealed the odd viscosity effect on the different unstable modes induced by shear-imposed viscous liquid down a slippery incline. The significant results for this considered fluid flow model are derived for a wide range of flow characteristic parameters by solving the OS BVP based on the numerical method Chebyshev spectral collocation in MATLAB. Due to the presence of insoluble surfactants, there exist two modes, such as surface mode and surface surfactant mode. In addition, another mode exits when the inertia force is very high with a small angle of inclination, called shear mode. We have renamed these three different modes, surface, surface surfactant, and shear modes as, SM, SSM, and SHM, respectively. The primary goal of our work is to explore the combined impact of imposed shear and induced odd viscous coefficient on the different modes of the shear-imposed surfactant-laden viscous liquid falling over a slippery substrate. In Fig.~\ref{fig3}, we have plotted the eigenspectrum with fixed flow characteristic parameters to identify the existence of different modes and also to examine the effect of the odd viscosity coefficient $\mu$ on the unstable SM, as in Fig.~\ref{fig3}(a), SSM as in Fig.~\ref{fig3}(b), and SHM, as in Fig.~\ref{fig3}(c). The eigenmodes maintain an inequality relation $c_r|_{SM}\geq c_r|_{SSM}\geq c_r|_{SHM}$ based on the phase speed (\cite{bhat2019linear}, \cite{bhat2020linear}, \cite{sani2020effect}). The primary instability of the surface mode reduces as soon as $\mu$ increases (Fig.~\ref{fig3}(a)). The coefficient $\mu$ restricts the surface wave energy by increasing the base pressure of the fluid flow. In Fig.~\ref{fig3}(b), the surfactant mode instability attenuates for higher odd viscosity parameter $\mu$ since higher odd viscosity reduces the deformation of surfactant concentration at the surface of the liquid. The shear mode instability is dominant when the inertia force becomes very large, including a small inclination angle. The odd viscosity coefficient advances the primary instability generated by the shear mode by raising the phase speed.        
In the following subsections, we have exposed the detailed linear instability behaviour of each existence mode, as shown in Fig.~\ref{fig3}, in a wide range of flow characteristic parameters. 

\subsection{\bf{Surface mode} }
  Here, we have discussed the external shear impact on the surface mode instability when the viscosity ratio $\mu$ is present. The numerical approach is performed to plot marginal stability and temporal growth rate results by choosing some fluid characteristic parameters as fixed. 
  \begin{figure}[ht!]
    \begin{center}
        \subfigure[]{\includegraphics*[width=7.2cm]{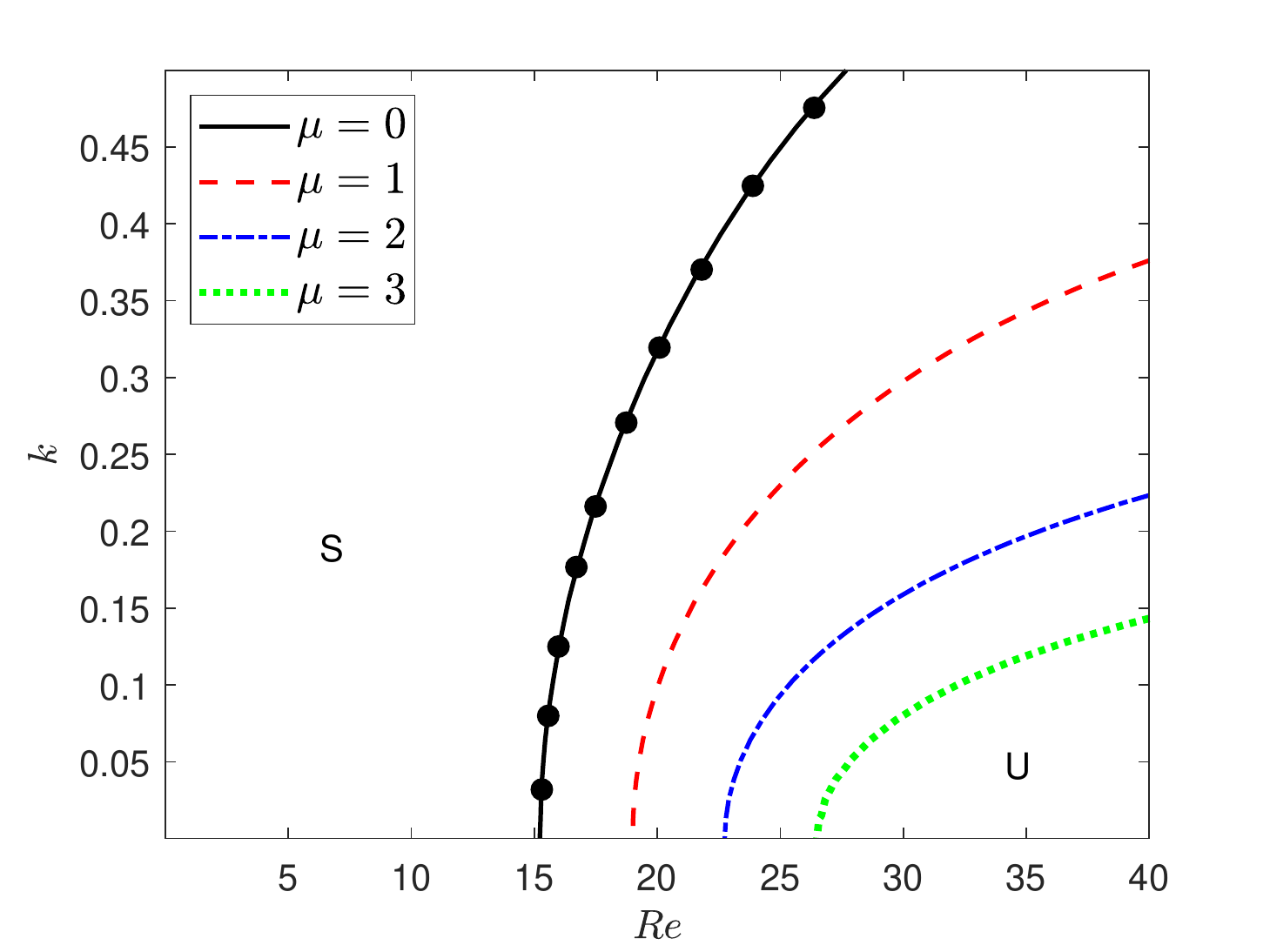}}
         \subfigure[]{\includegraphics*[width=7.2cm]{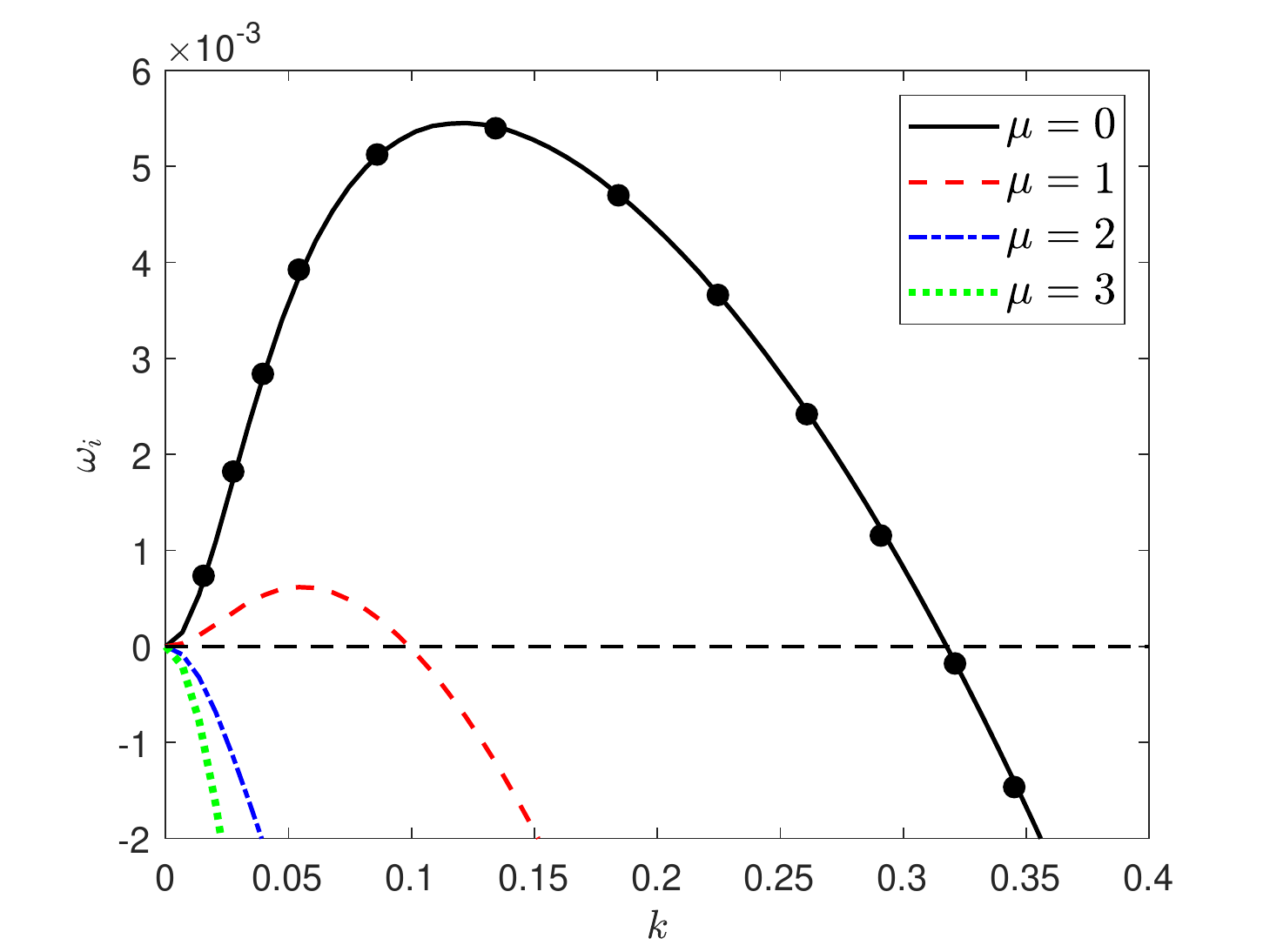}}
          \end{center}\vspace{-0.5cm}
    \caption{(a) Marginal stability curves of the surface mode for various odd viscosity $\mu$. (b) Corresponding growth rate $\omega_i$ versus wavenumber $k$ when $Re=20$. The common parameters are $\tau=0.5$, $\alpha=0$, $\beta=4^{\circ}$, $Ca=2$, $Ma=1$, and $Pe=1000$. The black solid circles are the result \citet{bhat2019linear} (Fig.~8(a) of their paper).}\label{fig4}
\end{figure}
  
  Fig.~\ref{fig4}(a) depicts the marginal stability curves of the unstable surface mode for various values of the odd viscosity coefficient $\mu$, when the fixed values are $\tau=0.5$, $\beta=4^{\circ}$, $\alpha=0$, $Ca=2$, $Ma=1$, and $Pe=1000$. Here, $S$ and $U$ symbols are used to denote the linear stable and unstable regimes, respectively. The main focus is to check the validity and accuracy of our numerical result of surface mode with the previous result obtained by \citet{bhat2019linear}.
  Notably, the unstable surface mode result fully recovers the result of \citet{bhat2019linear}. The unstable zone reduces for the higher value of $\mu$ by enhancing the critical Reynolds number and reducing the unstable wavenumber range. Further, the corresponding temporal growth rate result is plotted in Fig.~\ref{fig4}(b), which strengthens the outcomes obtained in Fig.~\ref{fig4}(a). The higher odd viscosity $\mu$ reduces the maximum growth rate, which confirms that odd viscosity $\mu$ has a stabilizing nature in the surface waves.  
\begin{figure}[ht!]
    \begin{center}
        \subfigure[$Ma=0$]{\includegraphics*[width=5.4cm]{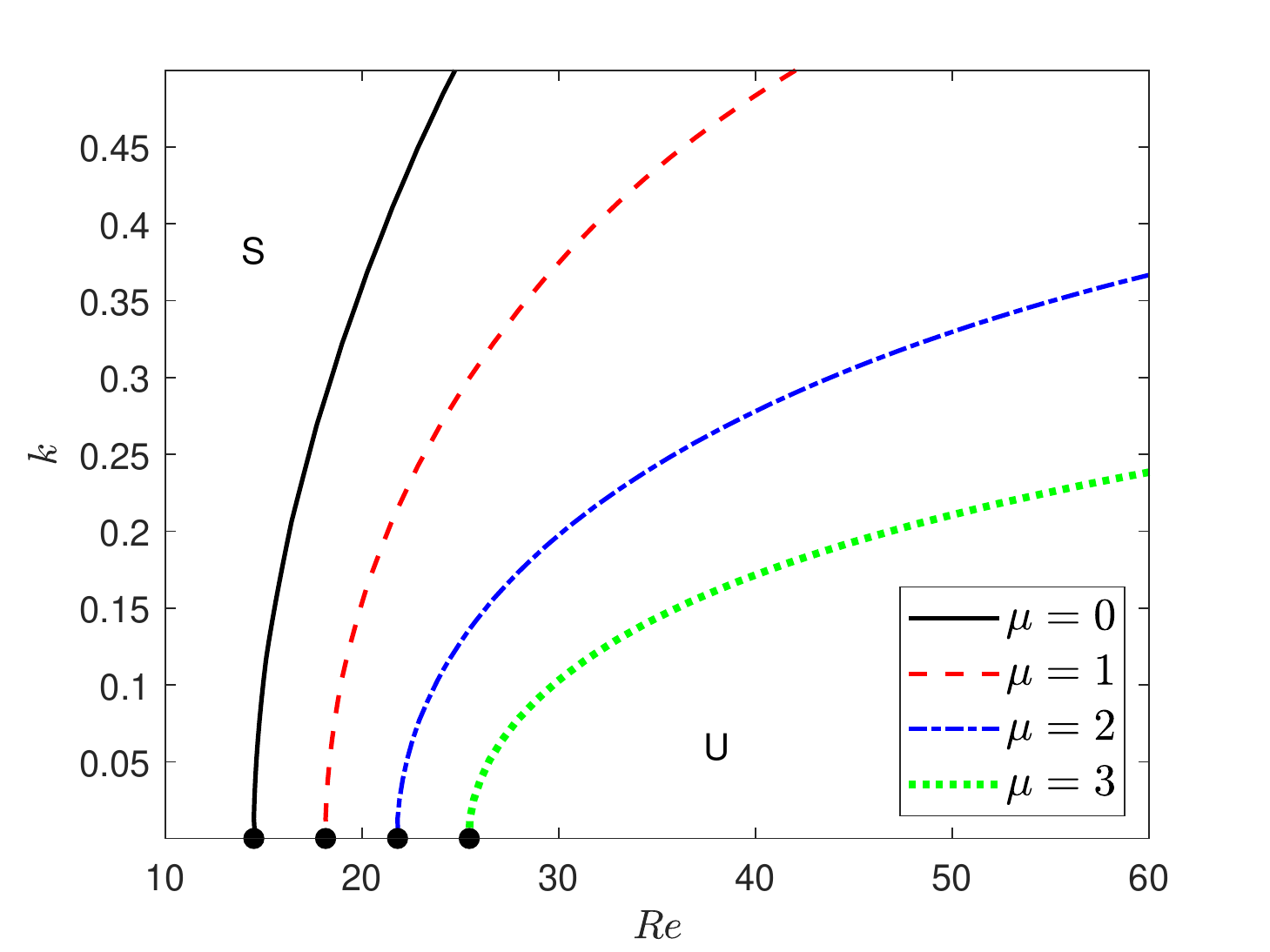}}
         \subfigure[$Ma=1$]{\includegraphics*[width=5.4cm]{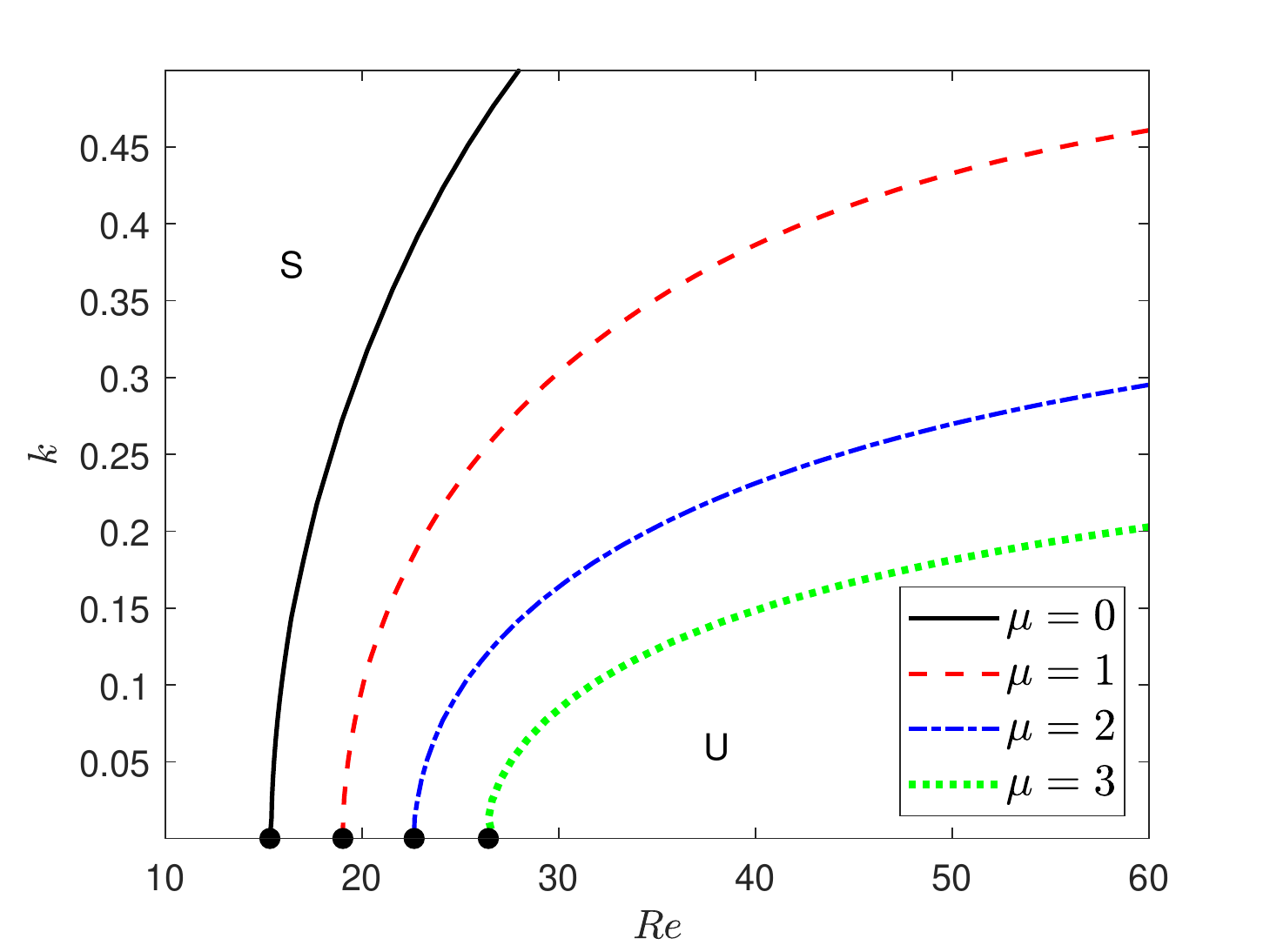}}
         \subfigure[$Ma=2$]{\includegraphics*[width=5.4cm]{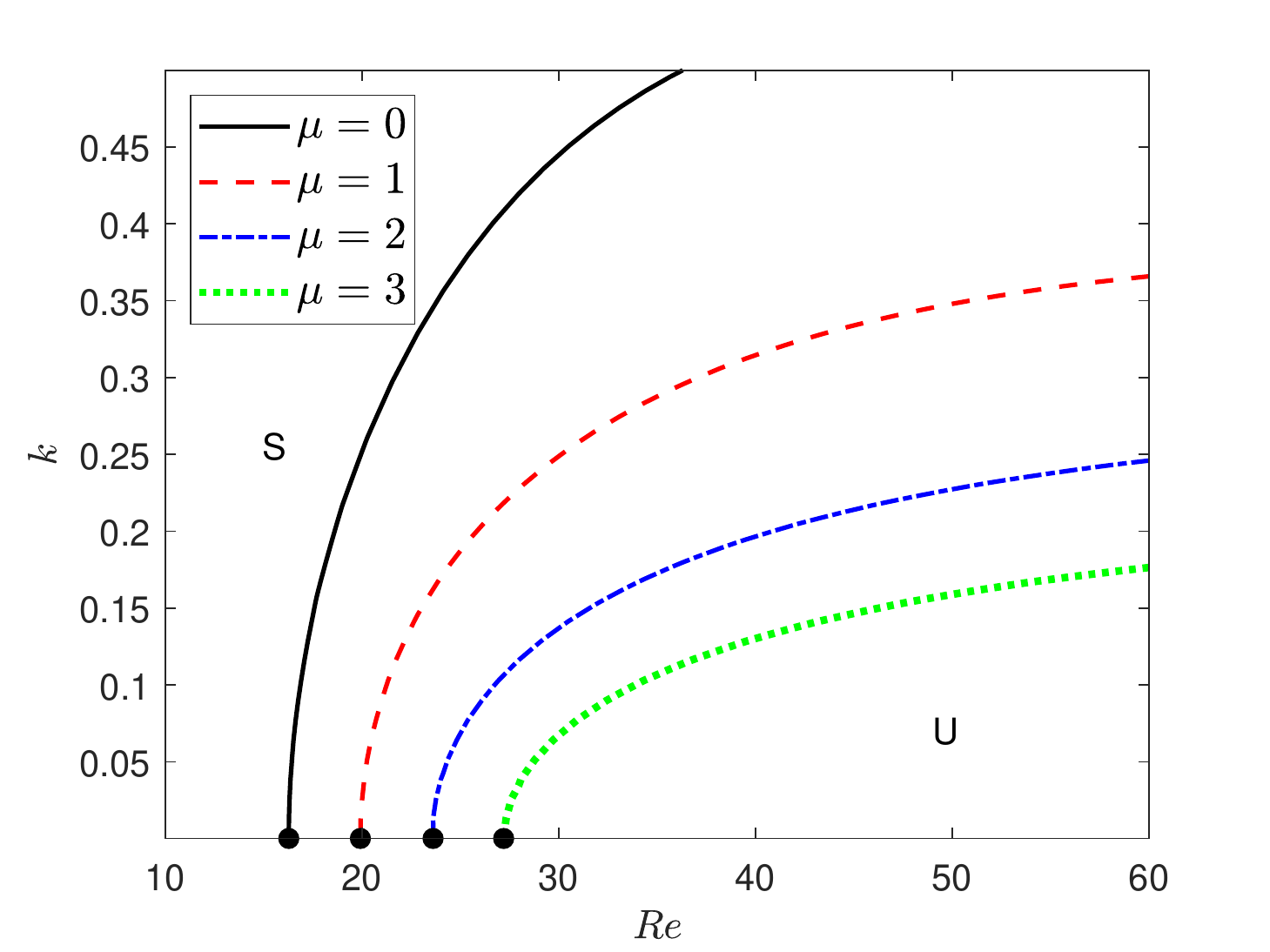}}
          \end{center}\vspace{-0.5cm}
    \caption{The variation of marginal stability curves of the surface mode for different odd viscosity $\mu$ when (a) $Ma=0$ (clean surface) and ((b) and (c)) $Ma\neq0$ (contaminated surface) with fixed $\tau=0.4$, $\alpha=0.01$, $Re=20$, $\beta=4^{\circ}$, $Ca=2$, and $Pe=1000$. }\label{fig5}
\end{figure}

In Fig.~\ref{fig5}, the marginal curves of uncontaminated ($Ma=0$, as in Fig.~\ref{fig5}(a)) and contaminated ($Ma\neq0$, as in Figs.~\ref{fig5}(b) and (c)) surface waves are demonstrated when the viscosity ratio $\mu$ alters. For both clean and contaminated fluid surfaces, the bandwidth of the unstable regime shrinks at higher values of $\mu$ due to the continuous increment in the critical Reynolds number. So, one can easily predict the stabilizing effect of the odd viscosity coefficient $\mu$ on both clean and contaminated surface waves. The odd viscosity $\mu$ strengthens the base pressure, which weakens the inertia force to the convective flow and results in a stabilizing behaviour in the unstable surface mode. Further, it is found from Figs.~\ref{fig5}(a), (b), and (c) that the higher Marangoni number $Ma$ attenuates the film flow instability instigated by the surface mode by decreasing the unstable zone bandwidth and confirms the stabilizing nature of Marangoni force $Ma$. The presence of the damped surfactant mode in the considered flow characteristic parameter range is mainly responsible for this stabilizing nature of the surface mode. The surfactant concentration gradient causes the surface tension gradient, and the resulting Marangoni stress reduces the transfer of energy to the perturbed surface wave. Consequently, the Marangoni force plays a vital role in controlling the development of surface mode instability.
So, one can conclude that the presence of the odd viscosity coefficient in the viscous falling film strengthens the stabilizing effect of the Marangoni force.
\begin{figure}[ht!]
    \begin{center}
        \subfigure[$\mu=0$]{\includegraphics*[width=7.2cm]{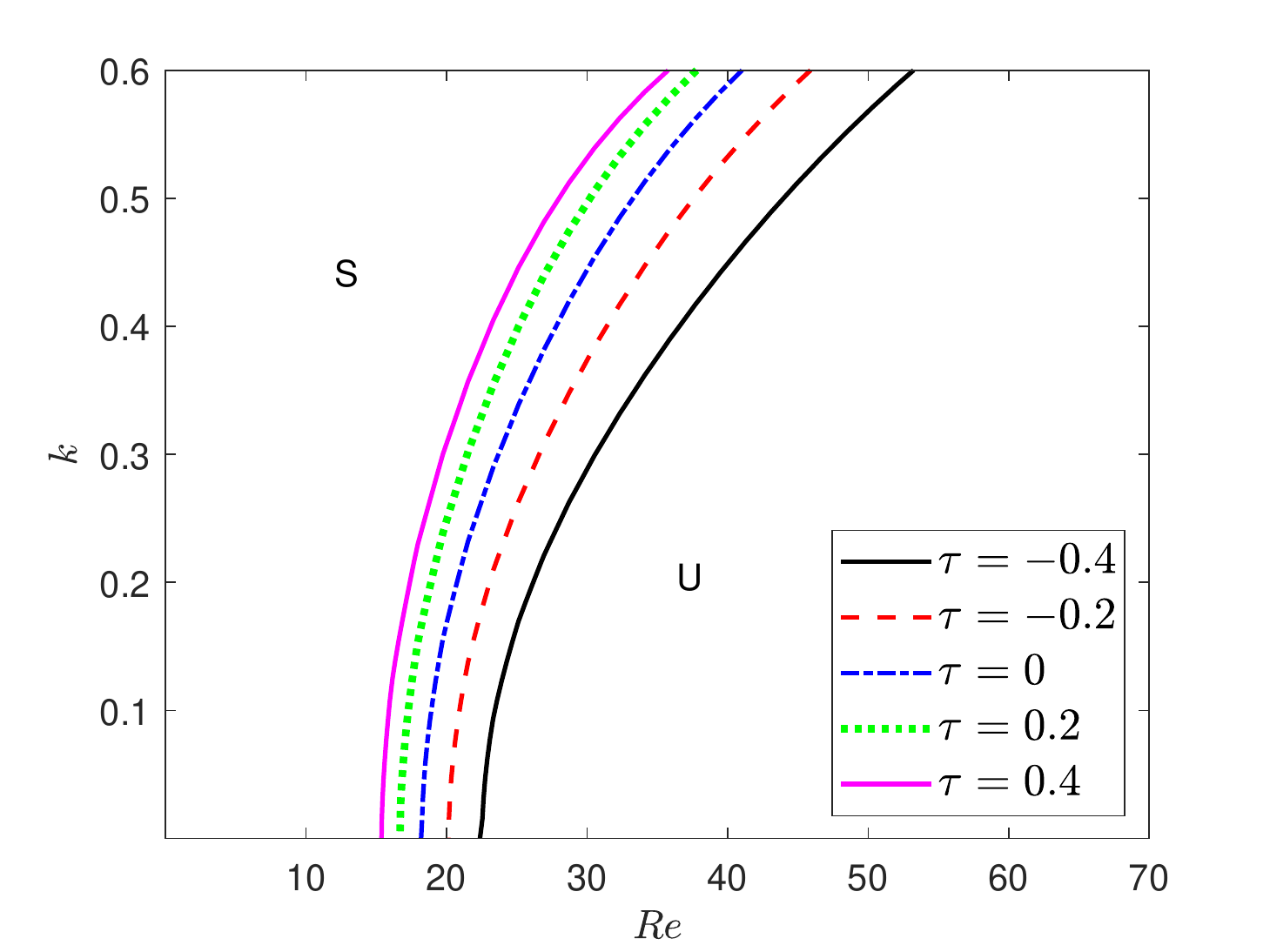}}
         \subfigure[$\mu=1$]{\includegraphics*[width=7.2cm]{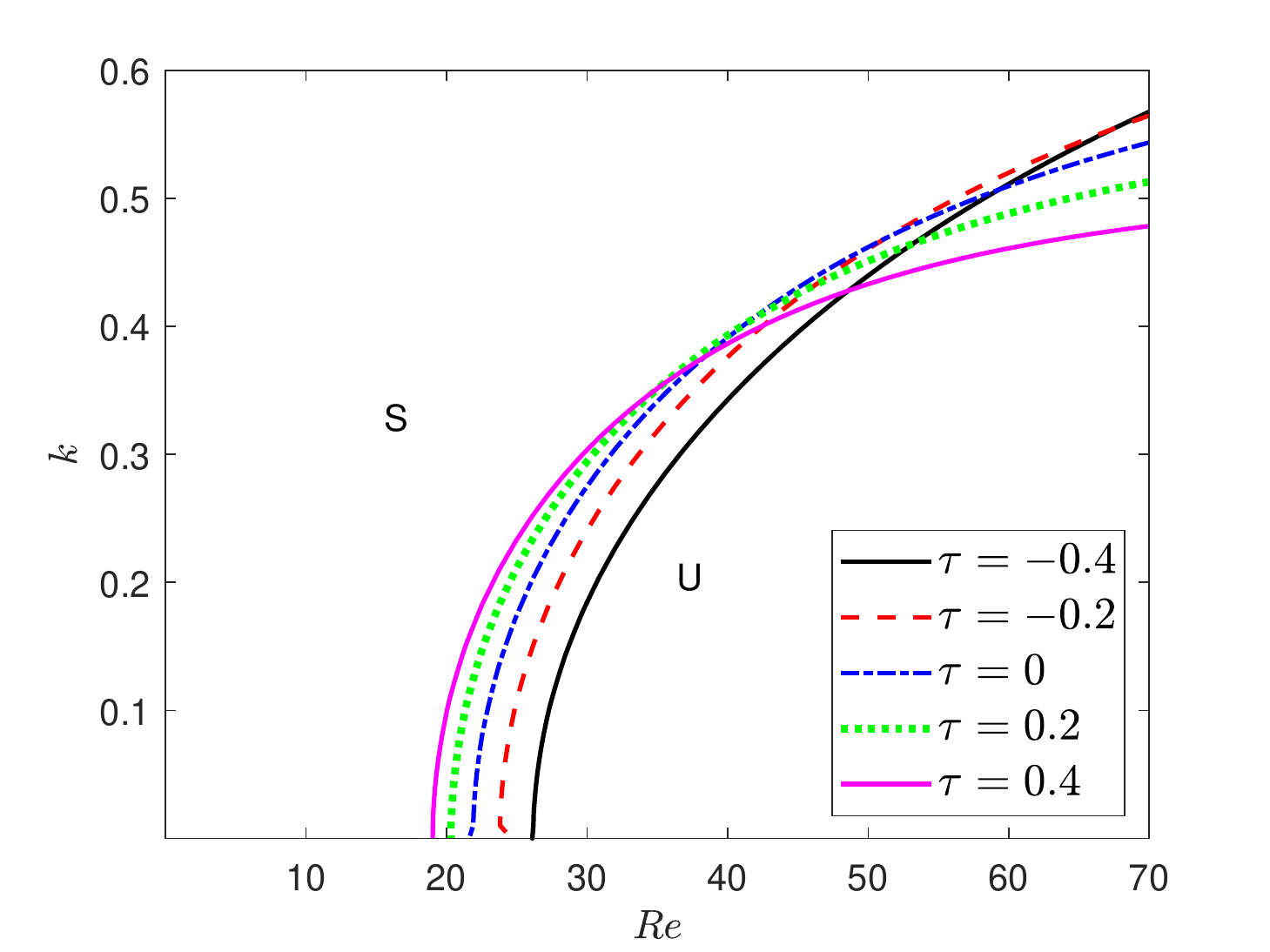}}
         \subfigure[$\mu=2$]{\includegraphics*[width=7.2cm]{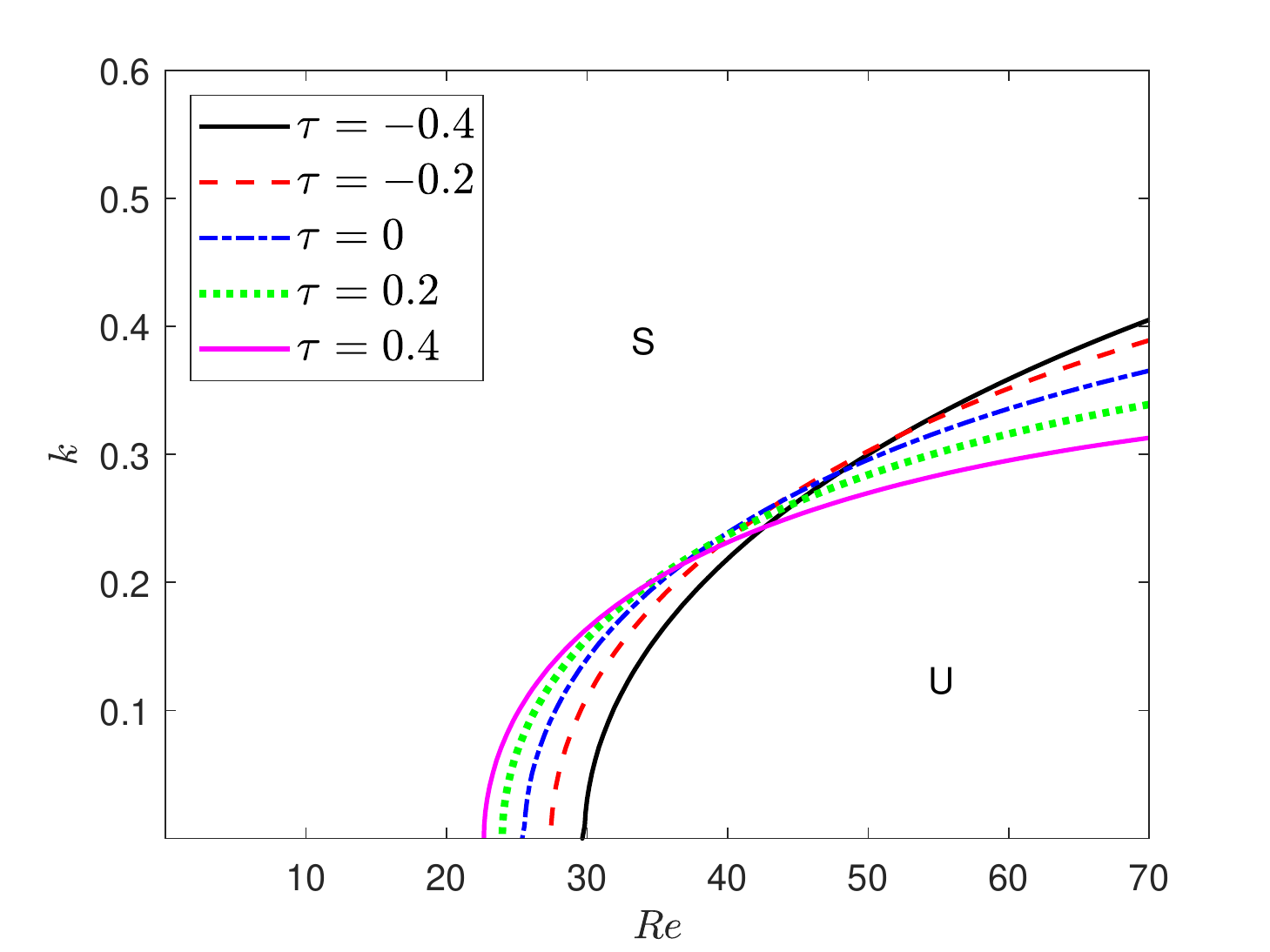}}
         \subfigure[$\mu=3$]{\includegraphics*[width=7.2cm]{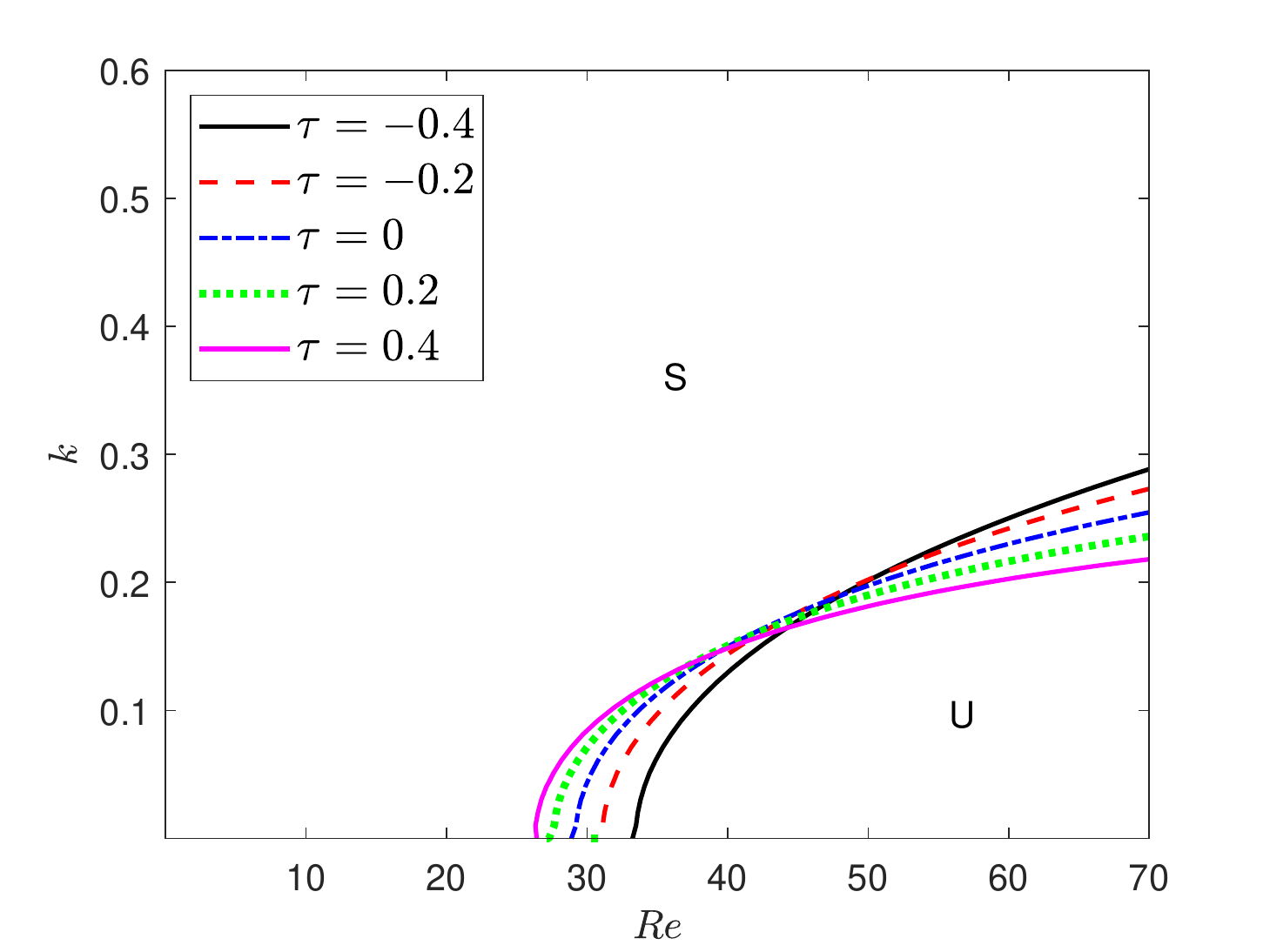}}
          \end{center}\vspace{-0.5cm}
    \caption{The variation of marginal stability curves related to the surface mode for different imposed shear $\tau$ with (a) $\mu=0$ (absence of broken time-reversal symmetry) and ((b), (c), and (d)) $\mu\neq0$ (presence of broken time-reversal symmetry). The common parameters are $\alpha=0.01$, $\beta=4^{\circ}$, $Ca=2$, $Ma=1$, and $Pe=1000$. }\label{fig6}
\end{figure}

\begin{figure}[ht!]
    \begin{center}
        \subfigure[$Re=35$]{\includegraphics*[width=7.2cm]{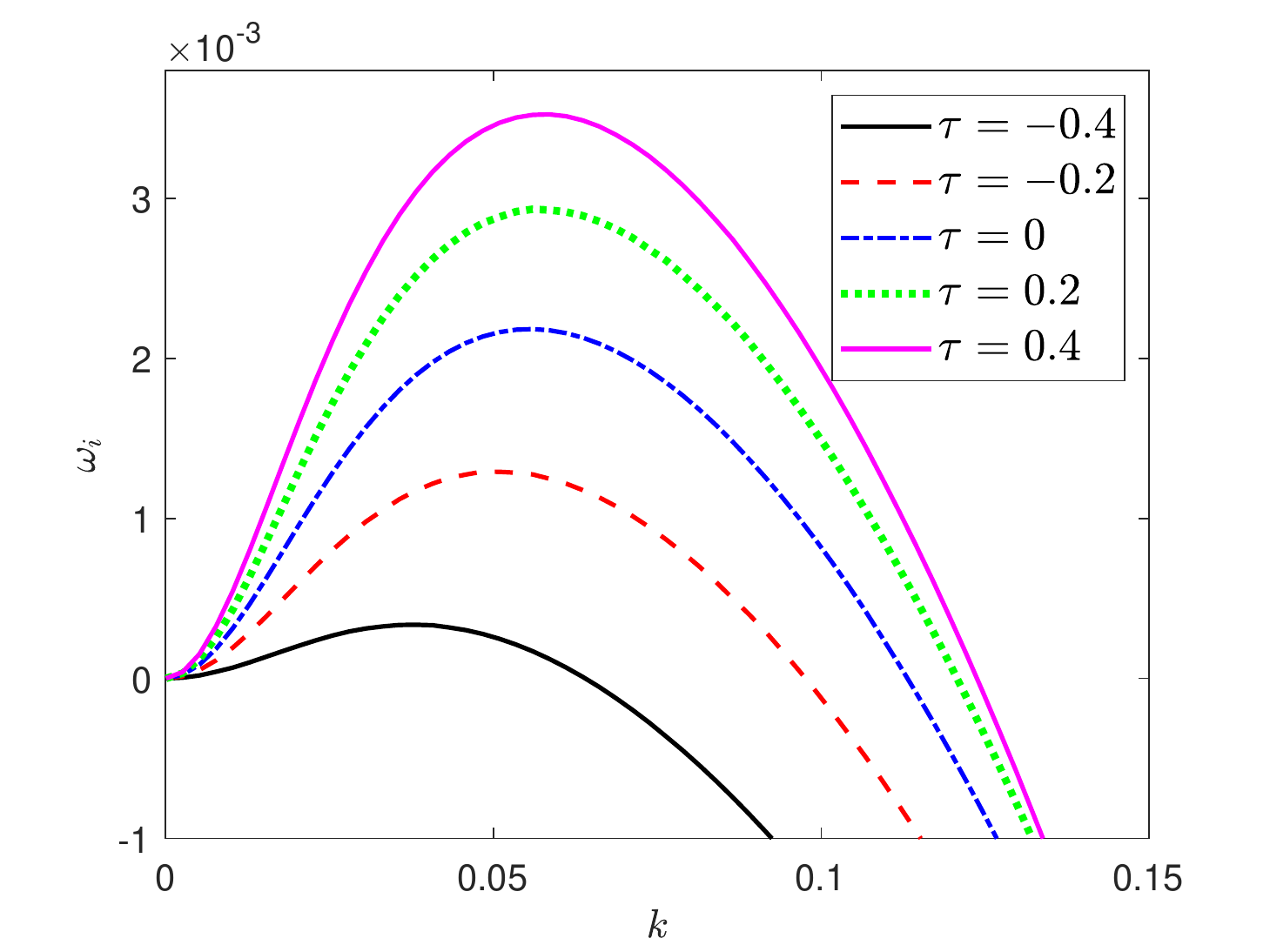}}
         \subfigure[$Re=70$]{\includegraphics*[width=7.2cm]{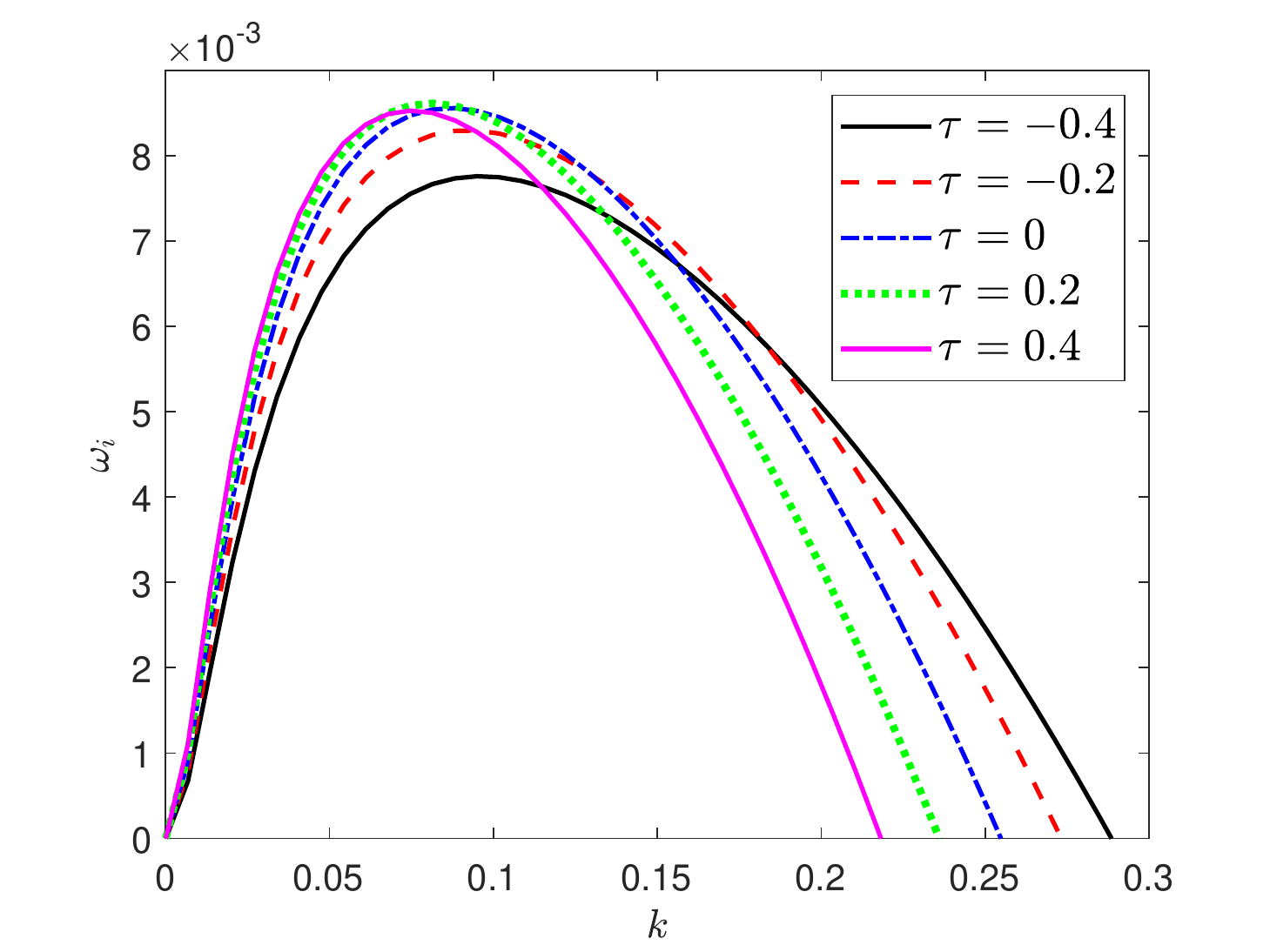}}
        \end{center}\vspace{-0.5cm}
    \caption{The effect of imposed shear ($\pm\tau$) on the temporal growth rate curve of the surface mode when (a) $Re=35$  and (b) $Re=70$. The common parameters are $\mu=3$, $\alpha=0.01$, $\beta=4^{\circ}$, $Ca=2$, $Ma=1$, and $Pe=1000$. }\label{fig7}
\end{figure}
\begin{figure}[ht!]
    \begin{center}
        \subfigure[]{\includegraphics*[width=7.2cm]{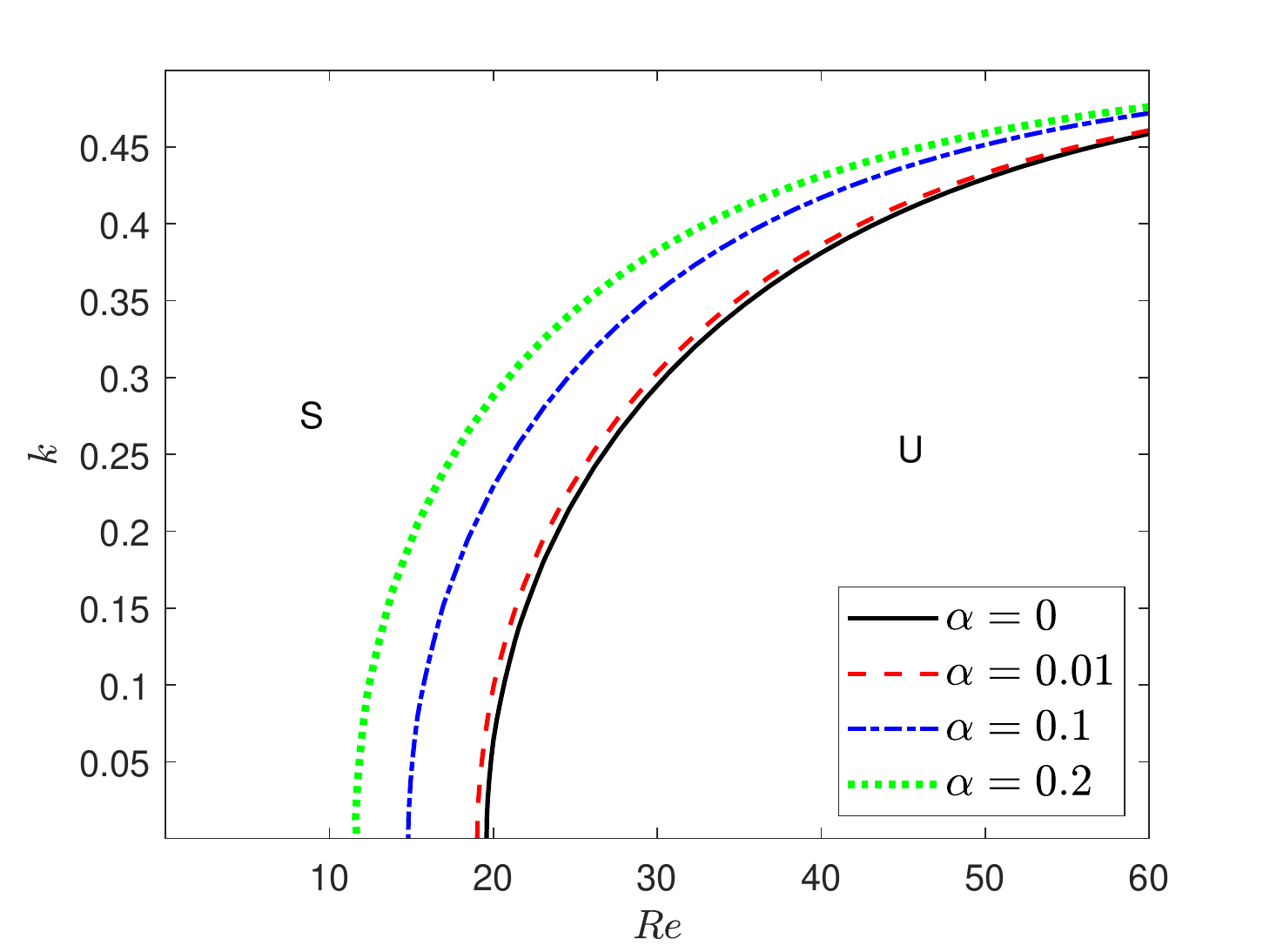}}
         \subfigure[]{\includegraphics*[width=7.2cm]{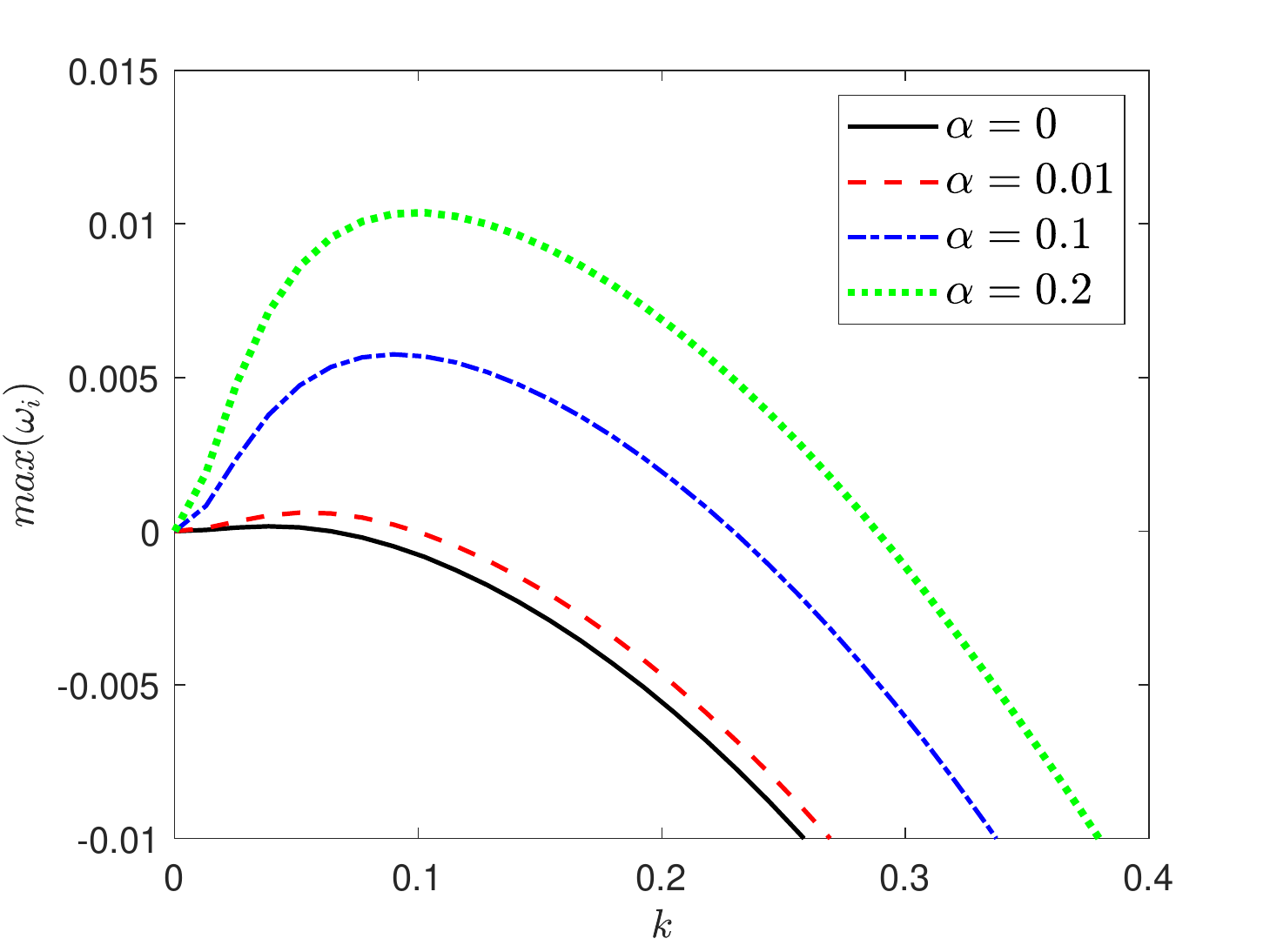}}
        \end{center}\vspace{-0.5cm}
    \caption{(a) The change in stability boundary with respect to the slip parameter ($\alpha$). (b) The change in the maximum growth rate of the surface mode when $Re=20$. The common parameters are $\mu=1$, $\tau=0.4$, $\beta=4^{\circ}$, $Ca=2$, $Ma=1$, and $Pe=1000$. }\label{fig8}
\end{figure}

Fig.~\ref{fig6} depicts the influence of external shear ($\tau$) along both the downstream and upstream directions on the surface mode instability when the odd viscosity coefficient ($\mu$) differs. The unstable bandwidth of the surface mode (see Fig.~\ref{fig6}(a)) amplifies/diminishes as long as external shear ($\tau$) increases in the downstream/upstream direction in the absence of viscosity ratio (i.e., $\mu=0$). So, the positive imposed shear has the potential to develop surface mode instability when the reversal symmetry of time does not break in the flow system (\cite{bhat2019linear}). But, when the viscosity ratio $\mu\neq0$, the dual behaviour of external shear on the surface mode instability is observed in Figs.~\ref{fig6}(b), (c), and (d). The unstable zone corresponding to the surface mode significantly enhances/reduces near the onset of instability, whereas it reduces/enhances far away from the onset of instability as long as imposed shear increases in the downstream/upstream flow direction. The stronger external shear advances the base pressure in the vicinity of the free surface because of the odd viscosity influence. This results in an increment in the hydrostatic force of the fluid. Consequently, the stabilizing effect of potent flow-directed imposed shear is possible due to an interplay between odd viscosity and external shear. 

To strengthen the above point, the temporal growth rate results are plotted in Fig.~\ref{fig7} for the fixed $\mu=3$, where the $Re$ value is selected from the unstable region. Accordingly, $Re=35$ (Fig.~\ref{fig7}(a)) is chosen from the
zone near the criticality and $Re=70$ (Fig.~\ref{fig7}(b)) is chosen from the zone located far away from the criticality.
When $Re=35$, the surface mode growth rate enhances in the longwave region as the positive external shear ($+\tau$) increases, while the growth rate of surface mode attenuates in the finite wavenumber region with the increasing value of positive imposed shear ($+\tau$) when $Re=70$. This fact confirms the dual behaviour of shear force on the surface mode. Conclusively, it is important to mention that one can use the external shear in the downstream direction to reduce the liquid surface instability and restrict the film flow rate by choosing an appropriate viscosity ratio value in the viscous falling film.

Fig.~\ref{fig8}(a) depicts that the higher value of the slip parameter $\alpha$ expands the instability boundary by randomly decreasing the critical Reynolds number. Hence, the slippery surface promotes the instability influence on the surface wave. From the physical point of view, as slip length increases, the velocity gradient decreases at the bottom of the fluid layer. Therefore, the friction of the fluid decreases, resulting in a destabilizing effect on the film flow system. Further,
the maximum growth rate curve over the full range of the external shear $\tau$ ($-0.4\leq \tau \leq 0.4$) of the surface mode, as in Fig.~\ref{fig8}(b), becomes more for 
higher slip length. That means the slippery surface boosts the shear-induced surface mode instability.

\begin{figure}[ht!]
    \begin{center}
     \subfigure[$\mu=0$]{\includegraphics*[width=5.4cm]{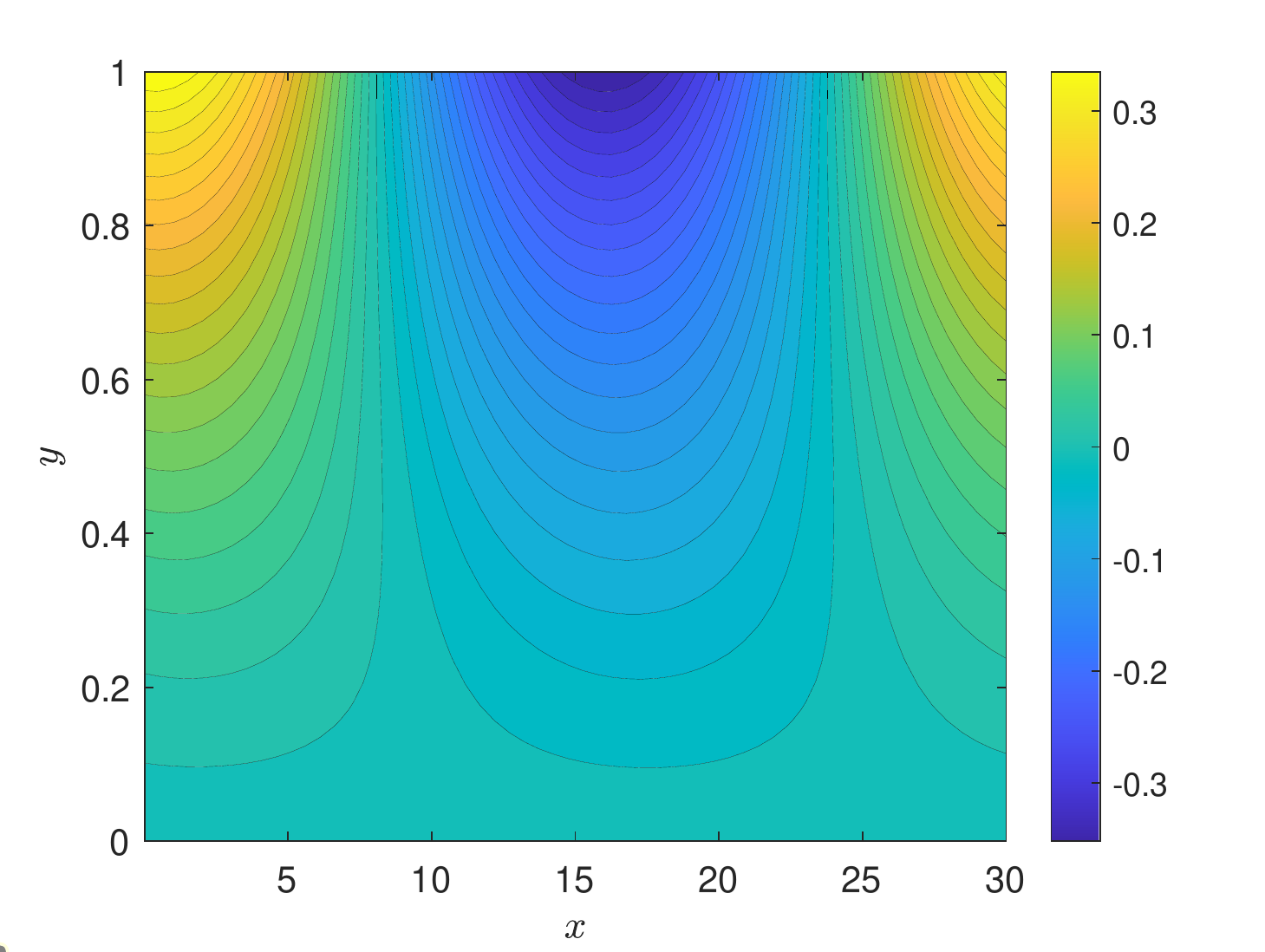}}
    \subfigure[$\mu=1$]{\includegraphics*[width=5.4cm]{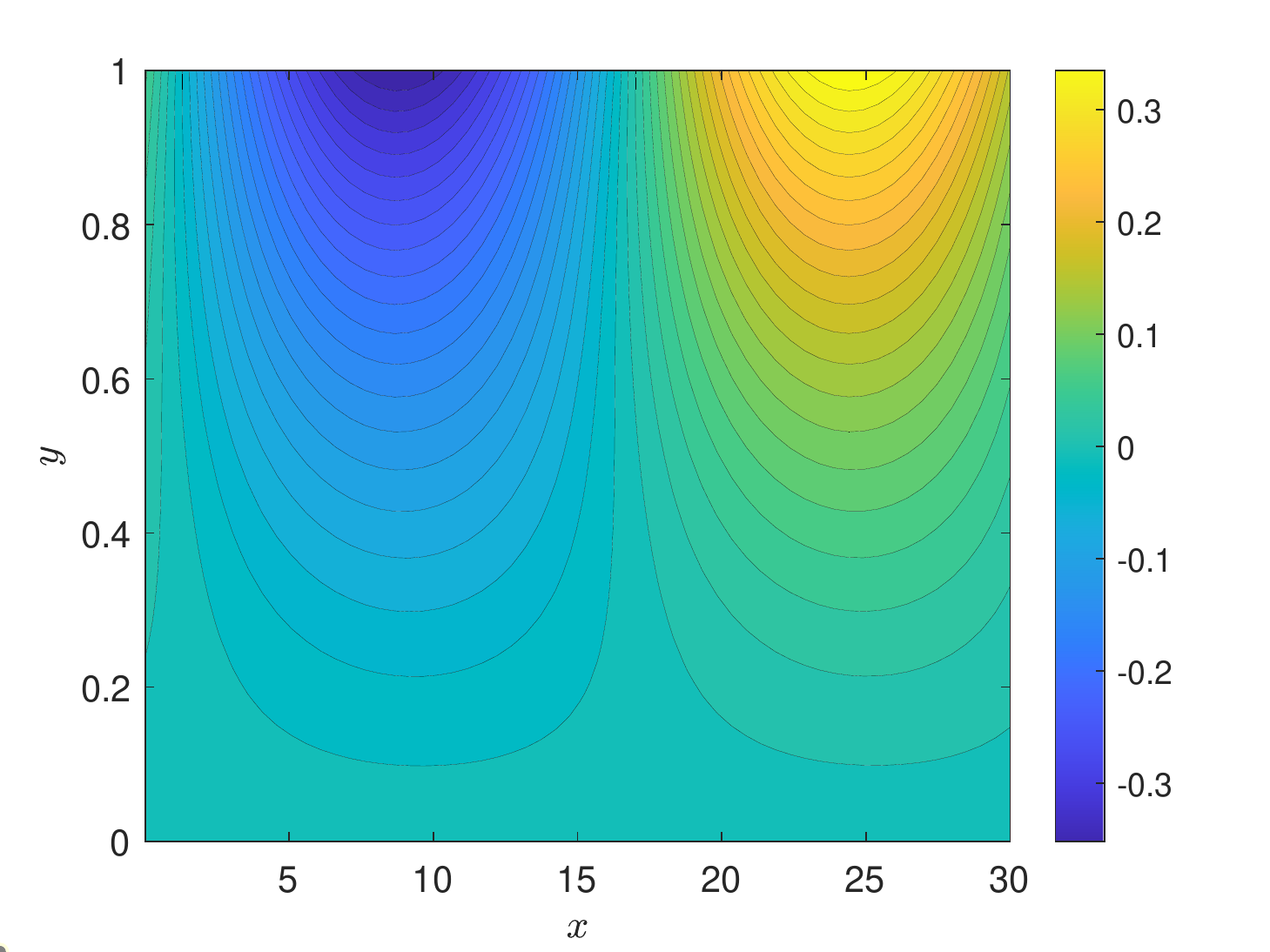}}
    \subfigure[$\mu=2$]{\includegraphics*[width=5.4cm]{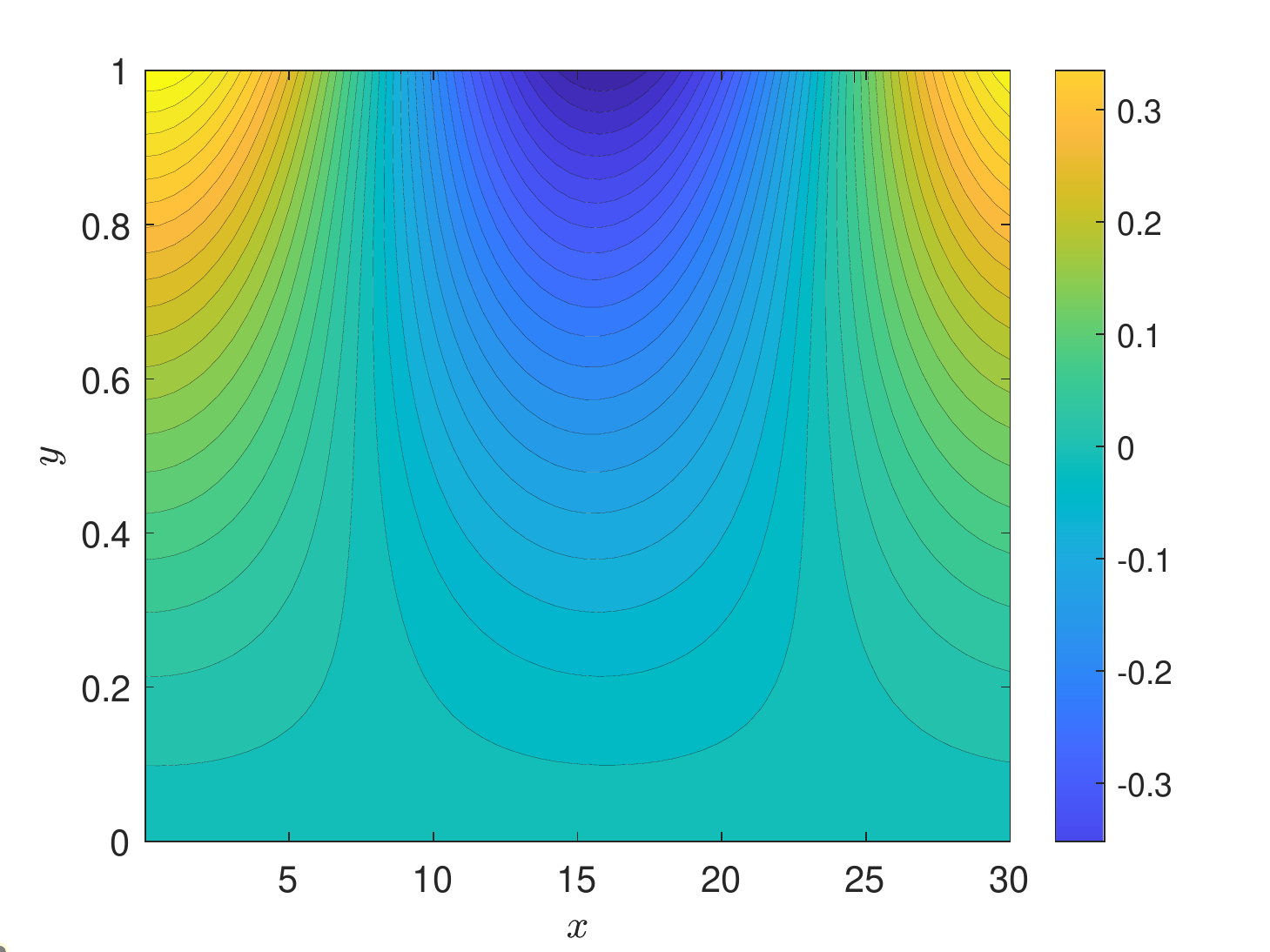}}
        \end{center}\vspace{-0.5cm}
    \caption{Velocity distribution associated with the surface mode for different values of odd viscosity $\mu$ with $\tau=0.4$ and $\alpha=0.1$. The common parameters are  $k=0.2$, $Re=20$, $\beta=4^{\circ}$, $Ca=2$, $Ma=1$, and $Pe=1000$. }\label{fig9a}
\end{figure}
\begin{figure}[ht!]
    \begin{center}
       \subfigure[$\tau=-0.4$]{\includegraphics*[width=5.4cm]{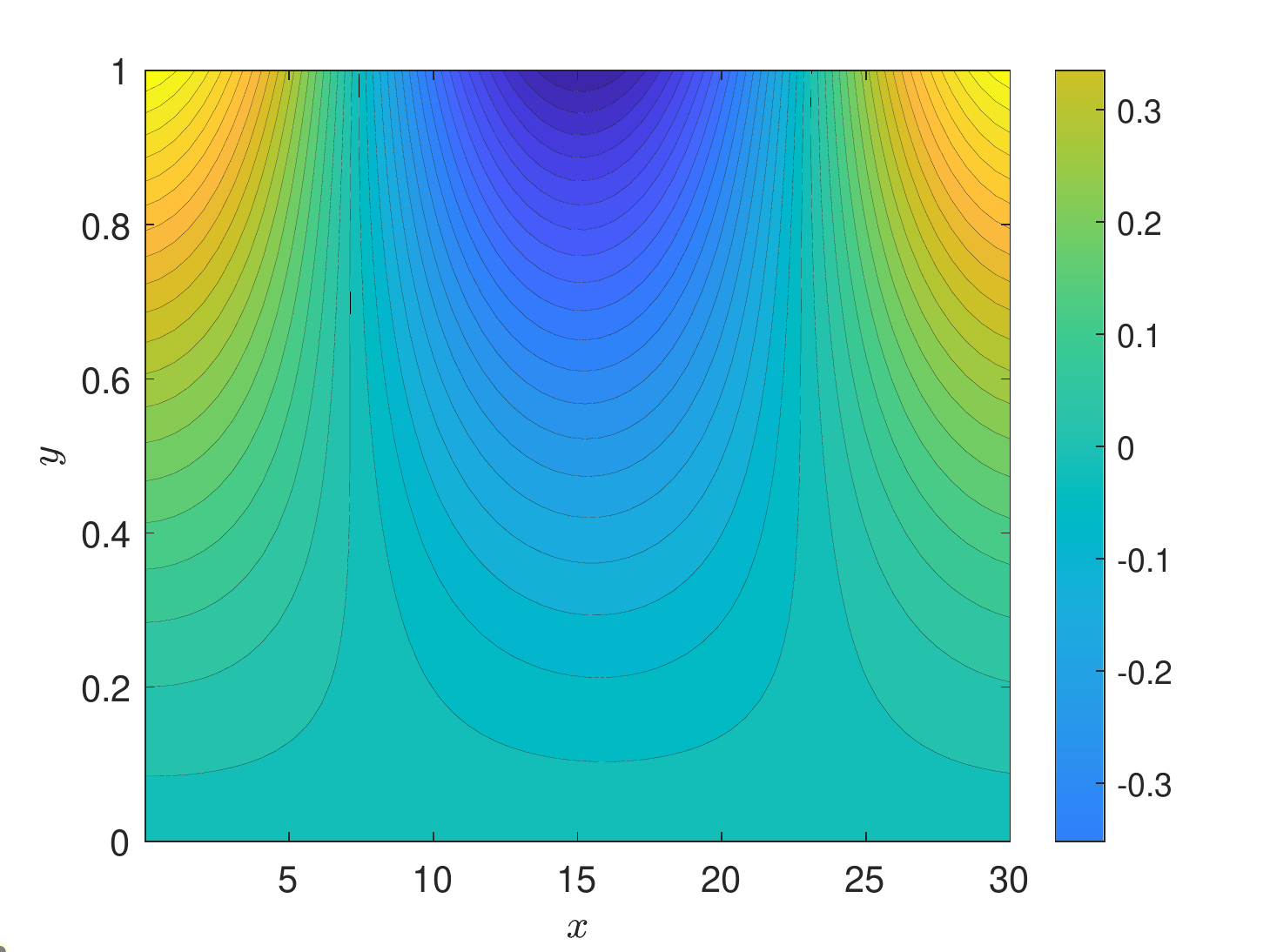}}
    \subfigure[$\tau=0$]{\includegraphics*[width=5.4cm]{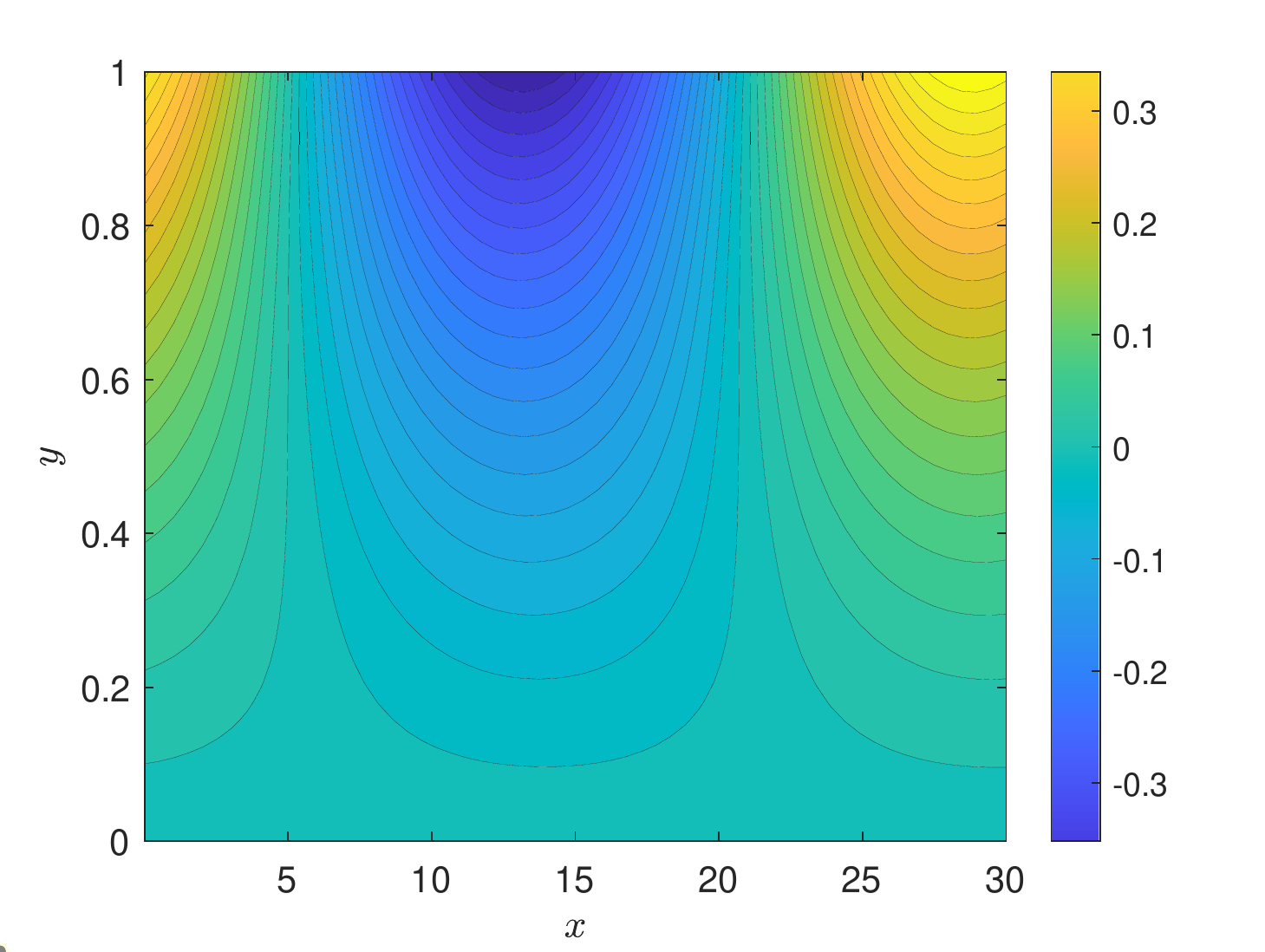}}
    \subfigure[$\tau=0.4$]{\includegraphics*[width=5.4cm]{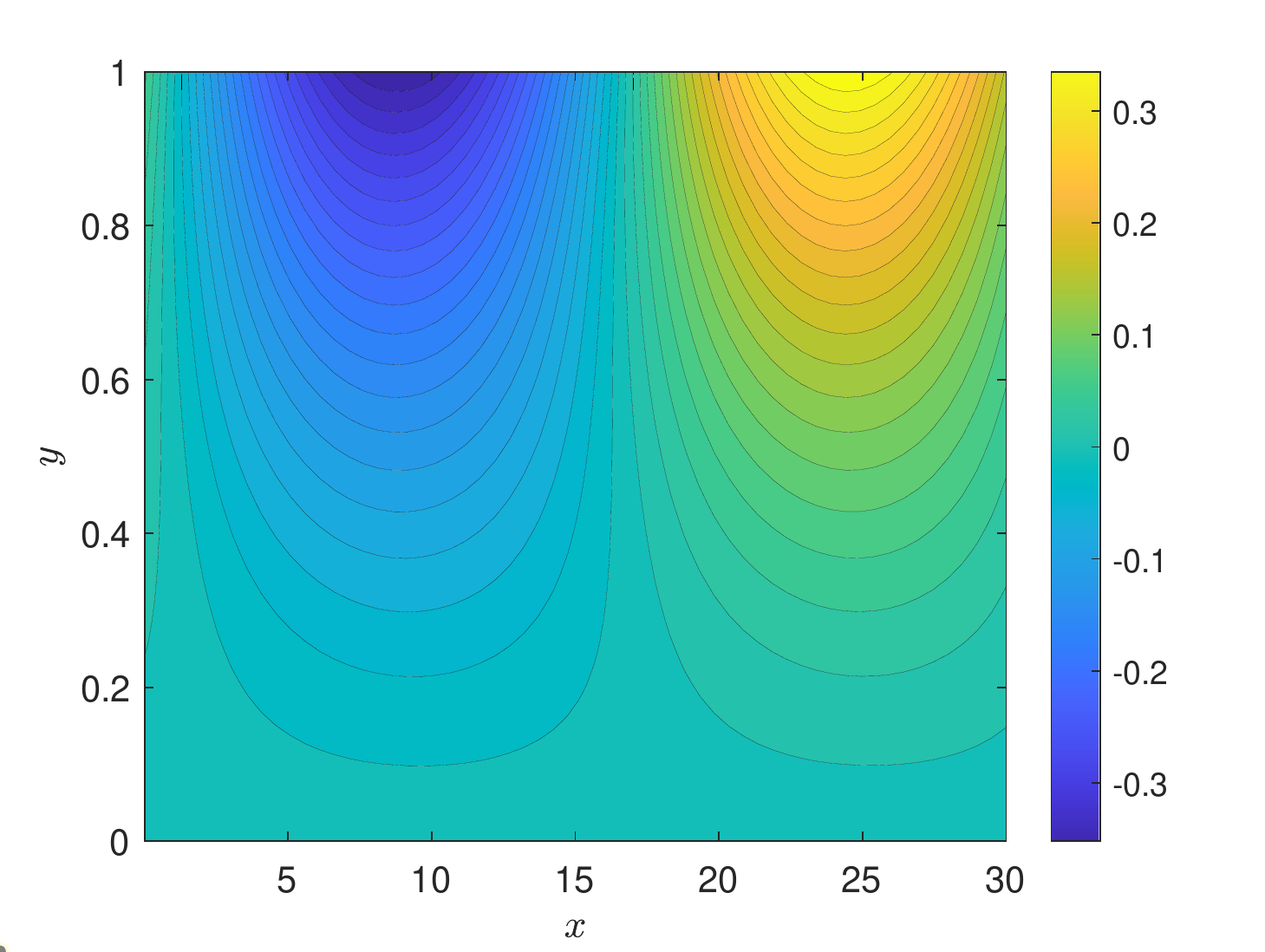}}
   \end{center}\vspace{-0.5cm}
    \caption{Velocity distribution associated with the surface mode for different values of externally imposed shear $\tau$  with $\mu=1$ and $\alpha=0.1$. The common parameters are  $k=0.2$, $Re=20$, $\beta=4^{\circ}$, $Ca=2$, $Ma=1$, and $Pe=1000$. }\label{fig9d}
\end{figure}

\begin{figure}[ht!]
    \begin{center}
    \subfigure[$\alpha=0$]{\includegraphics*[width=5.4cm]{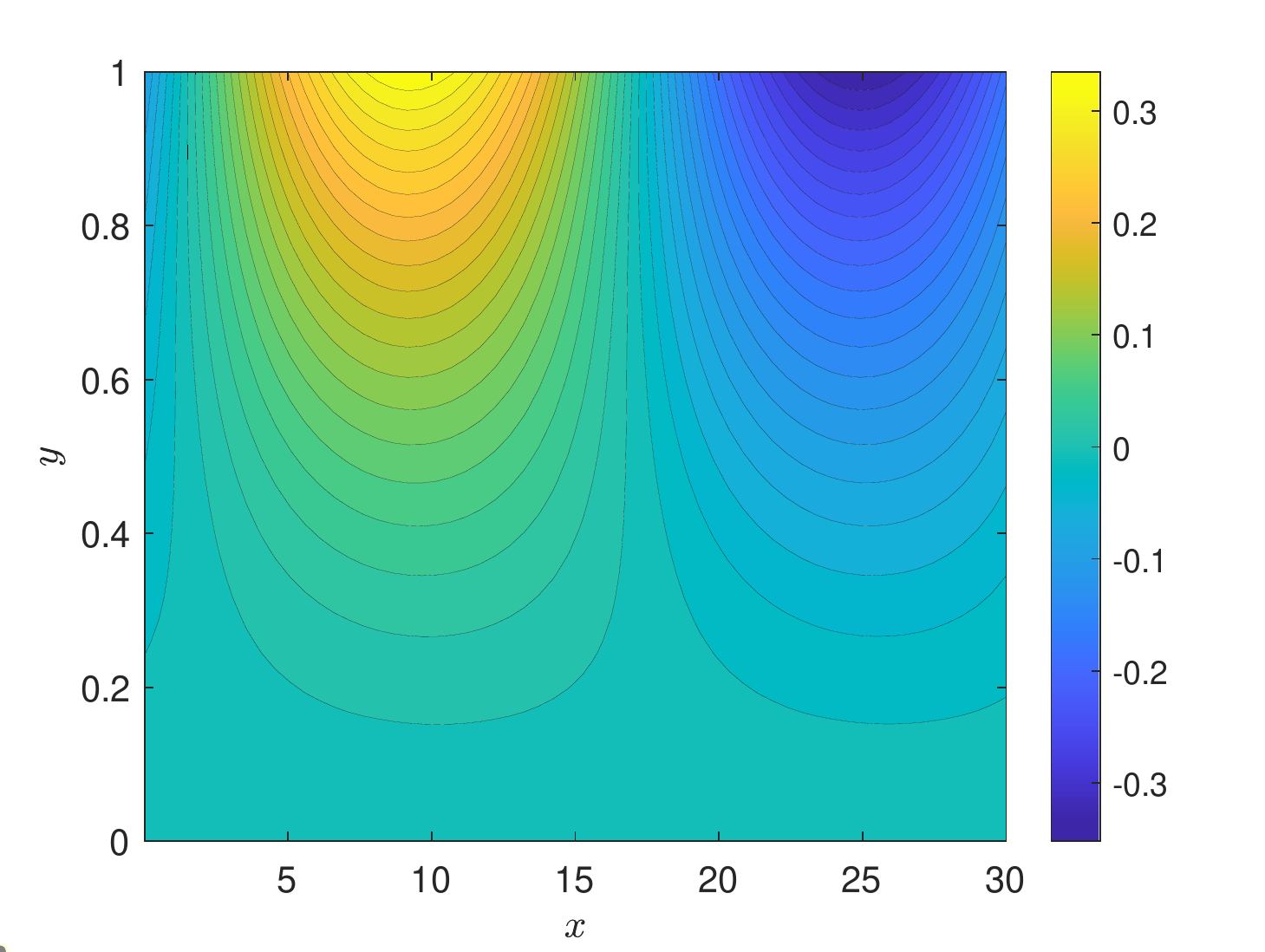}}
    \subfigure[$\alpha=0.1$]{\includegraphics*[width=5.4cm]{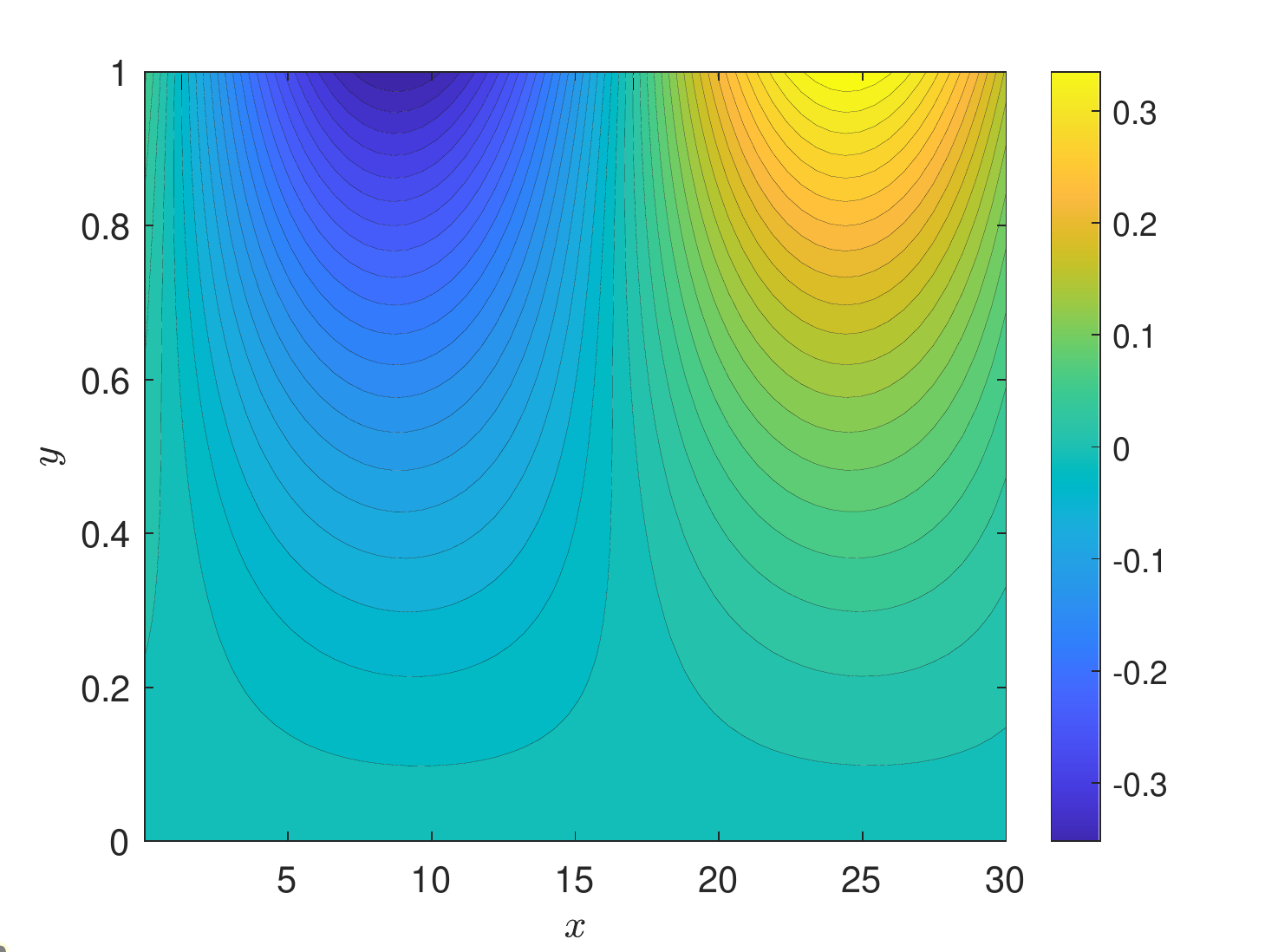}}
    \subfigure[$\alpha=0.2$]{\includegraphics*[width=5.4cm]{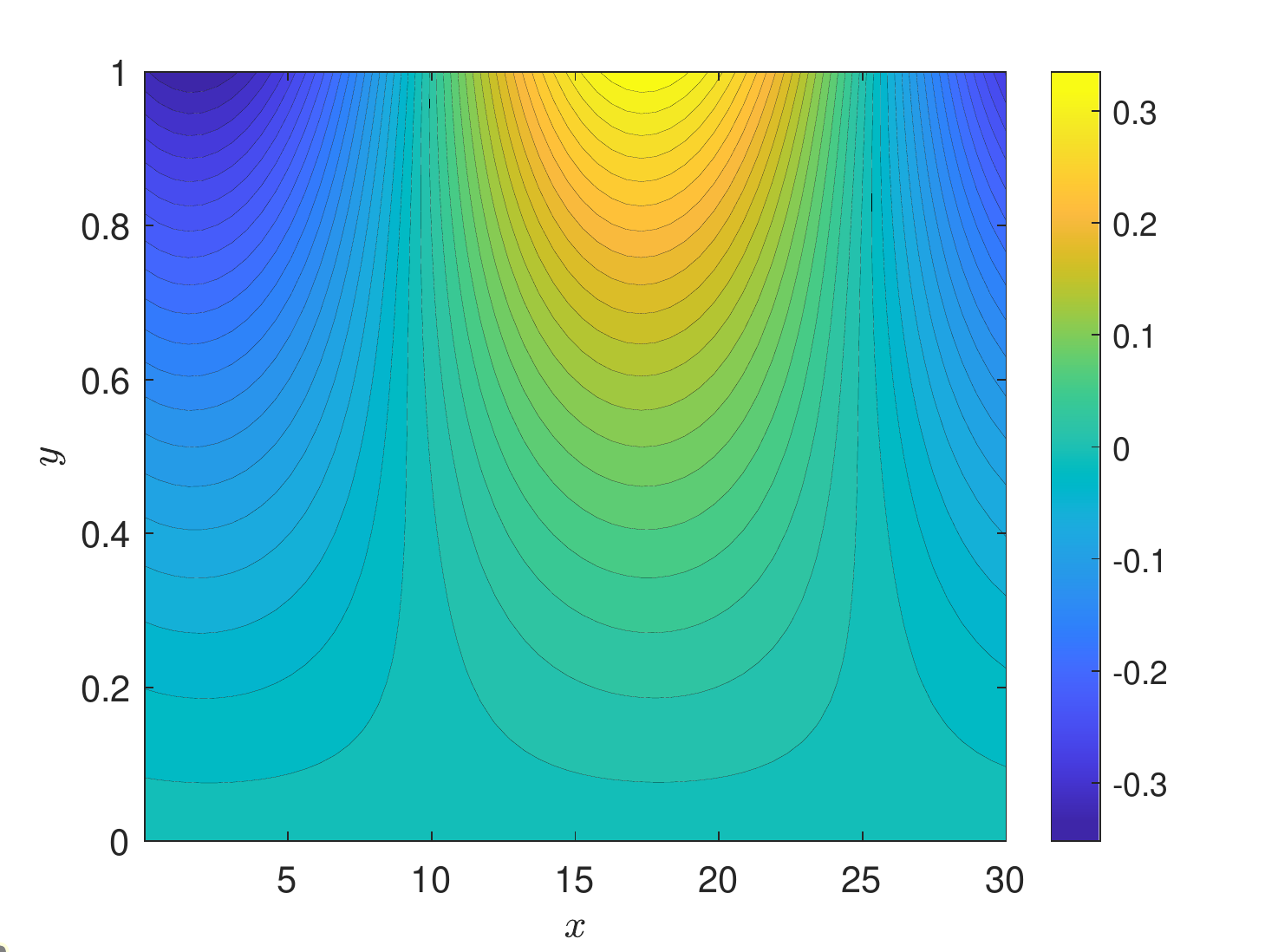}}
    \end{center}\vspace{-0.5cm}
    \caption{Velocity distribution associated with the surface mode for different values of slip parameter $\alpha$ with $\mu=1$ and $\tau=0.4$. The common parameters are  $k=0.2$, $Re=20$, $\beta=4^{\circ}$, $Ca=2$, $Ma=1$, and $Pe=1000$. }\label{fig9g}
\end{figure}

Figs.~\ref{fig9a}-\ref{fig9g} display the isolines of the horizontal perturbed velocity part of the surface wave for different viscosity ratios $\mu$ (Figs.~\ref{fig9a}), external shear $\tau$ (Figs.~\ref{fig9d}), and slip parameters (Figs.~\ref{fig9g}). The perturbation vorticity is centered in the vicinity of the liquid surface and spreads up to the bottom surface. It is found in Fig.~\ref{fig9a} that a higher value of the odd viscosity coefficient restricts the perturbation wave movement, which is followed by the fact that the maximum and minimum vortexes shift in the opposite to the spatial direction. Also, the sturdy external force ($+\tau$) in the downstream direction (Figs.~\ref{fig9d}(b) and (c)) expedites the velocity distribution in the streamwise direction and encourages the perturbation wave movement. On the other hand, the potent external force ($-\tau$) in the upstream direction (see Figs.~\ref{fig9d}(a) and (b)) delays the perturbation wave movement. Moreover, an increase in the velocity distribution along the flow direction can be observed in Fig.~\ref{fig9g} as the slip parameter goes up.

\subsection{\bf{Surfactant mode}}
\begin{figure}[ht!]
    \begin{center}
        \subfigure[]{\includegraphics*[width=7.4cm]{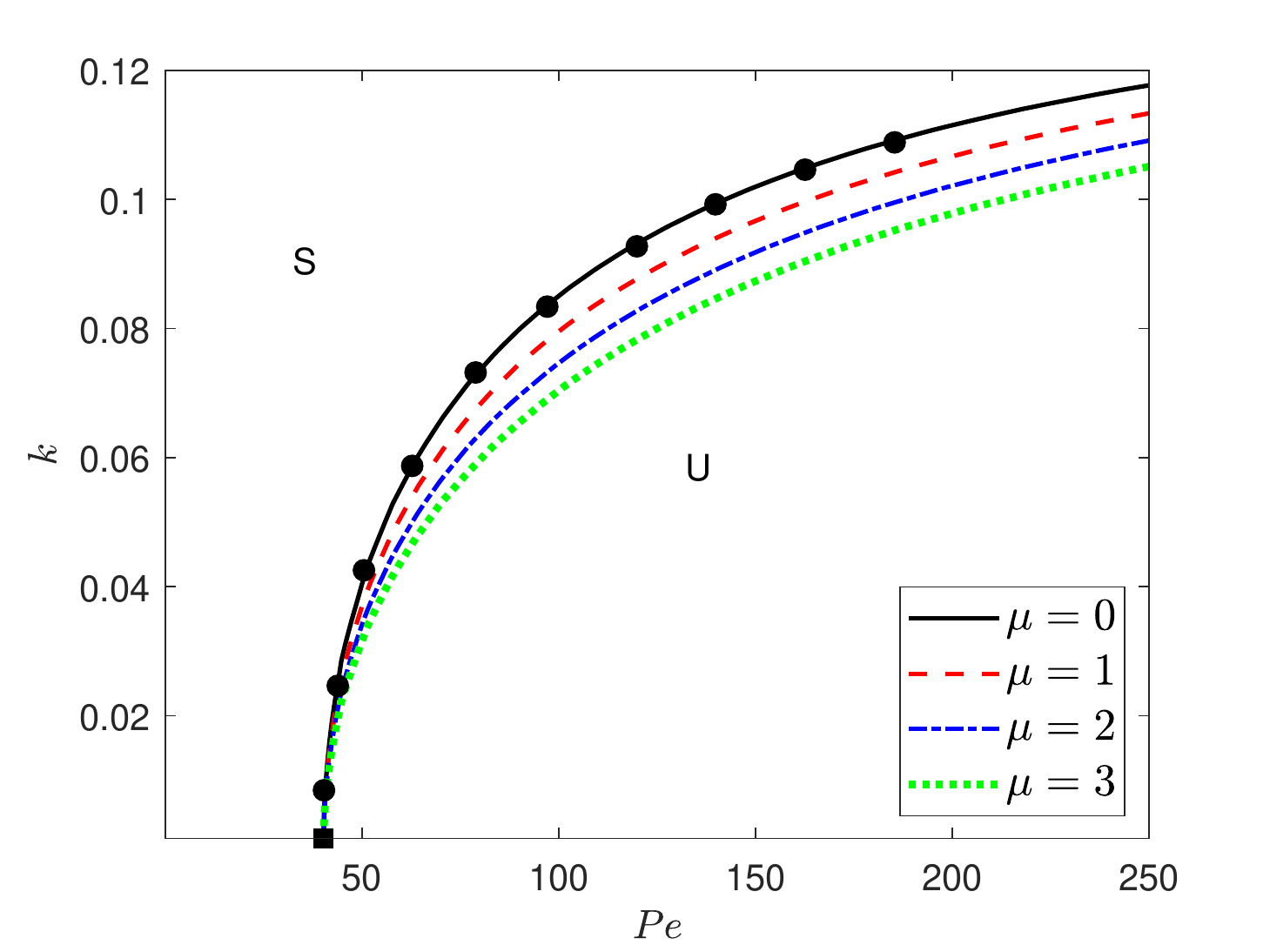}}
         \subfigure[]{\includegraphics*[width=7.4cm]{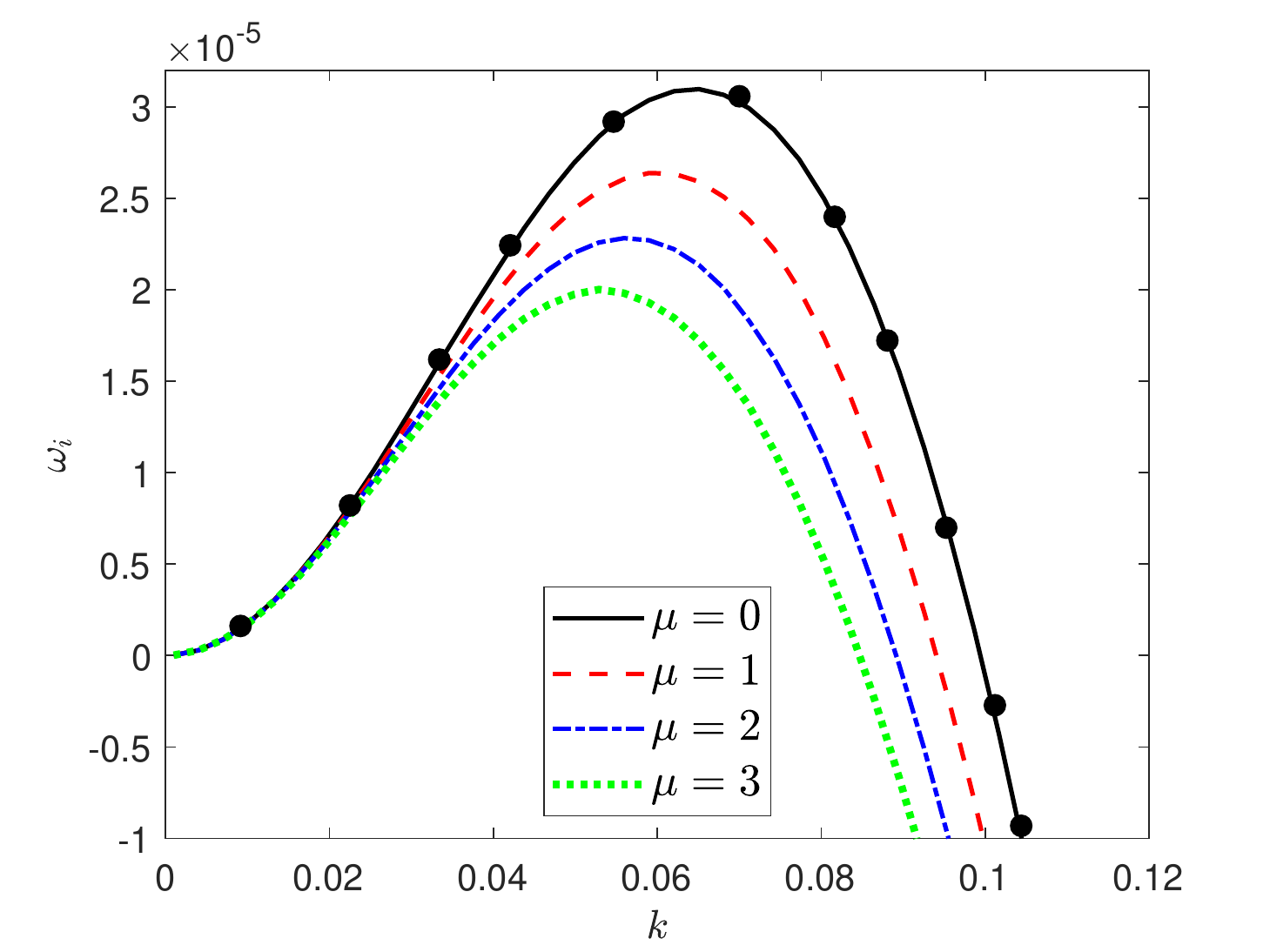}}
          \end{center}\vspace{-0.5cm}
    \caption{(a) The influence of odd viscosity coefficient $\mu$ on the marginal stability curves, in the wavenumber $k$ versus the P\'eclet number $Pe$ plane. (b) The corresponding growth rate $\omega_i$ versus the wavenumber $k$ when $Pe=140$. Here the fixed values are $Re=3$, $\alpha=0$, $\beta=4^{\circ}$, $Ca=2$, $Ma=0.1$, and $\tau=1$. The solid rectangular refers to the critical P\'eclet number $Pe_c$. The solid black circles are the result \citet{bhat2019linear} (Fig.~11(a) of their paper).}\label{fig10}
\end{figure}

In this section, we have discussed the important instability behaviour of the co-existent surfactant mode, as shown in Fig.~\ref{fig3}(b). To validate the surfactant mode result with \citet{bhat2019linear}, the numerical result (see Fig.~\ref{fig10}) is computed for $\alpha=0$, $\beta=4^{\circ}$, $Ca=2$, $Ma=0.1$, $Re=3$, and $\tau=1$.
Fig.~\ref{fig10}(a) displays that the current marginal stability curve fully matches with the result derived by \citet{bhat2019linear} in the limit $\tau=1$.
It is evident from the numerical outcomes that the critical P\'eclet number is an independent function of the odd viscosity coefficient $\mu$. Note that the finite wavenumber unstable range ($k\nrightarrow 0$) of Marangoni mode attenuates as the value $\mu$ increases and results in the stabilization impact of the viscosity ratio $\mu$. Besides, the corresponding growth rate curves plotted in Fig.~\ref{fig10}(b) totally agree with the outcome of Fig.~\ref{fig10}(a). The odd viscosity coefficient $\mu$  decelerates the maximum growth rate of the surfactant mode in the finite wavenumber range.

 
\begin{figure}[ht!]
    \begin{center}
        \subfigure[]{\includegraphics*[width=7.4cm]{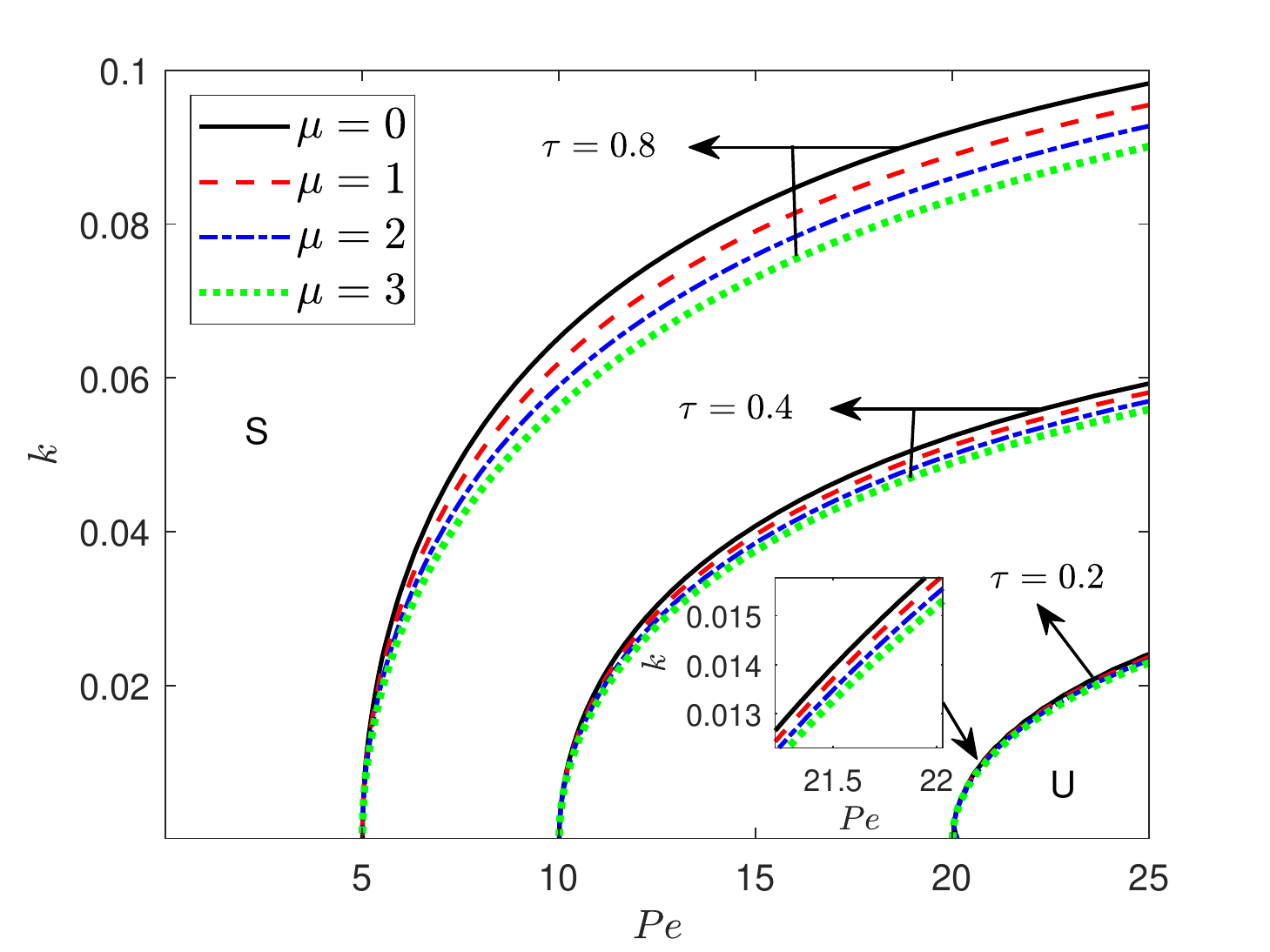}}
         \subfigure[]{\includegraphics*[width=7.4cm]{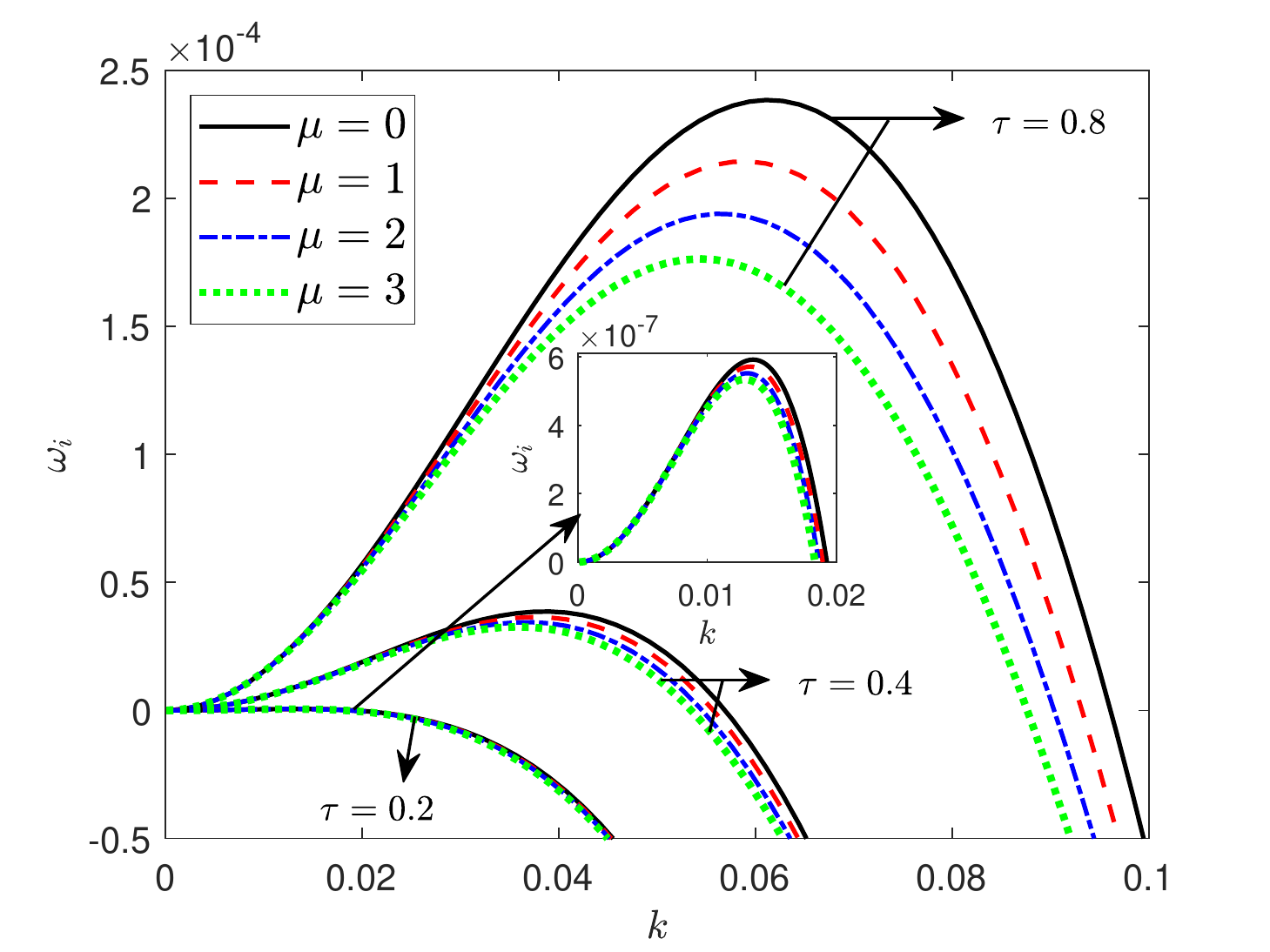}}
          \end{center}\vspace{-0.5cm}
    \caption{(a) The effect of odd viscosity on the marginal stability curve related to the surfactant mode in the ($Pe,k$) plane for different values of imposed shear $\tau$. (b) Corresponding growth rate $\omega_i$ as a function of wavenumber $k$ with $Pe=23$. The other parameter values are $\alpha=0.01$, $Re=20$, $\beta=4^{\circ}$, $Ca=2$, and $Ma=1$.}\label{fig11}
\end{figure}

The numerical approach is performed to study the odd viscosity influence on the primary instability of the Marangoni mode (see Fig.~\ref{fig11}). Fig.~\ref{fig11}(a) exhibits the marginal curves in the ($Pe, k$) plane of the surfactant mode for different odd viscosity ratios when imposed shear $\tau$ alters.
For each value of the odd viscosity coefficient $\mu$, the potent external shear ($+\tau$) acting in the streamwise direction magnifies the unstable bandwidth of the surfactant mode by reducing the critical P\'eclet number $Pe_c$. This outcome follows the destabilizing impact of external shear in the flow direction on the surfactant mode in both longwave and finite wavenumber ranges, which is similar to the result observed by \citet{bhat2019linear}. The main physical cause behind the destabilizing effect is that the positive external shear ($+\tau$) advances the perturbation energy of the insoluble surfactant. Indeed, the odd viscosity coefficient $\mu$ reduces the surfactant mode instability in the finite wavenumber zone but not in the longwave zone. To justify Fig.~\ref{fig11}(a), the corresponding growth rate is displayed in Fig.~\ref{fig11}(b) when the P\'eclet number $Pe$ is chosen from the unstable zone.
The surfactant mode growth rate decreases for higher odd viscosity $\mu$, and the opposite trend is observed for the stronger externally applied shear in the flow direction, which is entirely consistent with Fig.~\ref{fig11}(a).

\begin{figure}[ht!]
    \begin{center}
        \subfigure[$Ma=1$]{\includegraphics*[width=5.4cm]{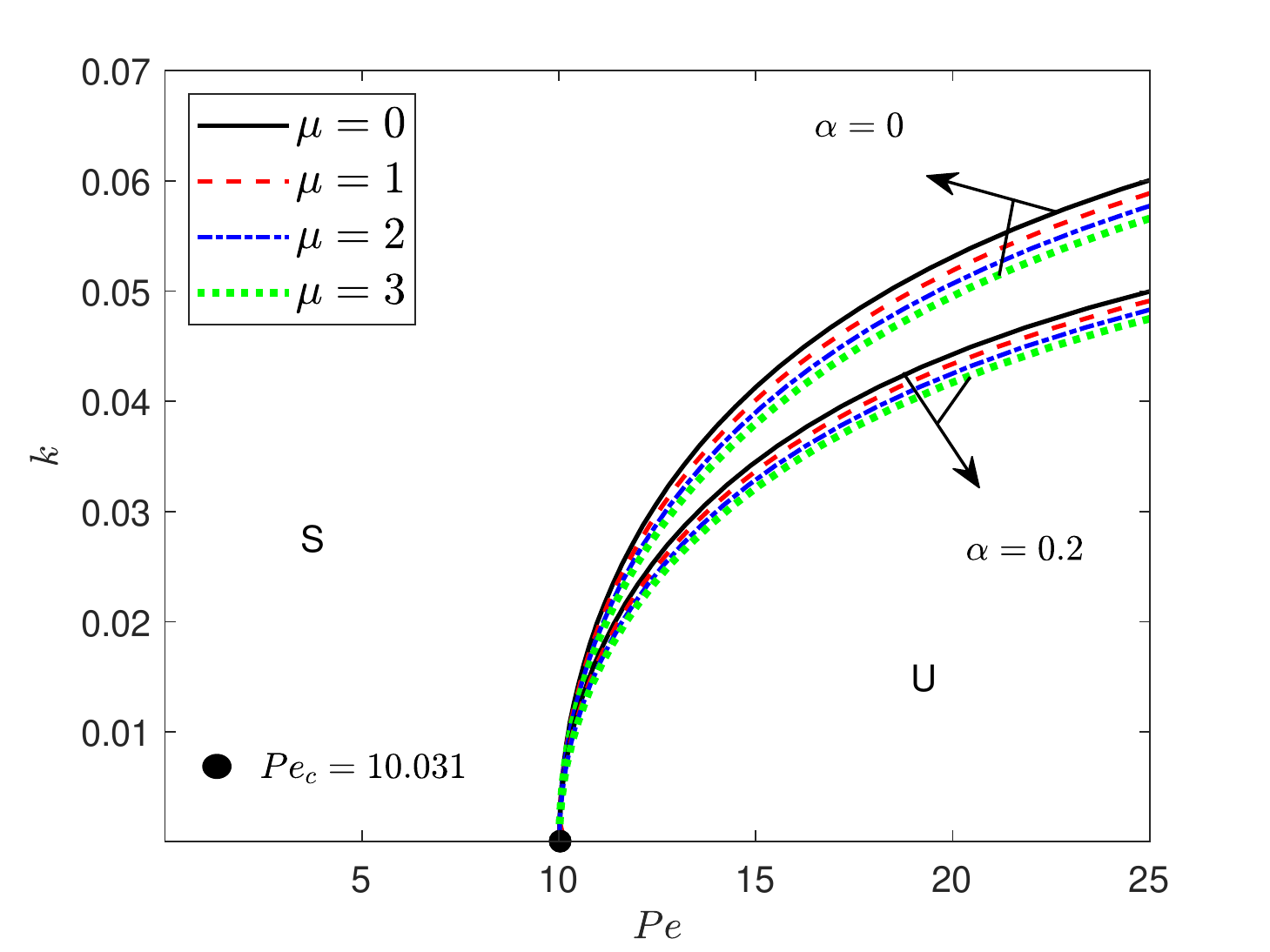}}
         \subfigure[$Ma=2$]{\includegraphics*[width=5.4cm]{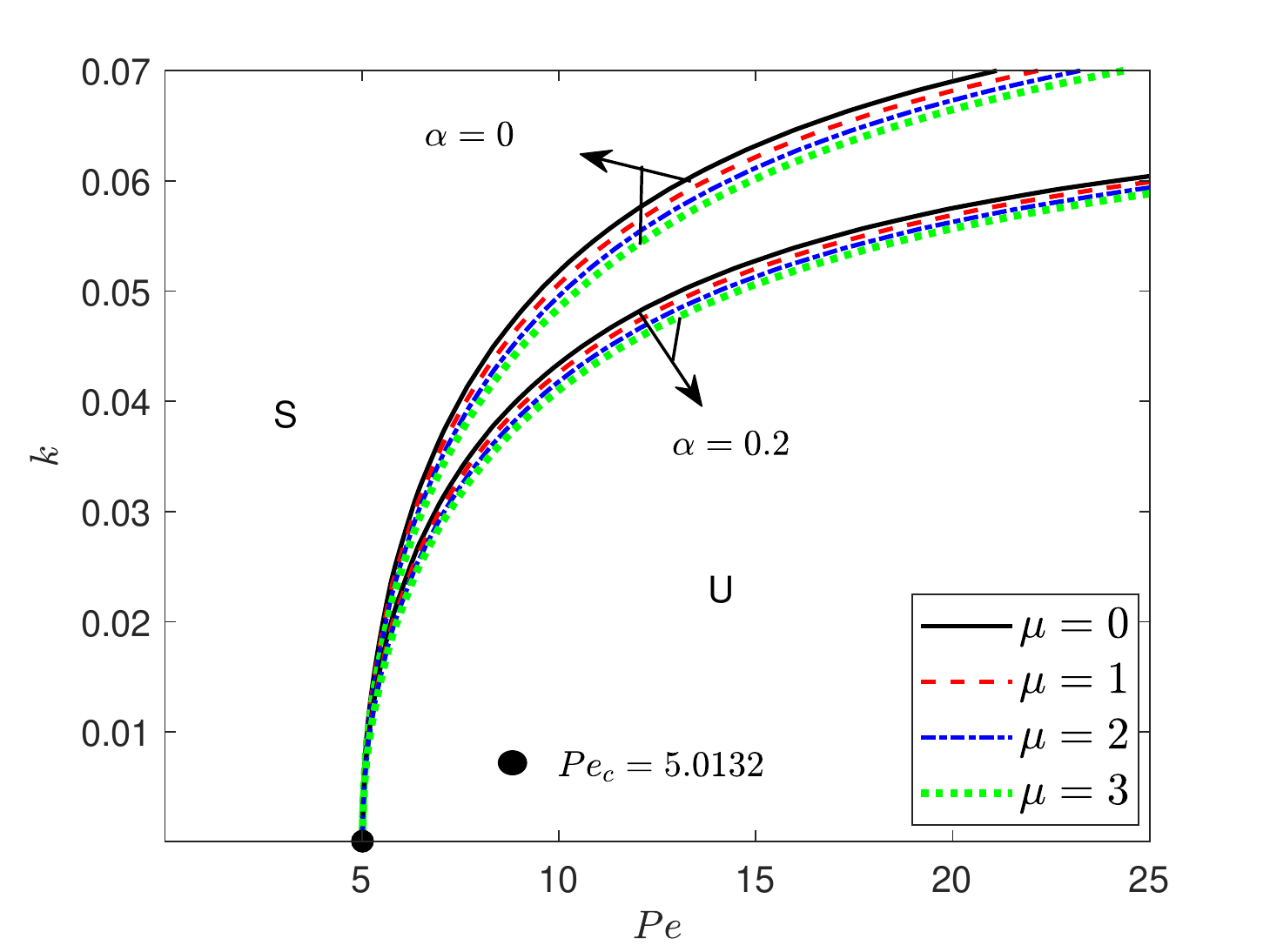}}
         \subfigure[$Ma=3$]{\includegraphics*[width=5.4cm]{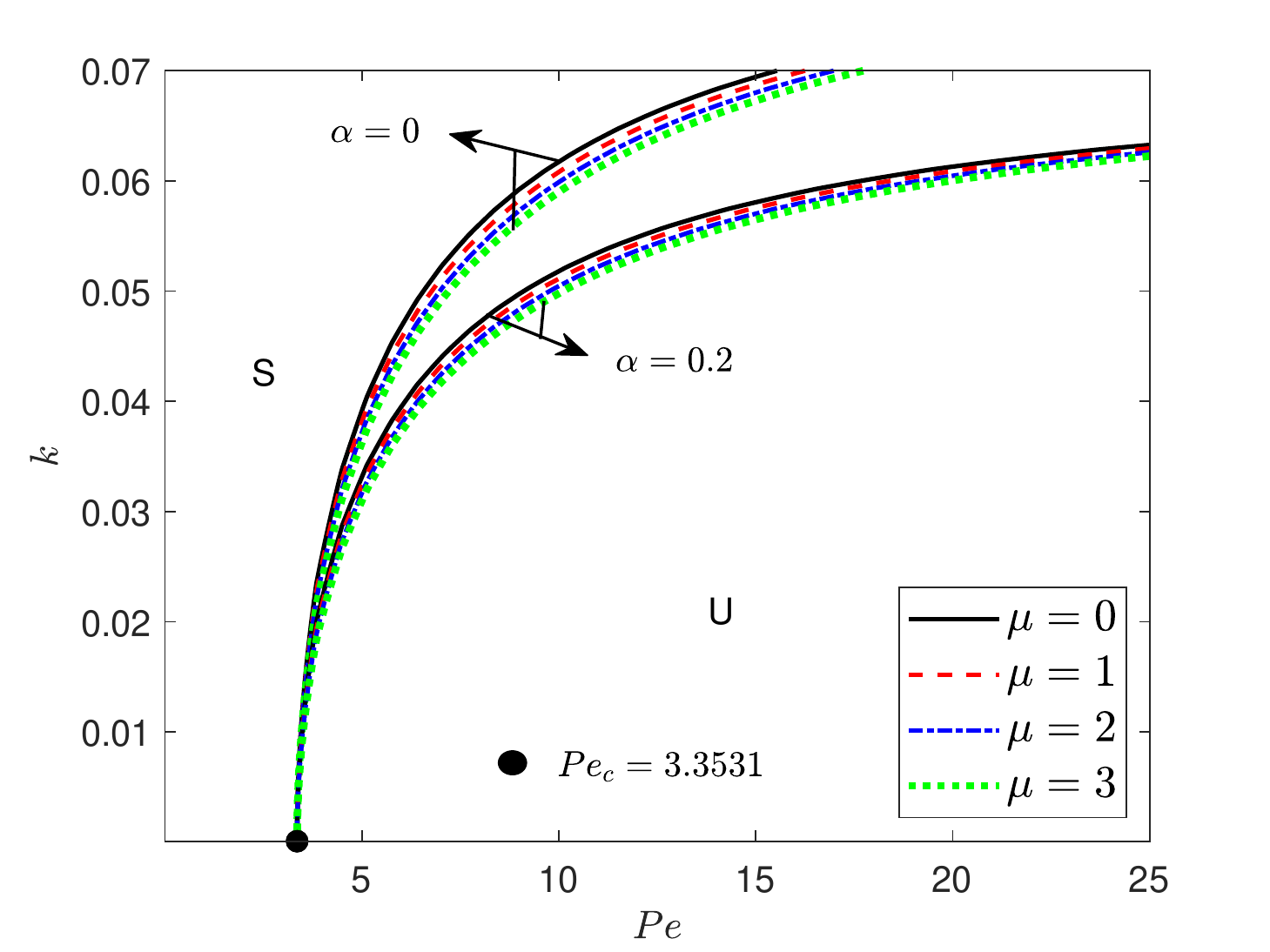}}
          \end{center}\vspace{-0.5cm}
    \caption{The variation of marginal stability curves associated with the surfactant mode for different values of odd viscosity $\mu$. The common parameters are $\tau=0.4$, $Re=20$, $\beta=4^{\circ}$, and $Ca=2$. The solid black circles indicate the corresponding critical P\'eclet number $Pe_c$.}\label{fig12}
\end{figure}
\begin{figure}[ht!]
    \begin{center}
        \subfigure[$\mu=0$]{\includegraphics*[width=5.4cm]{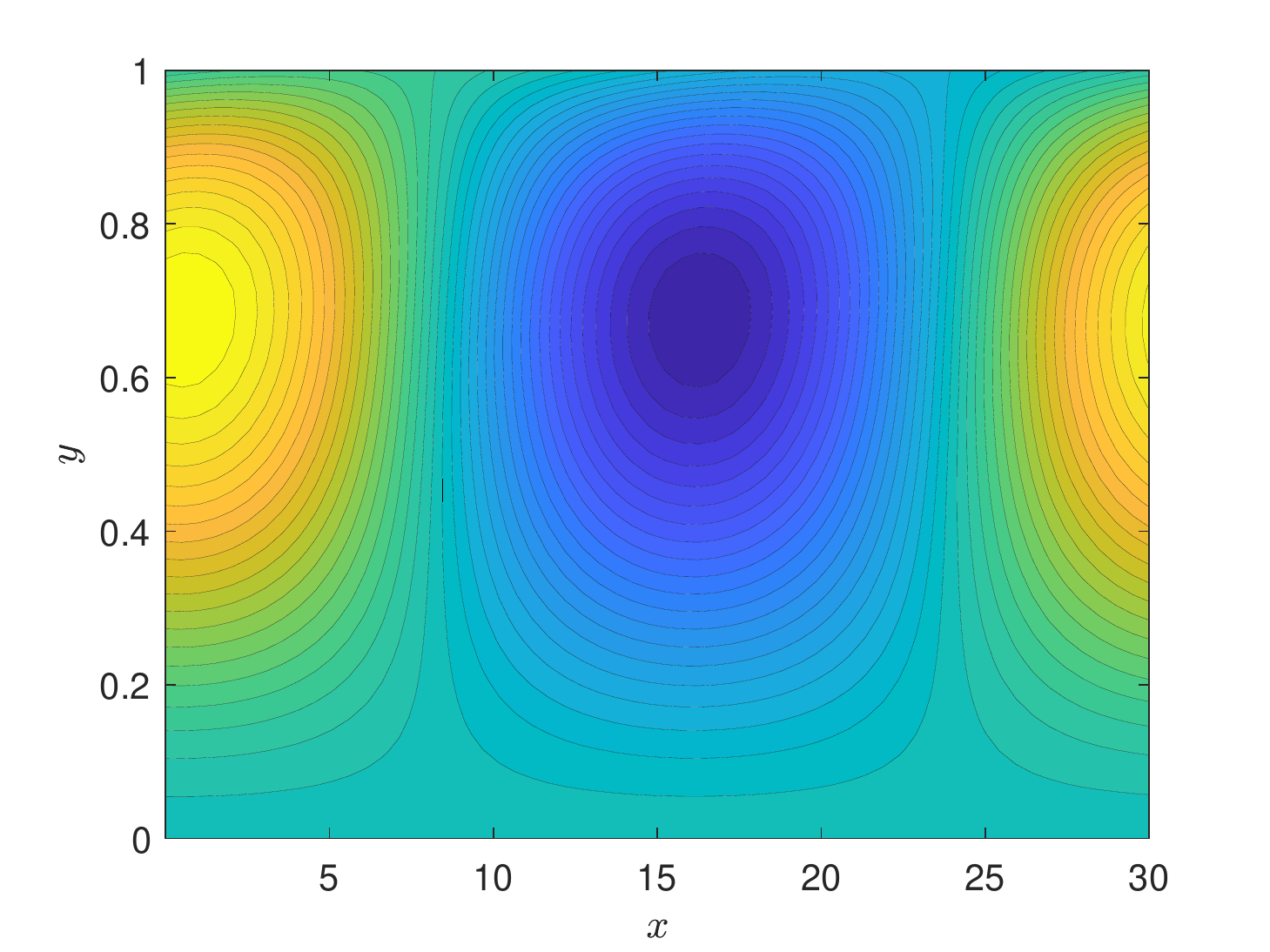}}
         \subfigure[$\mu=1$]{\includegraphics*[width=5.4cm]{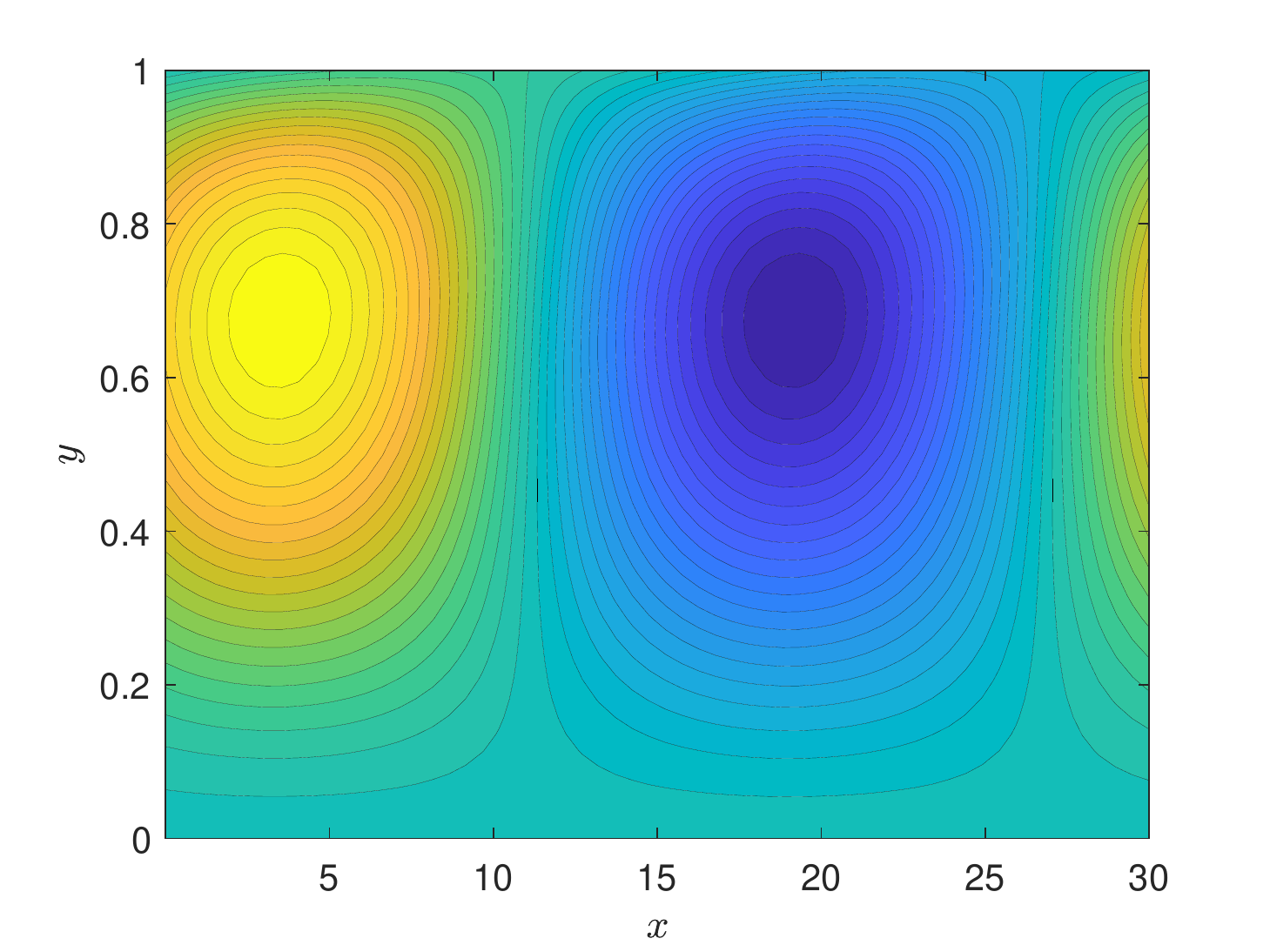}}
         \subfigure[$\mu=2$]{\includegraphics*[width=5.4cm]{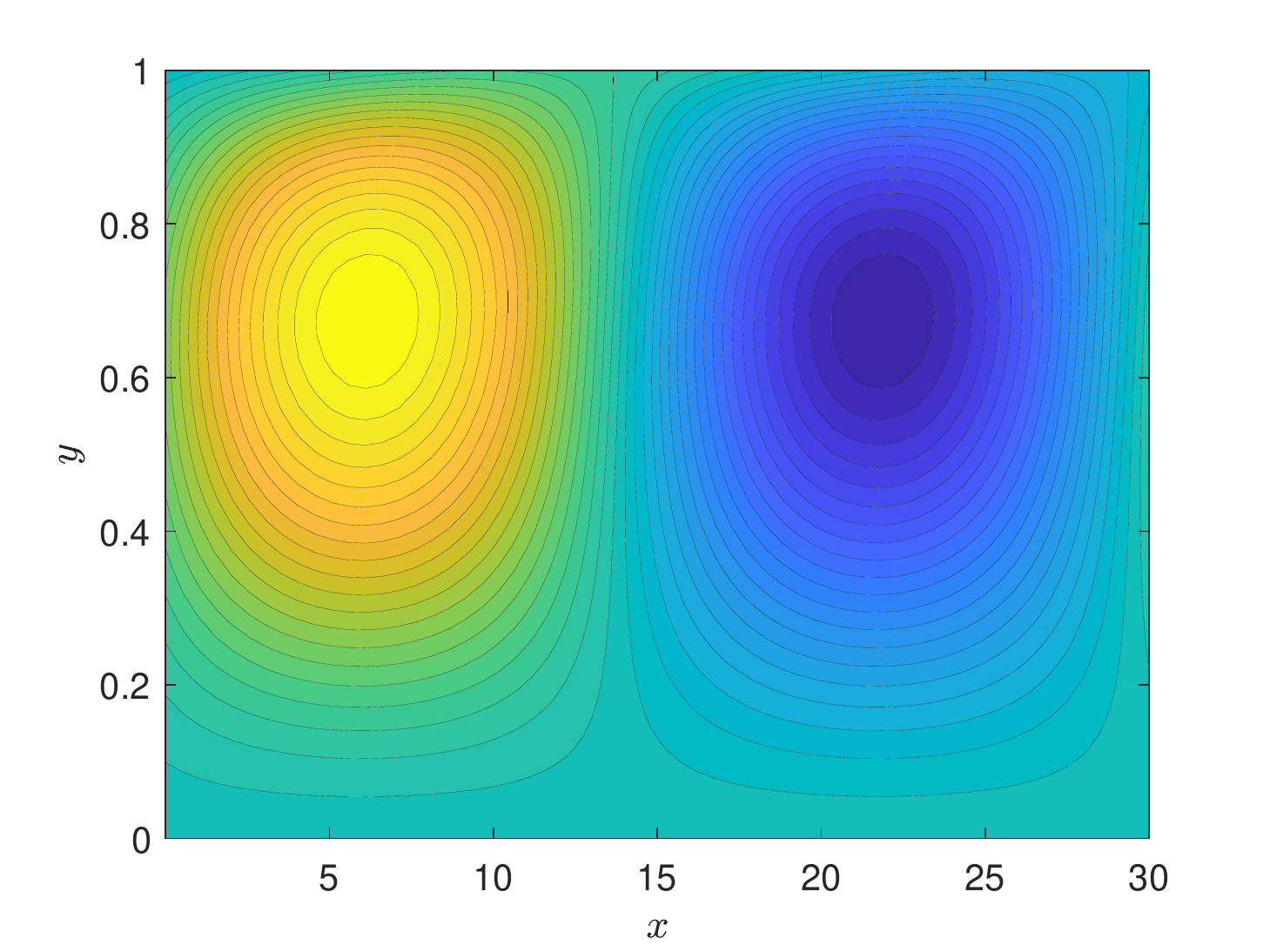}}
          \end{center}\vspace{-0.5cm}
    \caption{Velocity distribution associated with the surfactant mode for different values of odd viscosity $\mu$. The common parameters are $k=0.2$, $\tau=0.4$, $\alpha=0.01$, $Re=20$, $\beta=4^{\circ}$, $Ca=2$, $Ma=1$, and $Pe=20$. }\label{fig13}
\end{figure}

The variations of marginal curves in ($Pe,\,k$) plane for different $Ma$ values with different odd viscosity $\mu$ are illustrated in Fig.~\ref{fig12} for both rigid bottom ($\alpha=0$) and slippery bottom ($\alpha\neq0$). On comparing the curves in Figs.~\ref{fig12}(a) ( $Ma=1$), (b) ($Ma=2$), and (c) ($Ma=3$), it is found that the unstable surfactant mode bandwidth drastically changes as long as the Marangoni number $Ma$ changes. The unstable region of the surfactant mode amplifies for the higher $Ma$ value by decreasing the critical P\'eclet number $Pe_c$. So, one can conclude that the Marangoni number has the potential to boost surfactant mode instability. Physically, the surfactant concentration quickly evolves at the liquid surface as the Marangoni force increases.
Moreover, for each Marangoni number $Ma$, the odd viscosity $\mu$ reduces the surfactant mode instability in the finite wavenumber zone due to the increment in the unstable surfactant mode bandwidth. Another important observation from these figures is that for each $Ma$ and $\mu$ value, the slip parameter $\alpha$ reduces the instability of surfactant mode in the finite wavenumber range. The slip parameter does not influence the $Pe_c$ of the surfactant mode. Also, whether the reversal symmetry of time breaks or not (i.e., $\mu=0$ or $\mu\neq0$), the slippery bottom significantly restricts the propagation of surfactant concentration in the finite wavenumber range.  

The isolines for the horizontal velocity components of the surfactant mode, as  shown in Fig.~\ref{fig13}, show that the odd viscosity $\mu$ shifts maximum horizontal velocity perturbation in the opposite direction of the fluid flow. Conclusively, we can highlight two important circumstances of the surfactant mode: (i) a potent destabilization is possible if higher external shear imposes in the downstream direction with strong Marangoni force, and (ii) the destabilizing effect of Marangoni force can be mitigated by both the odd viscosity coefficient as well as the slippery bottom.

\subsection{\bf{Shear mode}}

In this subsection, we are interested in studying the behaviour of shear-imposed odd viscous induced falling film flow instability induced by identified unstable shear mode, as shown in Fig.~\ref{fig3}(c), which emerges numerically when $Re$ value becomes very high with a low inclination angle. 
\begin{figure}[ht!]
    \begin{center}
        \subfigure[$\mu=0$]{\includegraphics*[width=7.4cm]{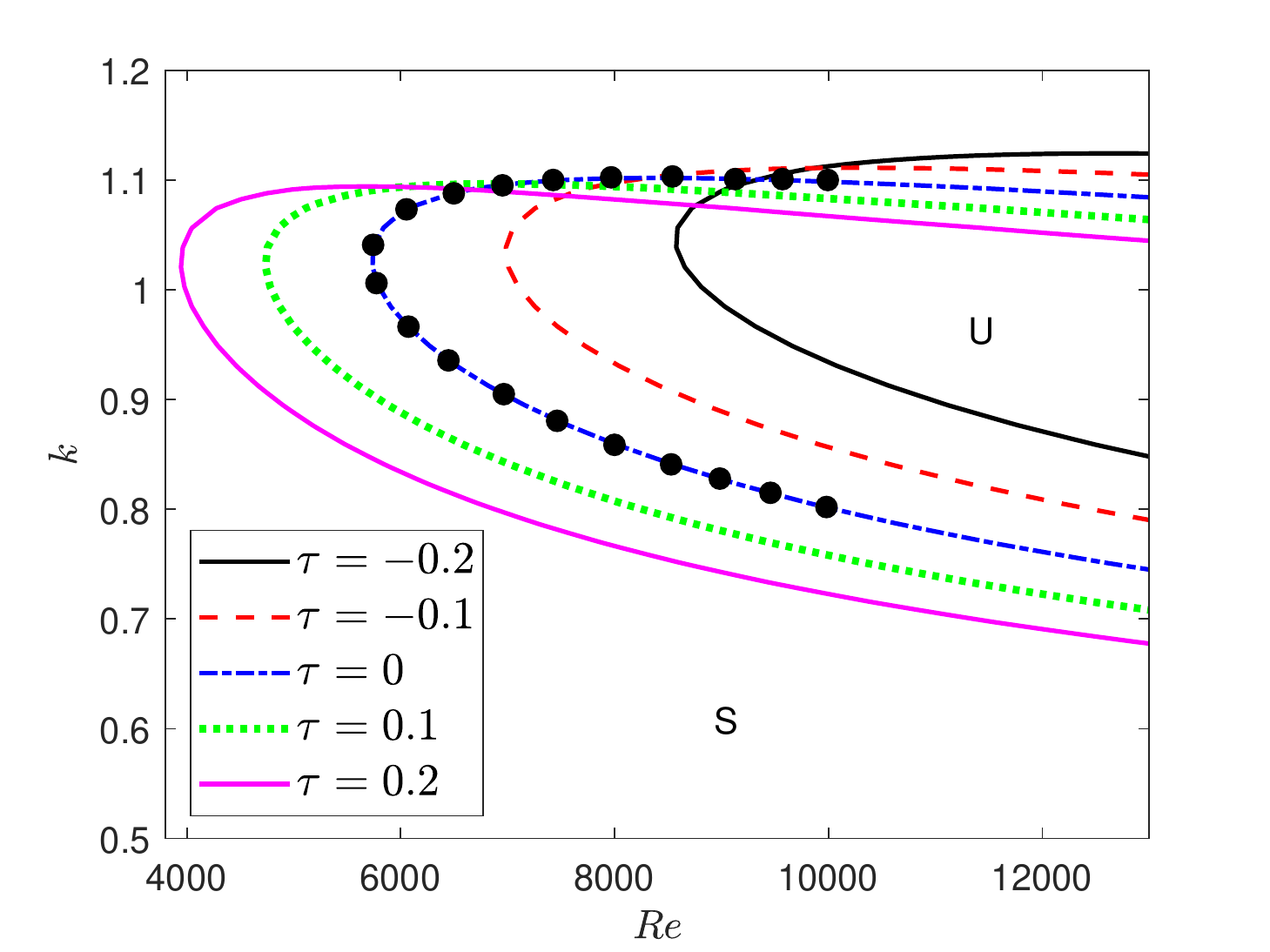}}
        \subfigure[$\mu=2$]{\includegraphics*[width=7.4cm]{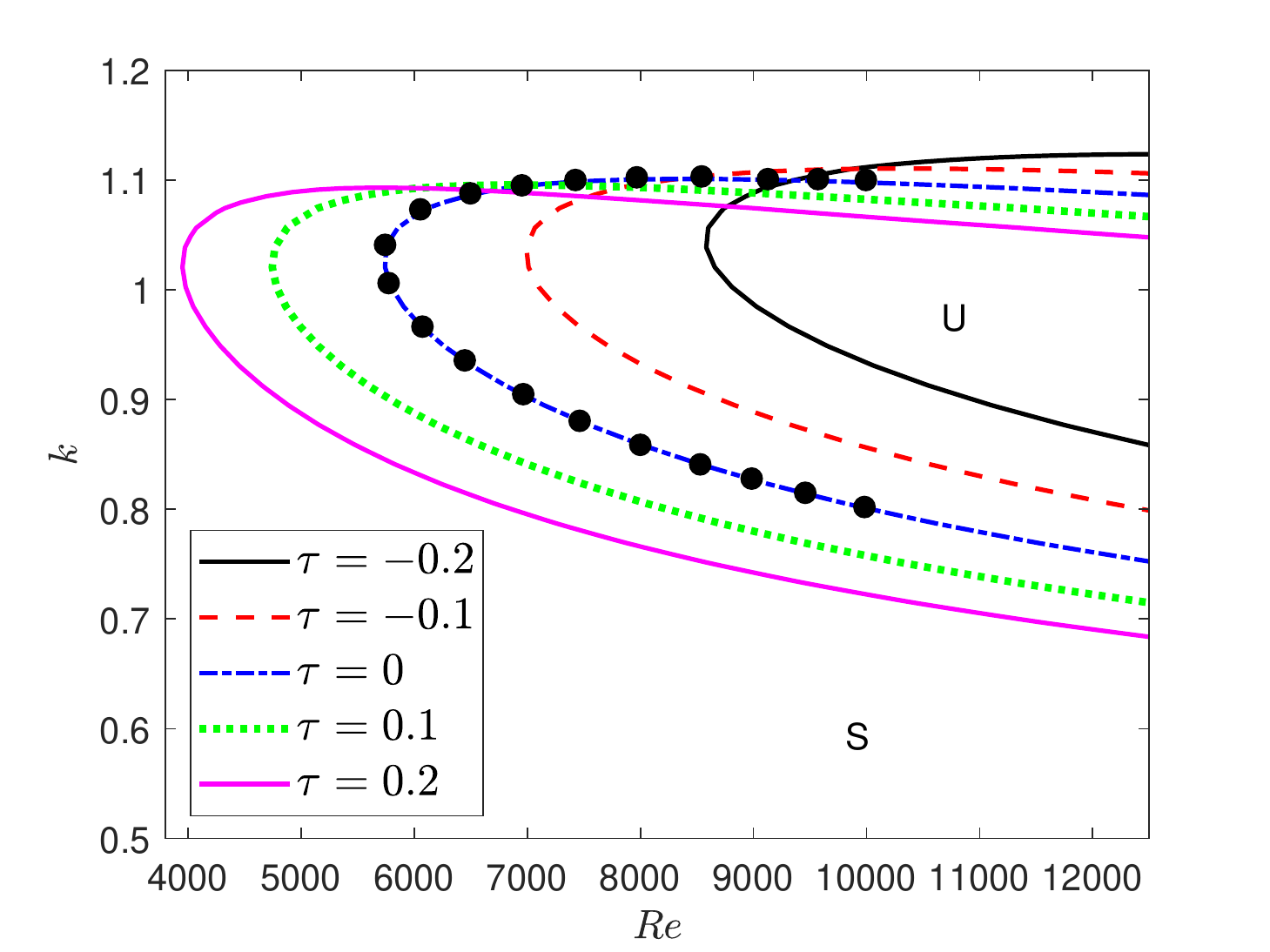}}
    \end{center}\vspace{-0.5cm}
    \caption{ The variation of marginal stability curves of the shear mode for different imposed shear $\tau$, when (a) the odd viscosity coefficient is absent (i.e., $\mu=0$), and (b) the odd viscosity coefficient $\mu$ is present (i.e., $\mu\neq0$) in the viscous falling film. The common parameters are $\alpha=0$, $\beta=4^{\circ}$, $Ca=2$, $Ma=1$, and $Pe=10000$. The solid black circles stand for the results of \citet{bhat2019linear} (Fig.~13 of their paper). }\label{fig14}
\end{figure}
To continue the discussion about the shear mode instability results, we have validated the numerical investigation by comparing our result with the previous outcomes derived by \citet{bhat2019linear}, when the fixed parameters are $\alpha=0$, $\beta=4^{\circ}$, $Ma=1$, $Pe=10000$, and $Ca=2$. \citet{bhat2019linear} considered a contaminated falling liquid down a rigid bed, where the external shear is applied on the contaminated surface, and the odd viscosity is avoided in the viscous fluid. The neutral stability curve in  Fig.~\ref{fig14}(a) totally matches with the results of \citet{bhat2019linear} when $\mu=0$, $\alpha=0$, and $\tau=0$.

\begin{figure}[ht!]
    \begin{center}
        \subfigure[]{\includegraphics*[width=7.4cm]{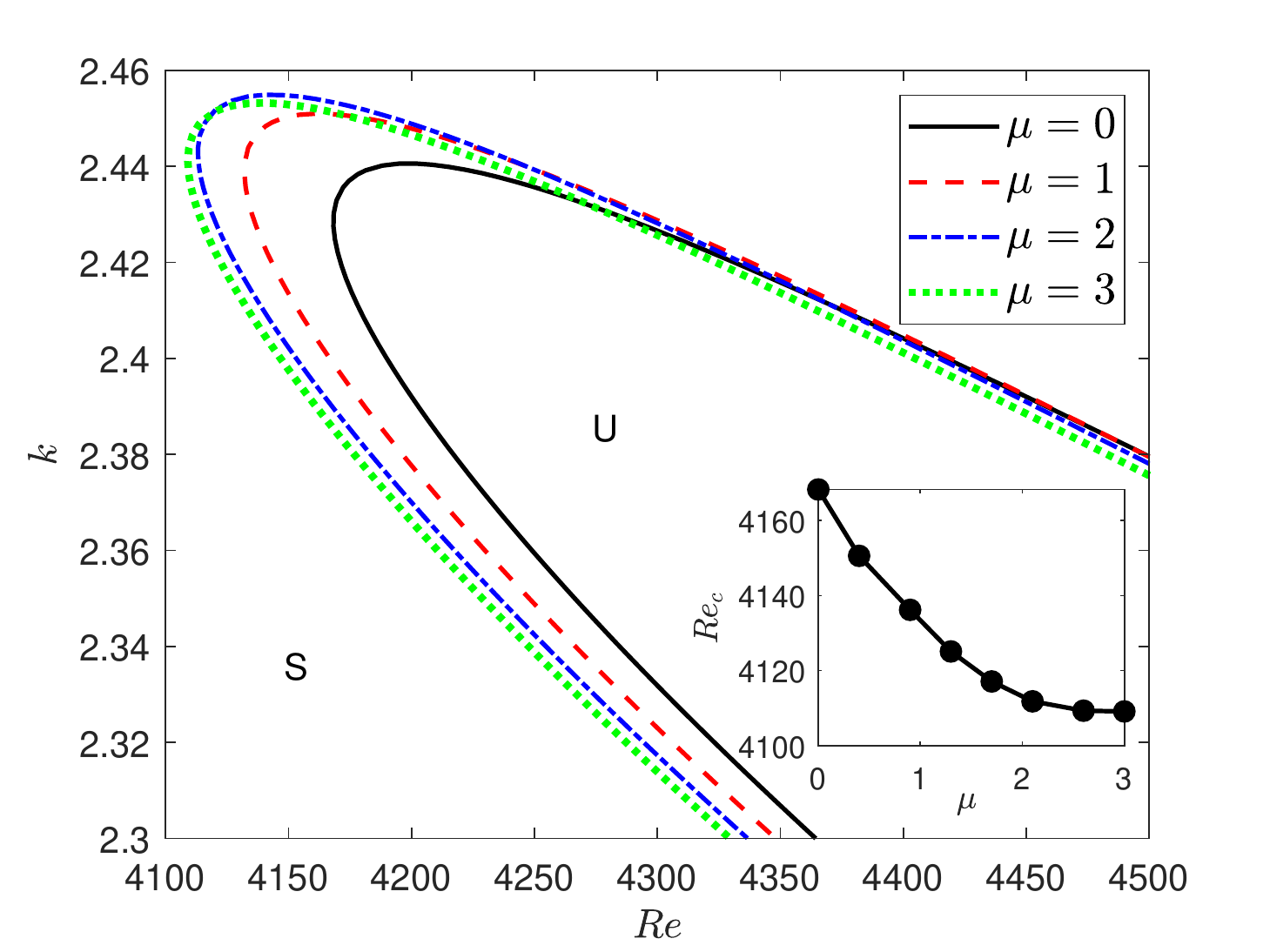}}
         \subfigure[]{\includegraphics*[width=7.4cm]{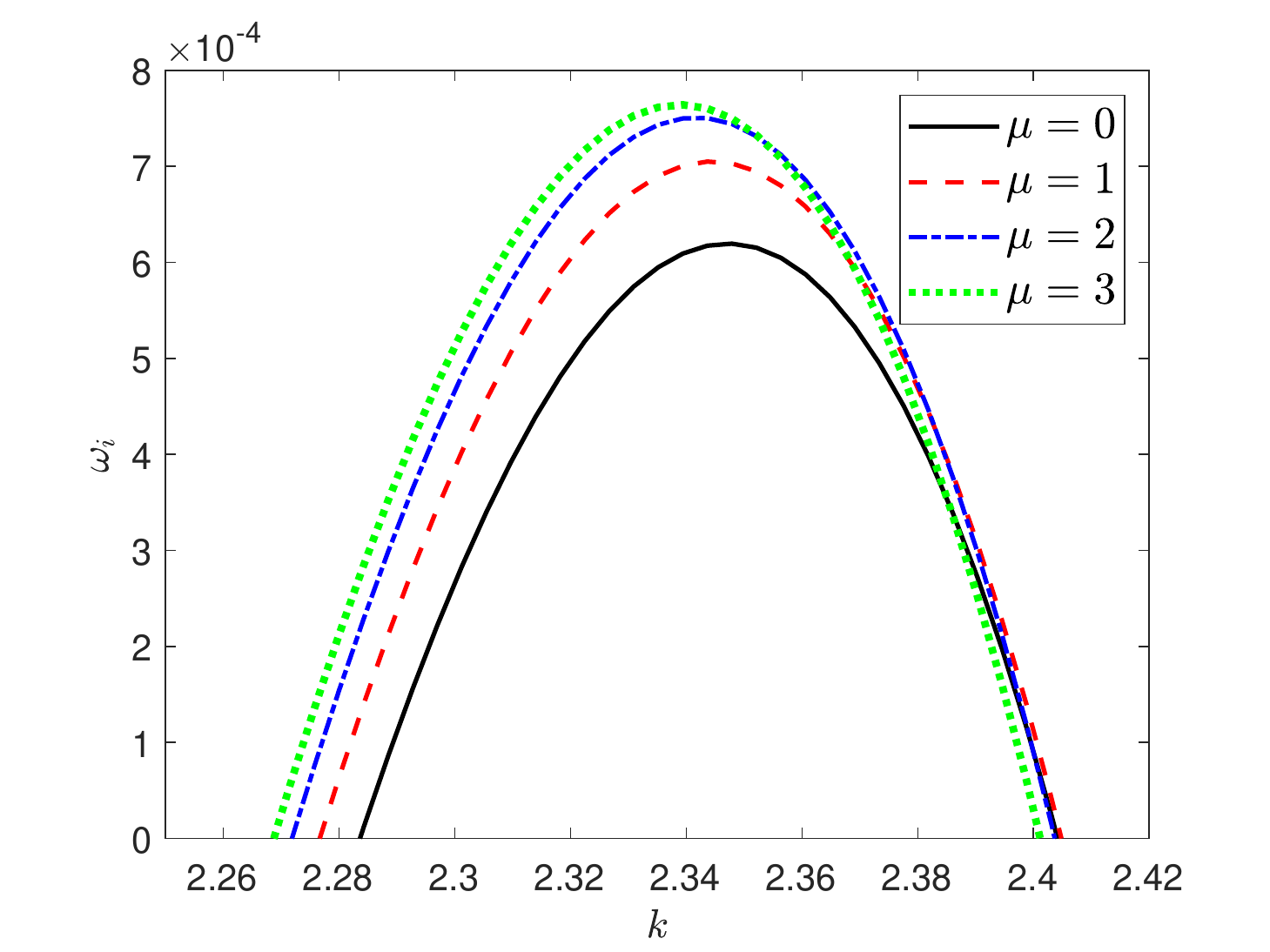}}
    \end{center}\vspace{-0.5cm}
    \caption{(a) The marginal curves associated with the shear mode for different odd viscosity ($\mu$) values. (b) The variation of temporal growth rate when $Re=4400$. The common parameters are $\tau=0.2$, $\alpha=0.01$, $\beta=1^{'}(=\frac{1}{60^{\circ}})$, $Ma=1$, $Ca=2$, and $Pe=1000$. }\label{fig15}
\end{figure}

In the finite wavenumber domain, the flow-directed external shear expands the bandwidth of the unstable zone excited by the shear mode up to a certain wavenumber $k$, and thereafter it shrinks that unstable region. Thus, the external shear plays a double role in the primary instability of the shear mode. Again, the numerical test is repeated to observe the effect of imposed shear (Fig.~\ref{fig14}(b)) when $\mu$ is present in the shear-imposed contaminated viscous falling film. Even though the broken-time reversal symmetry is present (i.e., the odd viscosity coefficient $\mu\neq0$), the similar influence of imposed shear $\tau$, as in Fig.~\ref{fig14}(a), on the shear mode induced unstable regime is also observed in Fig.~\ref{fig14}(b). Here the odd viscosity has a negligible impact on the shear mode unless very low angle of inclination (see in the subsequent Fig.~\ref{fig15}).

\begin{figure}[ht!]
    \begin{center}
        \subfigure[]{\includegraphics*[width=7.4cm]{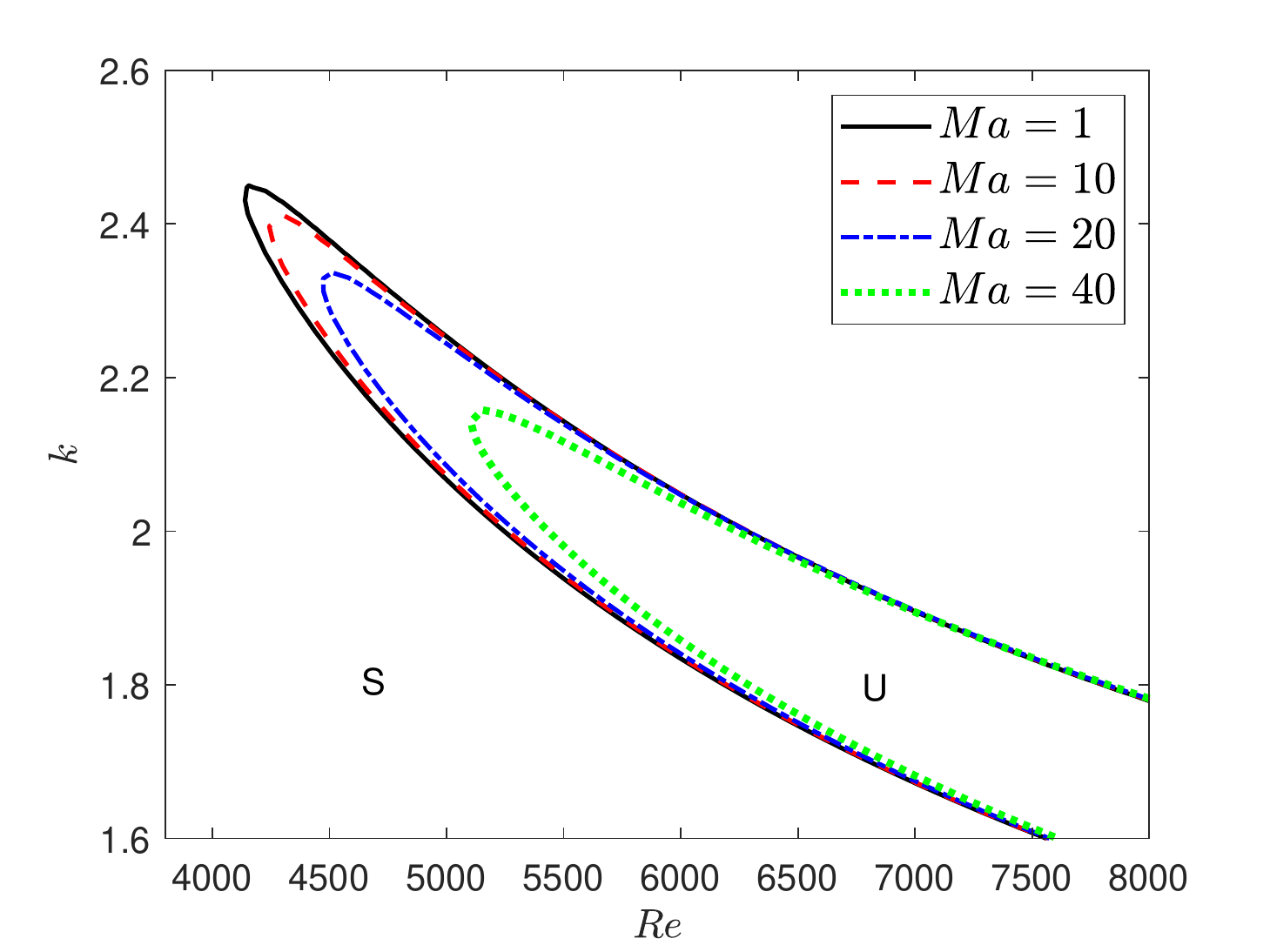}}
         \subfigure[]{\includegraphics*[width=7.4cm]{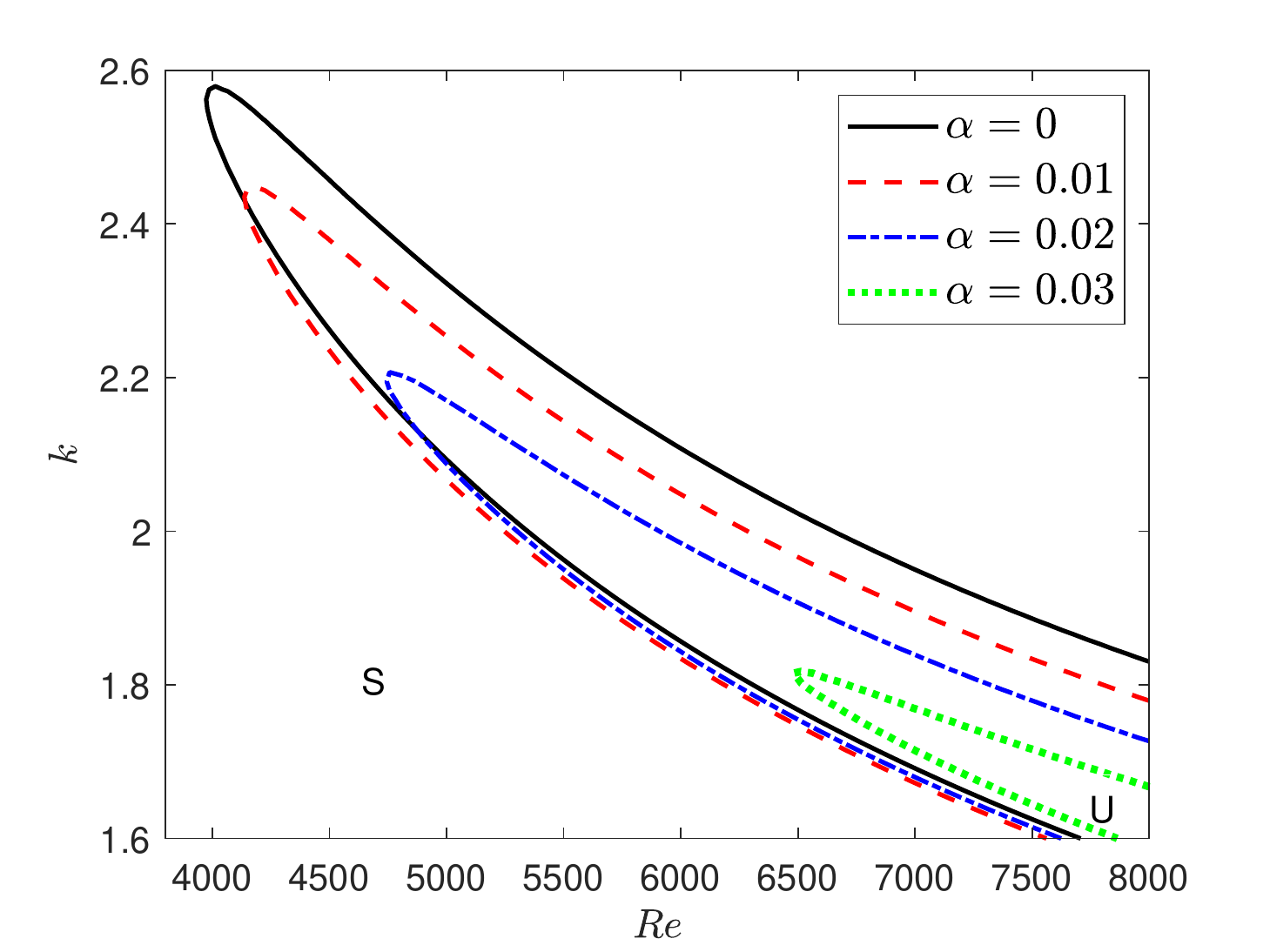}}
    \end{center}\vspace{-0.5cm}
    \caption{(a) The marginal curves associated with the shear mode for different $Ma$ values with $\alpha=0.01$. (b) The marginal curves for different $\alpha$ values with $Ma=1$. The common parameters are $\tau=0.2$, $\mu=1$, $\beta=1^{'}(=\frac{1}{60^{\circ}})$, $Ca=2$, and $Pe=1000$.}\label{fig16}
\end{figure}
In Fig.~\ref{fig15}, the variation of neutral curves of the shear mode and the corresponding growth rates for different $\mu$ values is illustrated when $\beta=1{'}$. The bandwidth of the unstable region in the higher wavenumber regime significantly amplifies for the higher odd viscosity ratio $\mu$ (see  Fig.~\ref{fig15}(a)). Therefore, the odd viscosity $\mu$ can increase the primary instability induced by the shear mode, which is further confirmed by the gradual depletion of the critical Reynolds number (see the inset plot of Fig.~\ref{fig15}(a)).  
To justify the above outcomes, the corresponding growth rate results are computed numerically and displayed in Fig.~\ref{fig15}(b). The higher value of the odd viscosity coefficient $\mu$ intensifies the maximum growth rate and yields the destabilizing nature of the shear mode. 

Fig.~\ref{fig16}(a) displays the variations of the marginal stability curve for different values of Marangoni number $Ma$, whereas the unstable bandwidth raised in the higher wavenumber region decreases gradually for higher $Ma$ value. Further, from Fig.~\ref{fig16}(b), it is found that a slight variation in the slip length $\alpha$ drastically changes the marginal stability curves in the higher wavenumber regime. The slip parameter depletes the instability instigated by the shear mode. Consequently, the slippery bottom has a stabilizing impact on the shear mode.

\subsection{\textbf{Mode competition}}
The main aim of this subsection is to observe the competition for dominance of different co-existence modes in various flow parameters ranges as discussed by \citet{bhat2018linear}, \citet{bhat2019linear} and \citet{ hossain2022linear}. 
\begin{figure}[ht!]
    \begin{center}
        \subfigure[]{\includegraphics*[width=7.4cm]{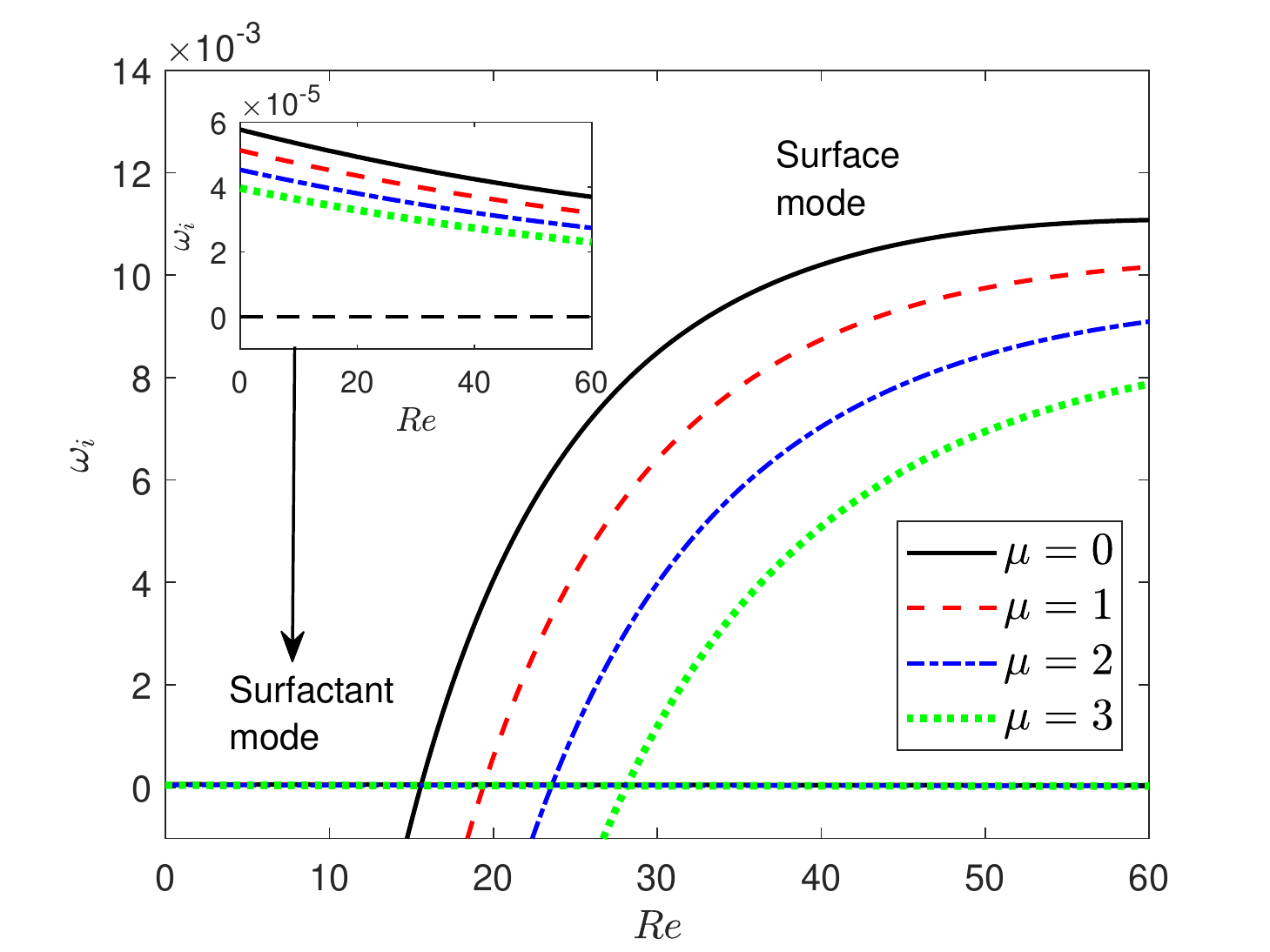}}
         \subfigure[]{\includegraphics*[width=7.4cm]{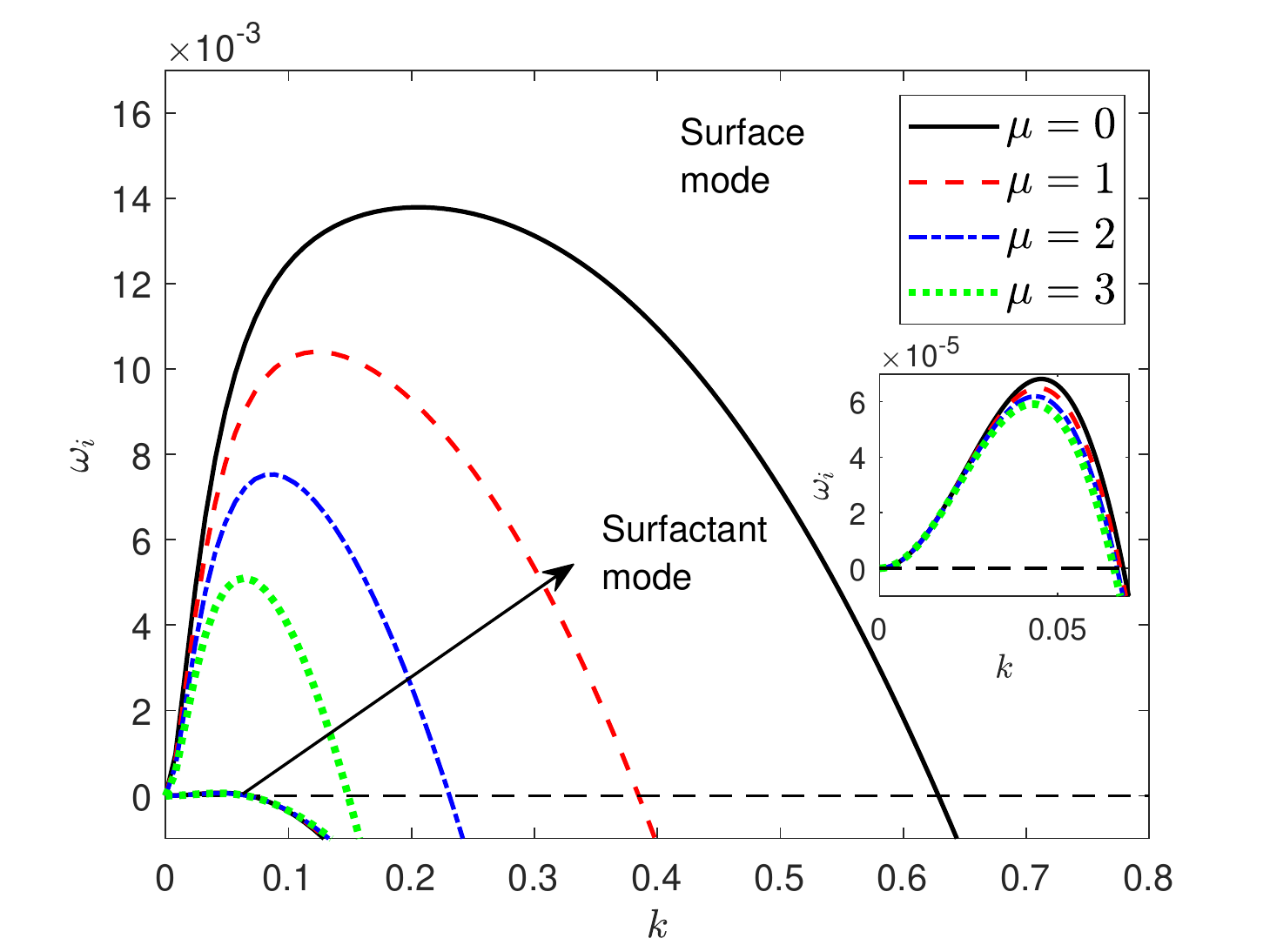}}
    \end{center}\vspace{-0.5cm}
    \caption{Competition for supremacy between the variation of unstable surface and surfactant modes for different values of odd viscosity $\mu$ when (a) $k=0.06$ and (b) $Re=40$. The common parameters are $\tau=0.4$, $\alpha=0.01$, $\beta=4^{\circ}$, $Ma=1$, $Ca=2$, and $Pe=40$. }\label{fig17}
\end{figure}

\begin{figure}[ht!]
    \begin{center}
    \subfigure[]{\includegraphics*[width=7.4cm]{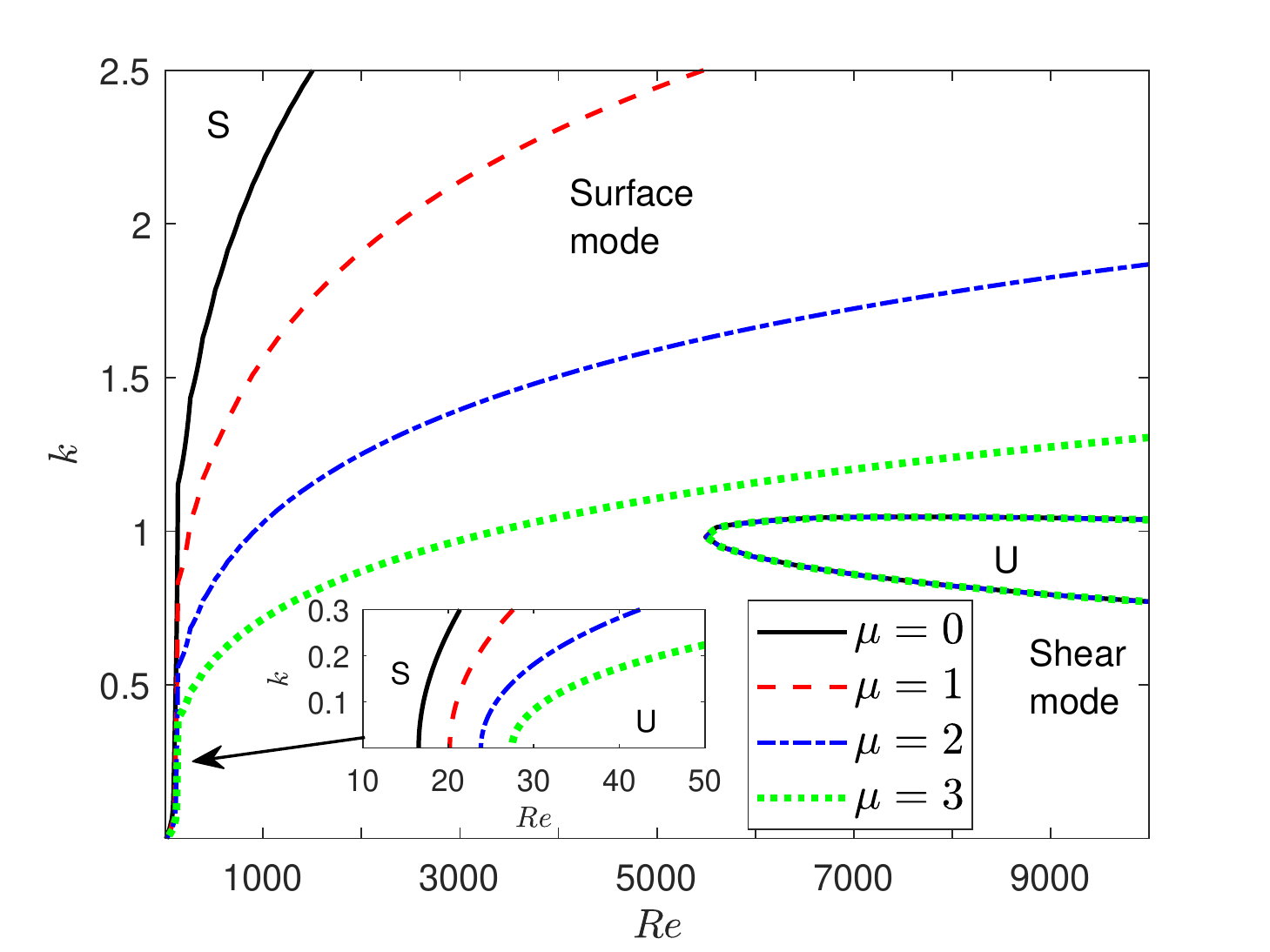}}
    \subfigure[]{\includegraphics*[width=7.4cm]{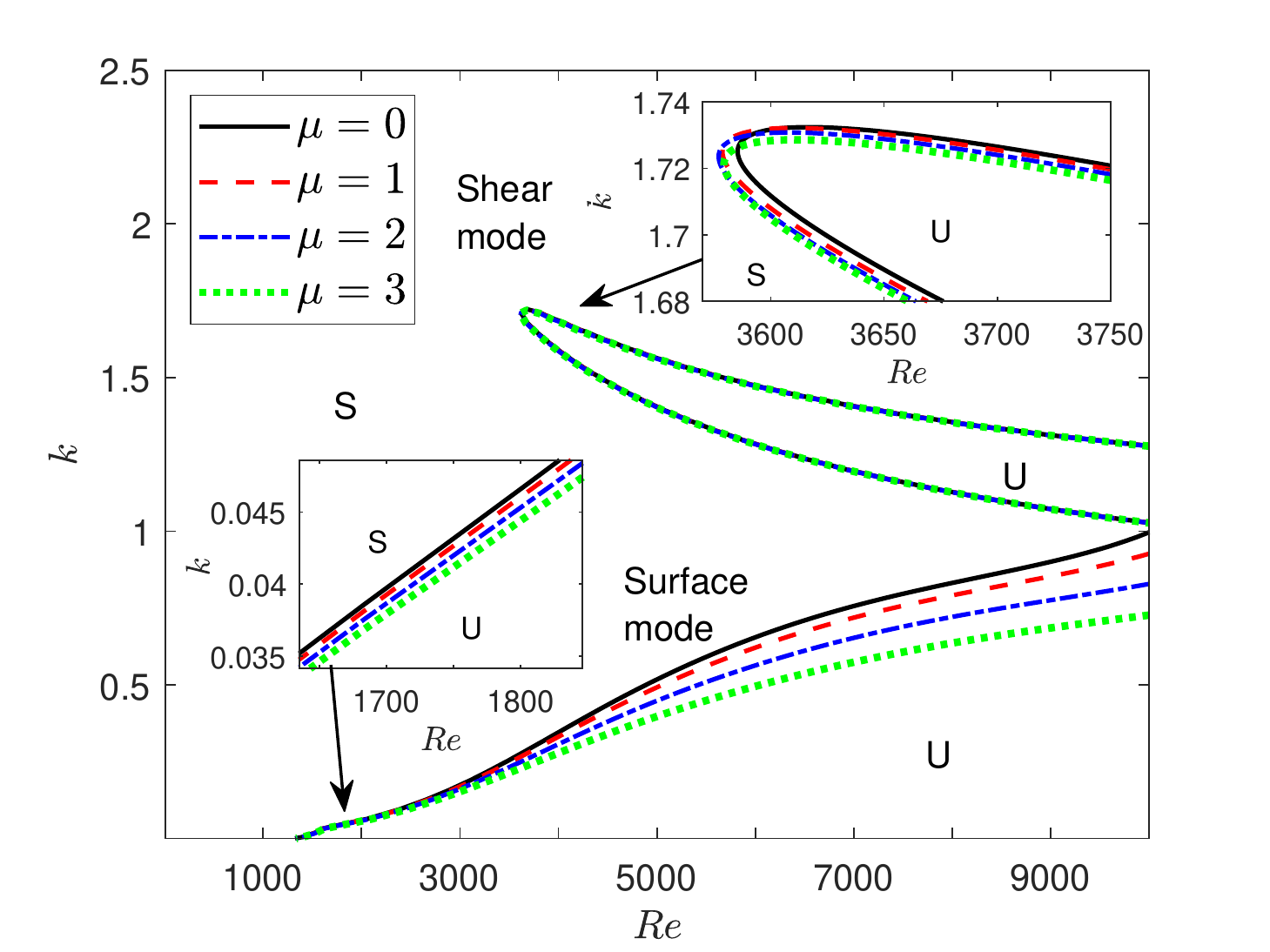}}
    \end{center}\vspace{-0.5cm}
    \caption{Competition for supremacy between the variation of unstable surface and shear modes for different values of odd viscosity $\mu$ when (a) $\beta=4^{\circ}$ and (b) $\beta=0.05^{\circ}$. The common parameters are $\tau=0.1$,  $\alpha=0.01$, $Ma=0.01$, $Ca=2$, and $Pe=1000$. }\label{fig18}
\end{figure}
The temporal growth rates versus Reynolds number of the surface, as well as surfactant modes for various viscosity ratios $\mu$, are plotted in Fig.~\ref{fig17}(a) when the fixed parameters are $\alpha=0.01$, $\tau=0.4$, $\beta=4^{\circ}$, $Pe=40$, $Ma=1$, and $Ca=2$.
The surface mode growth rate is negative, and the surfactant mode growth rate is positive up to the sufficiently small Reynolds number $Re$, which assures the dominance of surfactant mode up to the significantly small range of $Re$. Then on, as soon as the $Re$ increases, the surface mode growth rate ($\omega_i$) increases, and at a certain value of $Re$, the growth rate exceeds the marginal line ($\omega_i=0$) and causes the surface wave instability. The surface mode dominates the surfactant mode from that certain value of $Re$. The dominance of surface mode over the surfactant mode for sufficiently high Reynolds number is further confirmed from the temporal growth rate curves versus the wavenumber (Fig.~\ref{fig17}(b)) for different values of the odd viscosity coefficient when $Re=40$. The unstable surfactant mode exists only in the longwave zone, but the unstable surface mode exists in the longwave as well as the finite wavenumber regime. Therefore, the surface mode is fully dominant over the surfactant mode after a sufficiently small $Re$ value. Moreover, the higher values of the odd viscosity coefficient $\mu$ reduce the growth rate associated with the surface/surfactant mode instability. 
  
Now, the numerical technique is performed to observe the rivalry between the surface and shear modes for the primary instability when the  fixed values are $\tau=0.1$, $\alpha=0.01$, $Ma=0.01$, $Ca=2$, and $Pe=1000$. The marginal curves corresponding to the surface, as well as shear modes for different $\mu$, are exhibited in Fig.~\ref{fig18}. It is found that the higher $\mu$ value has a strong stabilization impact on the co-existing surface mode by reducing the unstable bandwidth but a weak stabilization/destabilization influence on the shear mode. As in Fig.~\ref{fig18}(a), the surface mode induced instability boundary lines fully occupy the boundary lines of the shear mode instability for all values of $Re$, when $\beta=4^{\circ}$. That means the surface mode totally dominates the shear mode in the whole $k$ domain (i.e., the longwave to higher wavenumber range). This fact assures that surface mode instability occurs quicker than shear mode instability and generates a higher growth rate for all $Re$ values. But, a different scenario is observed in Fig.~\ref{fig18}(b), when the inclination angle $\beta=0.05^{\circ}$. It is observed that the surface mode is dominant in the longwave regime, whereas the shear mode is dominant in the higher wavenumber regime. Indeed, compared to the surface mode, the shear mode is not much influenced by the odd viscosity $\mu$.
\section{Conclusions}\label{CON}
This study performs the wave dynamics of a surfactant-laden shear-imposed viscous fluid film over a slippery incline when the reversal symmetry of time breaks. The Orr-Sommerfeld eigenvalue problem corresponding to the fluid model is derived by employing the normal mode analysis to the linear perturbed equation and then solved based on the numerical Chebyshev spectral collocation technique. Basically, the linear stability/instability characteristics of the surface and surfactant modes are examined for the different flow parameter regions. Additionally, the linear response of the shear mode identified with a high Reynolds number and small angle of inclination is discussed. 

The odd viscosity coefficient shrinks the unstable longwave domain generated by the surface mode and promotes the stabilizing nature of the surface mode. Further, the surface mode instability can be advanced if one can apply the stronger external shear in the flow direction, and the opposite trend can be achieved by imposing the external shear opposite to the flow direction.
Although, surface mode instability can be weakened by introducing the stronger downstream-directed external shear at the liquid surface by choosing a significant value of the odd viscosity coefficient. Moreover, owing to the increment in base velocity, the slippery bottom raises the level of wave instability.

Besides, in the finite wavenumber range, the surfactant mode becomes more stable for higher odd viscosity. The effect of Marangoni force on all the modes identified in the present study is almost similar to the shear-imposed surfactant-contaminated falling film down a rigid substrate (\citet{bhat2019linear}), where odd viscosity is absent. Moreover, in the zone of finite wavenumbers, the slippery substrate helps to reduce the instability of the surfactant mode and has the capability of lowering the Marangoni impact on the damped mode. 
 
For a small inclination angle (i.e., when the gravitational force weakly acting upon the fluid is weak enough), a very high Reynolds number $Re$ triggers the inertia force and causes the emergence of the unstable shear mode in the higher wavenumber zone. 
The most excited waves caused by the shear mode can be stabilized by both the surfactant and the slippery bottom. Also, the viscosity ratio increases the shear wave speed and intensifies the flow rate of shear layered fluid by reducing the critical Reynolds number. Indeed, the influence of odd viscosity on the shear mode is comparatively very less than that on the surface mode.  

\section*{Conflict of Interest}
I have no conflict of interest. 

\section*{Declaration of competing interest}
The author declares that he has no known competing financial interests or personal relationships that could have appeared to influence the work reported in this paper.

\section*{Credit authorship contribution statement}
\noindent Md. Mouzakkir Hossain: Conceptualization, Methodology, Software, Writing - original draft, Validation, Formal analysis, Investigation.
\\
Sukhendu Ghosh: Conceptualization, review \& editing, Software, review \& editing.
\\
Harekrushna Behera: Conceptualization, Define Problem, Methodology, Supervision,  Software, review \& editing.

\bibliographystyle{unsrtnat}
\bibliography{REF}
\end{document}